\documentclass{article}
\usepackage[english,frenchb]{babel}
\usepackage[pdftex]{graphicx}
\title{A new class of curves of rational \mbox{B-spline} type }
\author{
Mohamed ALLAOUI\footnote{Unit\'e de 
Math\'ematiques IMSP/UAC, Email : allamath9@gmail.com}\\
Aur\'elien GOUDJO\footnote{D\'epartement de 
Math\'ematiques FAST/UAC, Email : aurelien.goudjo@uac.bj}\\
}
\date{}
\newcommand{\RR}{{\bf I\!\!R}}
\newcommand{\NN}{{\bf I\!\!\!N}}
\newcommand{\ZZ}{{\bf Z\!\!\!\!Z}}

\newcommand{\suite}[1]{{\left( {#1}  \right)}}
\newcommand{\couple}[2]{{\left( {#1} \, , \, {#2} \right)}}
\newcommand{\triplet}[3]{{\left( {#1} \, , \, {#2} \, , \, {#3} \right)}}
\newcommand{\quadruplet}[4]{{\left({#1}\, ,\,{#2}\, ,\,{#3}\, ,\,{#4} \right)}}

\newcommand{\implique}{ \, \Rightarrow \,}
\newcommand{\equivaut}{ \, \Leftrightarrow \,}

\newcommand{\module}[1]{\left\vert {#1} \right\vert}
\newcommand{\Preuve}{\noindent{\bf \underline{Proof} } \\}

\newcommand{\Bezier}[3]{{{}^{#3}}{\bf B}_{#1}^{#2}}
\newcommand{\Bspline}[3]{{{}^{#3}}{\bf G}_{#1}^{#2}}
\newcommand{\calH}[2]{{\cal H}\left( \left[{#1}\, , \, {#2}\right] \right)}
\newcommand{\Support}[1]{{{\bf supp} \, #1}}
\newcommand{\segment}[2]{{\left[{#1}\, , \, {#2}\right]}}
\newcommand{\osegment}[2]{{\left({#1}\, , \, {#2}\right)}}

\newtheorem{definition}{\bf Definition}[section]

\newtheorem{lemme}{\bf Lemma}[section]

\newtheorem{proposition}{\bf Proposition}[section]
\newtheorem{corollaire}{\bf Corollary}[section]
\newtheorem{remarque}{\bf Remark}[section]
\newtheorem{conjecture}{\bf Conjecture}[section]
\newtheorem{castest}{\bf Illustration}[section]

\begin{document}
\maketitle

\begin{abstract}
Une nouvelle classe de param\'etrisation rationnelle a \'et\'e d\'evelopp\'ee et on l'a 
utilis\'ee pour g\'en\'erer une nouvelle famille de fonctions \mbox{B-splines} rationnelles 
$\displaystyle{\suite{\Bspline{i}{k}{\alpha}}_{i=0}^{k}}$
qui d\'epend d'un indice $\alpha \in \osegment{-\infty}{0}\cup  \osegment{1}{+\infty}$.
Cette famille de fonctions v\'erifie entre autres,
les propri\'et\'es de positivit\'e, de partition de l'unit\'e 
et, pour un d\'egr\'e $k$ donn\'e, constitue une v\'eritable base d'approximation de 
fonctions continues. On perd cependant la r\'egularit\'e optimale classique li\'ee \`a la 
multiplicit\'e des n{\oe}uds, ce que l'on r\'ecup\`ere dans le  cas asymptotique, 
lorsque $\alpha\to \infty$.

Les courbes de type \mbox{B-splines} associ\'ees v\'erifient les propri\'et\'es traditionnelles
notamment celle d'enveloppe convexe et l'on voit appara{\^i}tre une certaine
"sym\'etrie conjugu\'ee" li\'ee \`a $\alpha$.

Le cas des vecteurs n{\oe}uds ouverts sans  n{\oe}ud int\'erieur conduit 
\`a une nouvelle famille de courbes de B\'ezier rationnelles qui fera, s\'epar\'ement,
l'objet d'une analyse approfondie.

\end{abstract}
\noindent
\textbf{Mots cl\'es} : Vecteur n{\oe}ud, Fonctions \mbox{B-splines} rationnelles, 
Relation de Cox de-Boor,  Algorithme de de-Boor, Graphisme Informatique.

\selectlanguage{english}

\begin{abstract}
A new class of rational parametrization has been developed and
it was used to generate a new family of rational functions \mbox{B-splines}
$\displaystyle{\suite{\Bspline{i}{k}{\alpha}}_{i=0}^{k}}$
which depends on an index $\alpha \in (-\infty,0)\cup (1,+\infty)$.
This family of functions verifies, among other things, the properties of positivity, of
partition of the unit and,  for a given degree $k$, constitutes a true basis
approximation of continuous functions. We loose, however, the regularity
classical optimal linked to the multiplicity of nodes, which we recover
in the asymptotic case, when $\alpha \longrightarrow \infty$.
The associated \mbox{B-splines} curves verify the traditional properties
particularly that of a convex hull and we see a
certain "conjugated symmetry" related to $\alpha$.
The case of open knot vectors without an inner node leads to a
new family of rational Bezier curves that will be separately,
object of in-depth analysis.
\end{abstract}
\noindent
\textbf{Keywords} :  Knot vector, Rational  \mbox{B-splines} functions , 
Cox- de Boor recursion, de-Boor Algorithm, Computer Graphics.

\tableofcontents

\section{Introduction}
A standard B-spline curve $G$ of degree $k \in \NN^*$
in $\RR^d$ with \mbox{$d \in \NN^*$}, 
 \mbox{$1\leq d \leq 3$} is defined by a polynomial basis 
$\suite{\Bspline{i}{k}{}}_{i=0}^{n}$ on a parametrization space
$ [a \, , \, b]$ subdivided by a knot vector
$U=\suite{t_i}_{i=0}^{m}$ with \mbox{$m=n+k+1$}.
The basis $\suite{\Bspline{i}{k}{}}_{i=0}^{n}$ is given by 
the recurrence relation of Cox/de Boor~\cite{rogers2001} as follows:
\begin{equation} \label{BasBsplineP}
\begin{array}{rcl}
 \Bspline{i}{0}{}(x)&=&\left\lbrace
   {
   \begin{array}{ll}
    1   
    &\textrm{if }  t_{i} \leq x < t_{i+1} \textrm{ for } i=0, \ldots, m-1 \\
   0 &\textrm{otherwise }
   \end{array}
   }
   \right. \\

   \Bspline{i}{k}{}(x)&=& w_{i}^{k}(x)\Bspline{i}{k-1}{}(x) +
    {\left( 1 - w_{i+1}^{k}(x)\right)}\Bspline{i+1}{k-1}{}(x)
    \\
    w_{i}^{k}(x)&=&\left\lbrace
   {
   \begin{array}{ll}
   \displaystyle{\frac{x-t_{i}}{t_{i+k}-t_{i}}} 
    &\textrm{if }  t_{i} \leq x < t_{i+k} \textrm{ for } i=0, \ldots, n \\
   0 &\textrm{otherwise }
   \end{array}
   }
   \right.
   \\
\end{array}
\end{equation}

If $\displaystyle{\suite{d_i}_{i=0}^{n}}$ are the control polygon vertices of 
$G$, $d_i \in \RR^d$ for all $i$ then
$$
G(x) = \displaystyle{\sum_{i=0}^{n} d_i\, \Bspline{i}{k}{} (x)} \, ,\,
\forall x \in [a \, , \, b]
$$

Likewise we have the rational B-spline basis 
$\suite{ R_i }_{i=0}^{n}$
of degree $k \in \NN^*$ associated to the knot vector $U$ and the weight
vector $W=\suite{\omega_i}_{i=0}^{n}$ which can be defined by 
$$
 R_i (x) =
\displaystyle{
\frac{\omega_i \Bspline{i}{n}{}(x)}{
\displaystyle{\sum_{j=0}^{n}\omega_j \Bspline{j}{n}{}(x)}
}
}
$$
where $\omega_i >0, \, \forall i=0, \ldots , n$.

We can then define the rational B-spline curves replacing the polynomial basis
by the rational basis~\cite{rogers2001},\cite{versprille1975}.

One has to notice that 
$\displaystyle{w_{i}^{k}(x)=\varphi\triplet{x}{t_{i}}{t_{i+k}}}$
where $\varphi$ is a real function defined on $\RR^3$ satisfying the following
properties:
\begin{enumerate}
\item 
$\displaystyle{\varphi\triplet{x}{a}{b} \in [0 \, , \, 1)}$ for all
$\triplet{x}{a}{b}\in \RR^3$
\item 
For all $a, b \in \RR$ such that $a<b$ the function
$\displaystyle{x\in \RR \mapsto \varphi\triplet{x}{a}{b}}$ 
is continuous, strictly increasing on  $[a \, , \, b)$ and we have:
\begin{itemize}
\item 
$\displaystyle{ \varphi\triplet{x}{a}{b} =0}$ for all $ x\notin (a \, , \, b)$ 
\item 
$\displaystyle{\lim_{x\to b^-} \varphi\triplet{x}{a}{b} =1}$ 
\end{itemize}
\end{enumerate}

The aim of this work is to maintain these properties while imposing that for all
 $a, b \in \RR$ such that $a<b$, the function
$\displaystyle{x\in \RR \mapsto \varphi\triplet{x}{a}{b}}$ 
is homographic in order to build a natural B-spline basis composed of rational functions.



\section{A class of rational parametrization\label{SecParametriser}}
\subsection{Definition}
The targeted class of parametrization is based on the following lemma which
gives the foundation of a new class of curves of rational B-spline type.

\begin{lemme}
Let $a,\, b \in \RR$ verifying $a < b$. There exists a family $\calH{a}{b}$
of homographic functions $f$ strictly increasing on $[a\, ,\, b]$ such that 
$f(a)=0$ and $f(b)=1$.

More precisely, for all $f\in \calH{a}{b}$ there exists
a unique\\
\mbox{$\alpha \in (-\infty \, , \, 0) \cup (1  \, , \,  \infty) $} such that
$$
f(x) =\displaystyle{\frac{\alpha (x-a)}{x+(\alpha - 1)b-\alpha a}}, \quad 
\forall x \in [a\, ,\, b]
$$
\end{lemme}

\Preuve 
\emph{( Existence )} 
Since $f$ is homographic with $f(a)=0$ there exists $\alpha \neq 0$ and 
$c\in \RR \backslash \{ -a,\, -b\}$  such that for all $x \in [a\, ,\, b]$ we get:
$\displaystyle{f(x) = \frac{\alpha (x-a)}{x+c}}$. 
As $f(b)=1$ then   
$\displaystyle{1 = \frac{\alpha (b-a)}{b+c}}$. 
This leads to $c=(\alpha - 1)b-\alpha a$. Using the fact that $c\notin \{ -a, \, -b\}$
then  we have $\alpha \notin \{ 0, \, 1\}$. The strict increase of $f$ yields $\alpha (\alpha -1) > 0$,
therefore $\alpha \in (-\infty \, , \, 0) \cup (1  \, , \,  \infty)$.

We then write
 $$
\calH{a}{b} =
\left\lbrace
\displaystyle{f_\alpha } \vert \,
\displaystyle{f_\alpha (x) }=
\displaystyle{\frac{\alpha (x-a)}{x+(\alpha - 1)b-\alpha a}}, \,
 \alpha \in (-\infty \, , \, 0) \cup (1  \, , \,  \infty),
\, x \in  [a\, ,\, b]
\right\rbrace
$$

\emph{( Uniqueness )} 

Let $\alpha, \, \beta \in (-\infty \, , \, 0) \cup (1  \, , \,  \infty)$ and
$f_\alpha, \, f_\beta \in \calH{a}{b}$ corresponding
$$
\begin{array}{lcl}
f_\alpha = f_\beta
& \equivaut&
f_\alpha(x) =f_\beta(x) \quad \forall x\in  [a\, ,\, b]\\
&\implique&
\displaystyle{\frac{\alpha (x-a)}{x+(\alpha - 1)b-\alpha a}}=
\displaystyle{\frac{\beta (x-a)}{x+(\beta - 1)b-\beta a}}
\quad \forall x\in  [a\, ,\, b]\\
&\implique&
\displaystyle{(\alpha-\beta) (x-b)}=0
\quad \forall x\in  [a\, ,\, b]\\
&\implique&
\displaystyle{(\alpha-\beta) (x-b)}=0
\quad \forall x\in  [a\, ,\, b[\\
&\implique&
\displaystyle{\alpha=\beta}
\end{array}
$$

\begin{remarque}
 Let  $x \in [a\, ,\, b]$ and
  $\alpha \in (-\infty \, , \, 0) \cup (1  \, , \,  \infty)$.\\
  One has
 $D=x+(\alpha - 1)b-\alpha a \neq 0$.
 
 Indeed, Observing that
 $D=x- b+\alpha(b- a) =x- a+(\alpha - 1)(b- a)$,
 we have $ (\alpha-1)(b-a) \leq D \leq \alpha (b-a) $. One can then deduce that
 $$
 \left\lbrace
 \begin{array}{ll}
 0 < \alpha(\alpha-1)(b-a) \leq \alpha D \leq \alpha^2 (b-a)& \textrm{ if } \alpha > 1\\
 0 < \alpha^2(b-a) \leq \alpha D \leq \alpha(\alpha-1) (b-a)& \textrm{ if } \alpha <0
 \end{array}
 \right.
 $$
Thus $D\neq 0 \, \forall x \in  [a\, ,\, b]$.
\end{remarque}

\begin{remarque} \label{RemValFalfa}
Let $\alpha \in (-\infty \, , \, 0) \cup (1  \, , \,  \infty)$ and $a<b$.\\
 $f_\alpha \in \calH{a}{b}$ is continuous and  strictly increasing
on $[a\, ,\, b]$ with  \mbox{$f_\alpha([a\, ,\, b])= [0\, ,\, 1]$}.

Moreover for $\lambda \in [0\, ,\, 1]$ and for $x = a+\lambda(b-a) \in [a\, ,\, b]$ 
we have
$$
f_\alpha(x)=\displaystyle{\frac{\lambda\alpha}{\lambda+\alpha -1} \in  [0\, ,\, 1]}
$$

We thus obtain the classical case as an asymptotic situation; indeed:
$$
\displaystyle{\lim_{\module{\alpha} \to \infty}
f_\alpha(x)=\lambda=\frac{x-a}{b-a} }
$$
\end{remarque}

\begin{remarque} \label{RemAlfaConjug}
Let $\alpha \in (-\infty \, , \, 0) \cup (1  \, , \,  \infty)$ and $a<b$.\\
Let $f_\alpha, f_{1-\alpha} \in \calH{a}{b}$. For all
 $x \in [a\, ,\, b]$, we have:
$$
\begin{array}{rcl} 
\displaystyle{ f_\alpha(a+b-x)}
&=&
\displaystyle{ 1-f_{1-\alpha}(x)}
\\
\displaystyle{ f_\alpha(x)}
&=&
\displaystyle{ 1-f_{1-\alpha}(a+b-x)}
\\
\end{array}
$$

Indeed, we observe that 
$x \in [a\, ,\, b]$ is equivalent to $a+b-x \in [a\, ,\, b]$ and
$$
\begin{array}{rcl} 
\displaystyle{ f_\alpha(a+b-x)}
&=&
\displaystyle{ 
\frac{
\alpha{\left( b-x\right)}}
{-x +\alpha b +{\left(1 -\alpha \right)}a}
}
\\
&=&
\displaystyle{ 
-\frac{
\alpha{\left( b-x\right)}}
{x -\alpha b -{\left(1 -\alpha \right)}a}
}
\\
\displaystyle{ 1-f_{1-\alpha}(x)}
&=&
\displaystyle{ 1-
\frac{{\left( 1-\alpha \right)}{\left(x-a\right)}}
{x +{\left[{\left(1 -\alpha \right)}-1 \right]} b -{\left(1 -\alpha \right)}a}
}
\\
&=&
\displaystyle{ 1-
\frac{{\left( 1-\alpha \right)}{\left(x-a\right)}}
{x -\alpha b -{\left(1 -\alpha \right)}a}
}
\\
&=&
\displaystyle{ 
-\frac{
\alpha{\left( b-x\right)}}
{x -\alpha b -{\left(1 -\alpha \right)}a}
}
\\
&=&
\displaystyle{ f_\alpha(a+b-x)}
\\
\end{array}
$$

We obtain the second relation by a simple change of variables.

From now on, we say that $\alpha$ and  $1-\alpha$ are conjugated.
\end{remarque}

\begin{definition}
Let $\alpha \in (-\infty \, , \, 0) \cup (1  \, , \,  \infty)$.
A parametrization of index $\alpha$ is any real function $\varphi_\alpha $ defined for all
$\triplet{x}{a}{b}\in \RR^3$ by 
$$
\varphi_\alpha \triplet{x}{a}{b} =
\left\lbrace
\begin{array}{ll}
f_\alpha (x) & \textrm{if }
a\leq x < b \textrm{ with } 
f_\alpha \in \calH{a}{b}
\\
0 &\textrm{otherwise}
\end{array}
\right.
$$

\end{definition}

\subsection{Properties of the parametrization}

\begin{proposition} \label{PropParam1}
Let $\alpha \in (-\infty \, , \, 0) \cup (1  \, , \,  \infty)$
and $\varphi_\alpha$ the associated parametrization. 
Let $T$ be an affine and bijective function of $\RR$.
The following properties hold:
For all $\triplet{x}{a}{b}\in \RR^3$ 
\begin{enumerate}
\item 
$\displaystyle{0\leq \varphi_\alpha \triplet{x}{a}{b} <1}$
\item 
If $T$ is strictly increasing then
$$
\displaystyle{
 \varphi_\alpha \triplet{T(x)}{T(a)}{T(b)} =
  \varphi_\alpha \triplet{x}{a}{b} 
}
$$
\item 
If $T$ is strictly decreasing then
$$
\displaystyle{
 \varphi_\alpha \triplet{T(x)}{T(b)}{T(a)} =
  1-\varphi_{1-\alpha} \triplet{x}{a}{b} 
}
$$
\end{enumerate}
\end{proposition}

\Preuve
Let $T$ be an affine and bijective function of $\RR$.
There exists $\couple{\lambda}{\delta}\in \RR^* \times \RR$ such that, for all $x \in \RR$,
we have
$\displaystyle{T(x) = \lambda x + \delta}$.

Let $\triplet{x}{a}{b}\in \RR^3$

\begin{enumerate}
\item
If $T$ is strictly increasing and $a<x<b$ then
$T(a)<T(x)<T(b)$
$$
\begin{array}{rcl}
\displaystyle{
 \varphi_\alpha \triplet{T(x)}{T(a)}{T(b)} 
}
&=&
\displaystyle{
f_{\alpha}{(T(x))} 
\textrm{ with }f_{\alpha} \in \calH{T(a)}{T(b)}
}
\\
&=&
\displaystyle{
\frac{\alpha{\left[T(x)-T(a)\right]}}
{T(x) +(\alpha-1)T(b) -\alpha T(a)} 
}
\\
&=&
\displaystyle{
\frac{\alpha{\left[(\lambda x + \delta)-(\lambda a + \delta)\right]}}
{(\lambda x + \delta) +(\alpha-1)(\lambda b + \delta) -\alpha (\lambda a + \delta)} 
}
\\
&=&
\displaystyle{
\frac{\alpha\lambda{\left[x-a\right]}}
{\lambda{\left[x +(\alpha-1)b -\alpha a\right]}} 
}
\\
&=&
\displaystyle{
\frac{\alpha{\left[x-a\right]}}
{x +(\alpha-1)b -\alpha a} 
}
\\
&=&
\displaystyle{
g_{\alpha}{(x)} 
\textrm{ with }g_{\alpha} \in \calH{a}{b}
}
\\
&=&
\displaystyle{
 \varphi_\alpha \triplet{x}{a}{b} 
}
\\
\end{array}
$$

\item
If $T$ is strictly decreasing and $a<x<b$ then
$T(b)<T(x)<T(a)$
$$
\begin{array}{rcl}
\displaystyle{
 \varphi_\alpha \triplet{T(x)}{T(b)}{T(a)} 
}
&=&
\displaystyle{
f_{\alpha}{(T(x))} 
\textrm{ with }f_{\alpha} \in \calH{T(b)}{T(a)}
}
\\
&=&
\displaystyle{
\frac{\alpha{\left[T(x)-T(b)\right]}}
{T(x) +(\alpha-1)T(a) -\alpha T(b)} 
}
\\
&=&
\displaystyle{
\frac{\alpha{\left[(\lambda x + \delta)-(\lambda b + \delta)\right]}}
{(\lambda x + \delta) +(\alpha-1)(\lambda a + \delta) -\alpha (\lambda b + \delta)} 
}
\\
&=&
\displaystyle{
\frac{\alpha\lambda{\left[x-b\right]}}
{\lambda{\left[x +(\alpha-1)a -\alpha b\right]}} 
}
\\
&=&
\displaystyle{
\frac{\alpha{\left[x-b\right]}}
{x +(\alpha-1)a -\alpha b} 
}
\\
&=&
\displaystyle{
\frac{\alpha{\left[x-b\right]}}
{x  -\alpha b-(1-\alpha)a} 
}
\\
&=&
\displaystyle{
1-\frac{(1-\alpha){\left[x-a\right]}}
{x  -\alpha b-(1-\alpha)a} 
}
\\
&=&
\displaystyle{
1- g_{1-\alpha}{(x)} 
\textrm{ with }g_{1-\alpha} \in \calH{a}{b}
}
\\
&=&
\displaystyle{
 1-\varphi_{1-\alpha} \triplet{x}{a}{b} 
}
\\
\end{array}
$$

\end{enumerate}

\begin{corollaire}\label{CorolParam1}
Let \mbox{$\alpha \in (-\infty \, , \, 0) \cup (1  \, , \,  \infty)$}
and $\varphi_\alpha$ be the associated para\-metri\-zation. 
Let $a,b \in \RR$ such that $a<b$. Let  $a<t_{1}<t_{2}<b$. 
For all $x\in [a\, , \, b]$, we have
$$
\displaystyle{
 \varphi_\alpha \triplet{a+b-x}{t_{1}}{t_{2}} =
 1- \varphi_{1- \alpha} \triplet{x}{a+b-t_{2}}{a+b-t_{1}} 
 }
$$
\end{corollaire}

\Preuve
We apply proposition \ref{PropParam1} by taking $\displaystyle{T(x) = a+b-x}$ on $\RR$.
We observe that $T$ is strictly decreasing and verifies $T\circ T (x)=x$ for all
$x\in \RR$. This gives the result.

\begin{remarque}\label{RemRegularParam}
Let $a, b \in \RR$ such that $a<b$. The function
$$
\displaystyle{x \in \RR \mapsto \varphi_\alpha \triplet{x}{a}{b}}
$$
is continuous on $\RR\backslash \lbrace b \rbrace$
and we have:
\begin{equation}\label{EqRegularParam1}
\begin{array}{rcl}
\displaystyle{\varphi_\alpha \triplet{a}{a}{b}}
&=& 0\\
\displaystyle{\lim_{x\to b^-}\varphi_\alpha \triplet{x}{a}{b}}
&=& 1\\
\displaystyle{\lim_{x\to b^+}\varphi_\alpha \triplet{x}{a}{b}}
&=& 0\\
\end{array}
\end{equation}

On the other hand, this function is of class ${\cal C}^\infty$ on 
$(-\infty\, , \, a)\cup (a\, , \, b)\cup (b \, , \, \infty)$
and one has:
\begin{equation}\label{EqRegularParam2}
\begin{array}{rcl}
\displaystyle{\lim_{x\to a^-}\frac{d \varphi_\alpha}{dx} \triplet{x}{a}{b}}
&=& 0\\
\displaystyle{\lim_{x\to a^+}\frac{d \varphi_\alpha}{dx} \triplet{x}{a}{b}}
&=& 
\displaystyle{\frac{\alpha}{\left(\alpha - 1 \right)\left( b- a \right)}}
\\
\displaystyle{\lim_{x\to b^-}\frac{d \varphi_\alpha}{dx} \triplet{x}{a}{b}}
&=& 
\displaystyle{\frac{\alpha -1}{\alpha \left( b- a \right)}}
\\
\displaystyle{\lim_{x\to b^+}\frac{d \varphi_\alpha}{dx} \triplet{x}{a}{b}}
&=& 0\\
\end{array}
\end{equation}
\end{remarque}


\begin{castest}

The figures \ref{figCourbePhi1} and  \ref{figCourbePhi2}  illustrate 
$\displaystyle{\varphi_{\alpha}\triplet{x}{0}{1}}$ 
for \mbox{$\displaystyle{x\in (-1\,,\,2)}$} with values of $\alpha$
conjugated respectively. 

We observe that on the subinterval $\displaystyle{(0\,,\,1)}$ 
which is the interior of its support, the function is convex
for  $\displaystyle{\alpha <0}$ and
concave for  $\displaystyle{\alpha >1}$.

The figure \ref{figCourbePhi3} which
illustrates
$\displaystyle{\varphi_{\alpha}\triplet{x}{1}{3}}$ 
for $\displaystyle{x\in (0\,,\,6)}$ confirms the previous observations  and
lets suspect
the symmetrical role that the conjugated $\alpha$ are to play. 
It also shows that the effect of $\alpha$ is crucial in the neighborhood of 
$0$ and of $1$.

\begin{figure}[h!]
\begin{center}
\includegraphics[width=11cm]{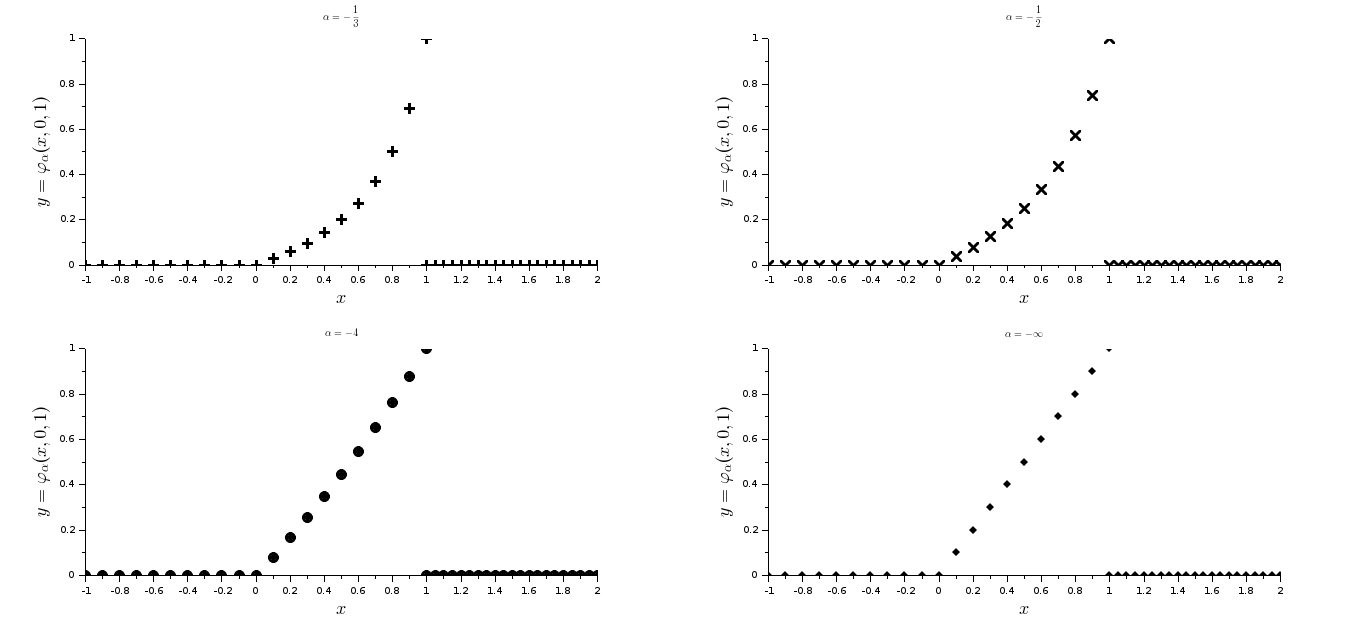}
\caption{The curves of $\displaystyle{\varphi_{\alpha}}$ for $\alpha \in \{-\frac{1}{3},
-\frac{1}{2},-4,\infty \}$}
\label{figCourbePhi1}
\end{center}
\end{figure}

\begin{figure}[h!]
\begin{center}
\includegraphics[width=11cm]{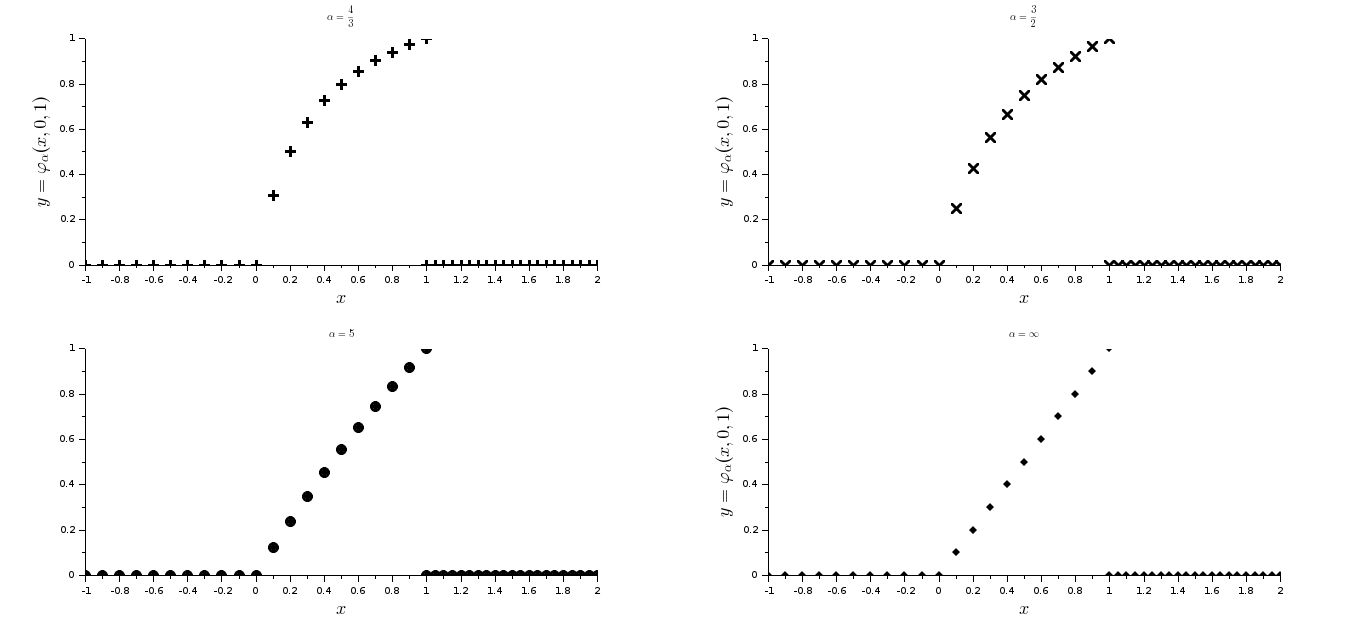}
\caption{The curves $\displaystyle{\varphi_{\alpha}}$ for $\alpha \in \{\frac{4}{3},
\frac{3}{2},5,\infty \}$}
\label{figCourbePhi2}
\end{center}
\end{figure}

\begin{figure}[h!]
\begin{center}
\includegraphics[width=11cm]{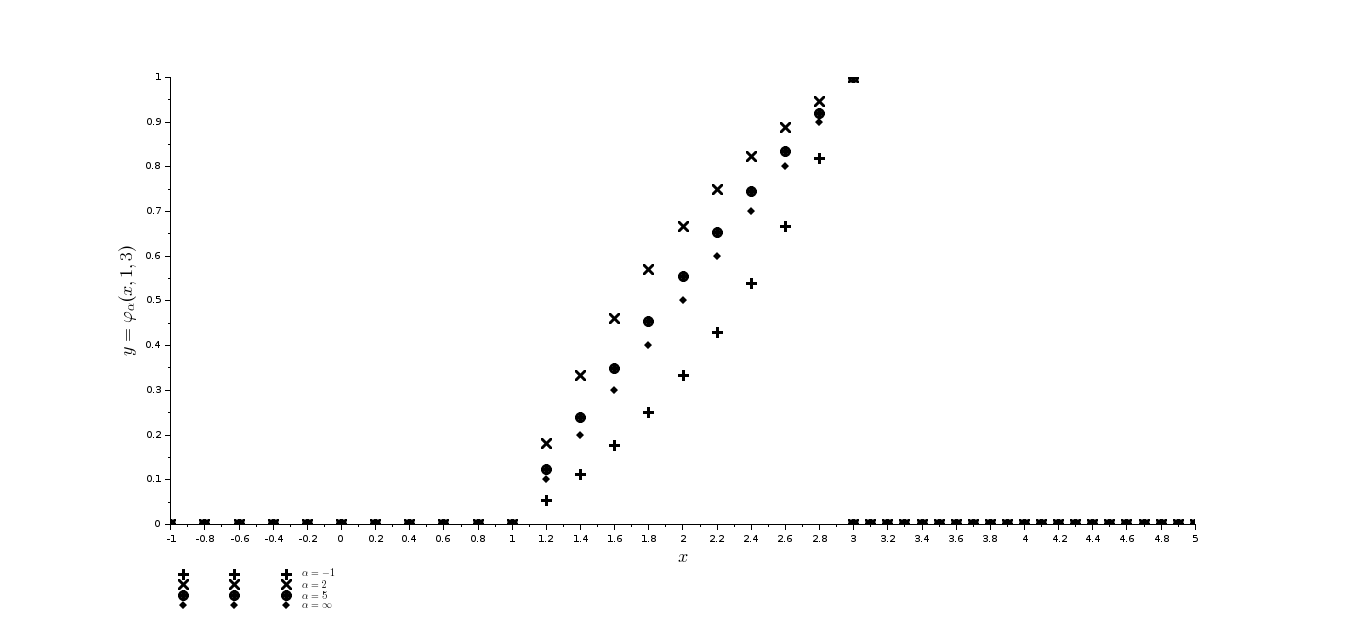}
\caption{Comparaison of $\displaystyle{\varphi_{\alpha}}$ for 
conjugated $\alpha$ and large $\alpha$} 
\label{figCourbePhi3}
\end{center}
\end{figure}

\end{castest}



\section{ New class of rational B-splines basis\label{SecNewClassBspline}}

\subsection{Definitions}

The B-splines curves are part of the family of curves obtained by concatenation of several generated pieces of curves using a family of basic functions of parametrization space $[a\, , \,b]$ subdivised by a knot vector 
$U$ and a set of points $\suite{d_i}_{i=0}^{n}$ of $\RR^d$
called control polygon.

The nature of chosen  knot vector  may strongly influence the properties
of B-spline basis generated as well as the resulting curve.
We must very quickly specify this object.

We follow the definitions of the book of D. F. Rogers   entitled "An Introduction to NURBS with historical perspective" \cite{{rogers2001}}.

\begin{definition}
Let $a, b \in \RR$ such that $a<b$.
A  knot vector  in $[a,b]$ is any increasing sequence 
$U=\suite{t_i}_{i=0}^{m}$ in $[a\, , \,b]$.
\end{definition}

The  knot vectors  fall into two categories:
the open  knot vectors and periodic  knot vectors.
Each category is divided in two variants: uniform and non-uniform.

\begin{definition}
Let $a,b\in\RR$ such that $a<b$ and $m,k\in\NN^*$ 
such that $m>2k$. We consider the  knot vector
 $U=\suite{t_i}_{i=0}^{m}$ such that $t_k=a$ and $t_{m-k}=b$.

\begin{enumerate}
\item
\emph{End nodes:}

The nodes $t_0,t_1,\ldots,t_k$ and the nodes $t_{m-k},t_{m-k+1},\ldots,t_m$ 
are called end nodes.

The nodes $t_{k+1},t_{k+2},\ldots,t_{m-k-1}$ are called interior nodes.

\item
\emph{Open knot vector: }

The knot vector is said to be open if its end nodes coincide; we then have 
$t_0=t_1=\ldots=t_k=a$ and 
$t_{m-k}=t_{m-k+1}=\ldots=t_m=b$.

Otherwise $U$ is said to be periodic.

\item 
\emph{Uniform knot vector : }

$U$ is uniform if its interior nodes are equidistant; that is, 
 there exists $h>0$  such that  $t_{i+1}-t_i=h$ for all $k\leq i\leq m-k-1$.

Otherwise $U$ is non-uniform.

\item 
\emph{Multiple node (multiplicity of a node) :}

Let $p\in \NN^*$ and $t_i$ be a node of $U$.
We say that $t_i$ is a node of multiplicity $p$ if there exists a unique 
$j\in [0, \ldots , m-1 ]\cap\NN$ such that the subsequence
$U_i=\suite{t_{j+l}}_{l=0}^{p-1}$ with $j\leq i \leq j+p-1$ is constant.

If $p>1$, we say that $t_i$ is multiple node.

\item
\emph{Breakpoints:}

The set $\suite{u_i}_{i=0}^{r}$ of distinct nodes of $U=\suite{t_i}_{i=0}^{m}$
constitutes the breakpoints. We have $u_0=t_0 < u_1 < \ldots < u_r=t_m$ and 
there exists a unique sequence of nonnegative integers $p=\suite{p_i}_{i=0}^{r}$
such that for all $i=0, \ldots , r$, $u_i$ is of multiplicity $p_i$.

We shall remark that $\displaystyle{\sum_{i=0}^{r} p_i = m+1}$.
On the other hand, these nodes define the different segments of studied curves
and the interior breakpoints define the transition between its segments.
\item
\emph{Symmetrical knot vector:}

$U=\suite{t_i}_{i=0}^{m}$ is a symmetrical knot vector if for all
$i=0, \ldots , m$, $t_{m-i}=t_{0}+t_{m}-t_{i}$.

\end{enumerate}
\end{definition}

\begin{definition}
Let $a, b \in \RR$ such that $a<b$ and $m,n, k \in \NN^*$ such that $n\geq k$ and $m=n+k+1$.
Let \mbox{$\alpha \in (-\infty \, , \,0 ) \cup  (1 \, , \,\infty )$} and
$\displaystyle{\varphi_\alpha}$ the parametrization of index $\alpha$.
Let $U=\suite{t_i}_{i=0}^{m}$ be a knot vector of the interval $ [ a\, , \,b ]$.

A B-spline basis of index $\alpha$ and of degree $k$ on the node vector $U$ is 
the real functions $\suite{\Bspline{i}{k}{\alpha}}_{i=0}^{n}$ defined by the recurrence relation:
\begin{equation} \label{BasBsplineR}
\begin{array}{rcl}
\displaystyle{
 \Bspline{i}{0}{\alpha}(x)
 }
 &=&
 \left\lbrace
   {
   \begin{array}{ll}
    1   
    &\textrm{if }  t_{i} \leq x < t_{i+1} \textrm{ for } i=0, \ldots, m-1 \\
   0 &\textrm{otherwise }
   \end{array}
   }
   \right. \\
\displaystyle{
   \Bspline{i}{k}{\alpha}(x)
}
&=& 
\displaystyle{
w_{i}^{k}(x)\Bspline{i}{k-1}{\alpha}(x) +
    {\left( 1 - w_{i+1}^{k}(x)\right)}\Bspline{i+1}{k-1}{\alpha}(x)
    }
    \\
\displaystyle{
    w_{i}^{k}(x)
    }
    &=&
\displaystyle{
    \varphi_\alpha \triplet{x}{t_{i}}{t_{i+k}}
    }
   \\
\end{array}
\end{equation}

This relation is said to be of Cox/de Boor.

\end{definition}

\begin{definition}
Let $a, b \in \RR$ such that $a<b$.
Let $m,n, k \in \NN^*$ such that $n> k$ and $m=n+k+1$.
Let \mbox{$\alpha \in (-\infty \, , \,0 ) \cup  (1 \, , \,\infty )$}.
Let $U=\suite{t_i}_{i=0}^{m}$ be a knot vector of interval $ [ a\, , \,b ]$.
Let $d \in \NN^*$ such that $d \leq 3$, 
and $\displaystyle{\Pi=\suite{d_i}_{i=0}^{n}\subset \RR^d}$.

Let $\suite{\Bspline{i}{k}{\alpha}}_{i=0}^{n}$ be the B-spline basis of index 
$\alpha$, of degree $k$ and of knot  vector $U$.

A B-spline curve of index $\alpha$, with knot vector $U$ and control points 
 $\displaystyle{\suite{d_i}_{i=0}^{n}}$ is the $\RR^d$ valued function 
$G_\alpha$ defined by:
$$
\displaystyle{x \in \segment{t_{0}}{t_{m}}
\mapsto
G_\alpha (x) =\sum_{i=0}^{n} d_{i}\Bspline{i}{k}{\alpha} (x)
}
$$

$\Pi$ is called control polygon of the curve $G_\alpha$.

\end{definition}

\subsection{Fundamental properties of the next class of basis }

\begin{proposition}

Let $m, k, n\in \NN^*$ such that  \mbox{$n \geq k$} and $m=n+ k+1$.
Let $\displaystyle{U=\suite{t_i}_{i=0}^{m}}$ be a knot vector  and  
 \mbox{$\alpha \in (-\infty \, , \, 0) \cup (1  \, , \,  \infty)$}. 

The rational B-spline basis of index $\alpha$ with knot vector 
$U$ and of degree $k$, $\suite{\Bspline{i}{k}{\alpha}}_{i=0}^{n}$, verifies the following 
properties: 
\begin{enumerate}
\item \emph{Local support property: }

For all $x \notin (t_{i}\, ,\, t_{i+k+1})$,  $\Bspline{i}{k}{\alpha} (x) = 0$

\item \emph{Positivity property: }

For all $i =0, \ldots , n$ and $x \in (t_{i}\, ,\, t_{i+k+1})$,  
$\Bspline{i}{k}{\alpha} (x) > 0$

\item \emph{Unit partition property:}

For all $j$ such that $t_{j} < t_{j+1}$, for all $x \in [t_{j}\, ,\, t_{j+1})$,  we have
$$
\displaystyle{ 
\sum_{i=0}^{n}\Bspline{i}{k}{\alpha}  (x) 
=\sum_{i=j-k}^{j}\Bspline{i}{k}{\alpha}  (x)  = 1
}
$$

\item \emph{Symmetry property:}

If $U$ is a symmetrical knot vector then for 
all $x \in [t_{0}\, ,\, t_{m}]$ and $i=0, \ldots , n$ we have
$$
\Bspline{i}{k}{\alpha}  (t_{0}+t_{m}-x)  = \Bspline{n-i}{k}{1-\alpha}  (x) 
$$

\end{enumerate}

\end{proposition}

\Preuve
Let $\alpha \in (-\infty \, , \, 0) \cup (1  \, , \,  \infty)$ and
$\displaystyle{\varphi_\alpha}$  be the parametrization of index $\alpha$.

We will proceed by recurrence on $k$.
\begin{enumerate}
\item (\emph{Local support and Positivity: })
\begin{itemize}
\item
For $k=0$, we have by definition: for all $i=0, \ldots , m-1$ 
$$
 \Bspline{i}{0}{\alpha}(x)=
 \left\lbrace
   \begin{array}{ll}
    1   
    &\textrm{if }  t_{i} \leq x < t_{i+1} \textrm{ for } i=0, \ldots, m-1 \\
   0 &\textrm{otherwise }
   \end{array}
  \right. 
$$
hence we have
$$
\begin{array}{ll}
\displaystyle{\Bspline{i}{k}{\alpha}(x)}=0
&
\textrm{ if } 
\displaystyle{x\notin (t_{i} \, , \, t_{i+k+1})}\\
\displaystyle{\Bspline{i}{k}{\alpha}(x)}>0
&
\textrm{ if } 
\displaystyle{x\in (t_{i} \, , \, t_{i+k+1})\neq \emptyset}\\
\end{array}
$$
\item
Let $k> 0$ and assume that for all $0\leq j <k$ we have
$$
\begin{array}{ll}
\displaystyle{\Bspline{i}{j}{\alpha}(x)}=0
&
\textrm{ if } 
\displaystyle{x\notin (t_{i} \, , \, t_{i+j+1})}\\
\displaystyle{\Bspline{i}{j}{\alpha}(x)}>0
&
\textrm{ if } 
\displaystyle{x\in (t_{i} \, , \, t_{i+j+1})\neq \emptyset}\\
\end{array}
$$

By definition we have
$$
\displaystyle{
   \Bspline{i}{k}{\alpha}(x)= w_{i}^{k}(x)\Bspline{i}{k-1}{\alpha}(x) +
    {\left( 1 - w_{i+1}^{k}(x)\right)}\Bspline{i+1}{k-1}{\alpha}(x)
    }
$$
with
$$
\begin{array}{l}
\left\lbrace
\begin{array}{ll}
\displaystyle{\Bspline{i}{k-1}{\alpha}(x)}=0
&
\textrm{ if } 
\displaystyle{x\notin (t_{i} \, , \, t_{i+k})}\\
\displaystyle{\Bspline{i}{k-1}{\alpha}(x)}>0
&
\textrm{ if } 
\displaystyle{x\in (t_{i} \, , \, t_{i+k})\neq \emptyset}\\
\end{array}
\right.
\\
\textrm{ and }\\
\left\lbrace
\begin{array}{ll}
\displaystyle{\Bspline{i+1}{k-1}{\alpha}(x)}=0
&
\textrm{ if } 
\displaystyle{x\notin (t_{i+1} \, , \, t_{i+k+1})}\\
\displaystyle{\Bspline{i+1}{k-1}{\alpha}(x)}>0
&
\textrm{ if } 
\displaystyle{x\in (t_{i+1} \, , \, t_{i+k+1})\neq \emptyset}\\
\end{array}
\right.
\\
\end{array}
$$

\begin{itemize}
\item
Let
$\displaystyle{x\notin (t_{i} \, , \, t_{i+k+1})=
(t_{i} \, , \, t_{i+k}) \cup (t_{i+1} \, , \, t_{i+k+1})
}
$. 
Then we have
$\displaystyle{x\notin (t_{i} \, , \, t_{i+k})}$
and
$\displaystyle{x\notin  (t_{i+1} \, , \, t_{i+k+1})}$ which gives
$\displaystyle{\Bspline{i}{k-1}{\alpha}(x)}=0$, 
$\displaystyle{\Bspline{i+1}{k-1}{\alpha}(x)}=0$
and $\displaystyle{\Bspline{i}{k}{\alpha}(x)}=0$

\item
Let
$\displaystyle{x\in (t_{i} \, , \, t_{i+k+1})=
(t_{i} \, , \, t_{i+k}) \cup (t_{i+1} \, , \, t_{i+k+1}) \neq \emptyset
}
$. 
Then we have 
$\displaystyle{x\in (t_{i} \, , \, t_{i+k}) \neq \emptyset}$
or
$\displaystyle{x\in (t_{i+1} \, , \, t_{i+k+1}) \neq \emptyset}$.

If $\displaystyle{x\in (t_{i} \, , \, t_{i+k}) \neq \emptyset}$ then one has
$\displaystyle{\Bspline{i}{k-1}{\alpha}(x)}>0$ and
$\displaystyle{\Bspline{i+1}{k-1}{\alpha}(x)}\geq 0$.
But from proposition \ref{PropParam1} we have
$$
\left\lbrace
\begin{array}{l}
\displaystyle{w_{i}^{k}(x)=\varphi_\alpha\triplet{x}{t_{i}}{t_{i+k}}\in (0\, , \, 1)}
\\
\displaystyle{w_{i+1}^{k}(x)=\varphi_\alpha\triplet{x}{t_{i+1}}{t_{i+k+1}}\geq 0}
\\
\end{array}
\right.
$$

We conclude that
$$
\displaystyle{
\Bspline{i}{k}{\alpha}(x)
\geq  w_{i}^{k}(x)\Bspline{i}{k-1}{\alpha}(x)
>0
}
$$

Similarly if 
$\displaystyle{x\in (t_{i+1} \, , \, t_{i+k+1}) \neq \emptyset}$ then
$\displaystyle{\Bspline{i}{k-1}{\alpha}(x)}\geq 0$ and
$\displaystyle{\Bspline{i+1}{k-1}{\alpha}(x)}> 0$.
By using once more proposition \ref{PropParam1} we have 
$$
\left\lbrace
\begin{array}{l}
\displaystyle{w_{i}^{k}(x)=\varphi_\alpha\triplet{x}{t_{i}}{t_{i+k}}\geq 0}
\\
\displaystyle{w_{i+1}^{k}(x)=\varphi_\alpha\triplet{x}{t_{i+1}}{t_{i+k+1}}\in (0\, , \, 1)}
\\
\end{array}
\right.
$$

We then conclude that
$$\displaystyle{\Bspline{i}{k}{\alpha}(x)
\geq
{\left( 1 - w_{i+1}^{k}(x)\right)}\Bspline{i+1}{k-1}{\alpha}(x)
>0
}
$$

Hence 
$\displaystyle{\Bspline{i}{k}{\alpha}(x)>0}$ if 
$\displaystyle{x \in (t_{i}\, , \, t_{i+k+1})}$

\end{itemize}
\end{itemize}

\item (\emph{Unit partition})

Let $m, k,  n \in \NN^*$ such that $n>k$ and $m=n+k+1$. 
\begin{itemize}
\item
Let $j$ such that $t_{j}< t_{j+1}$. Let $i=0,\ldots , n$.
$$
[t_{i}\, , \, t_{i+k+1}) \cap [t_{j}\, , \, t_{j+1}) \neq \emptyset
\equivaut
j-k \leq i \leq j
$$

\item
Let $x \in [t_{j}\, , \, t_{j+1})$ and $i=0,\ldots , n$.
$$
\Bspline{i}{k}{\alpha} (x)\neq 0
\equivaut j-k \leq i \leq j
$$
Thus we have 
$\displaystyle{
\sum_{i=0}^{n} \Bspline{i}{k}{\alpha}(x)
=\sum_{i=j-k}^{j}  \Bspline{i}{k}{\alpha}(x)
}
$. 

As
$
\displaystyle{
\Bspline{i}{k}{\alpha}(x) =
w_{i}^{k}(x) \Bspline{i}{k-1}{\alpha}(x)+
{\left[ 1 - w_{i+1}^{k}(x) \right]}\Bspline{i+1}{k-1}{\alpha}(x)
}
$
then
$$
\begin{array}{rcl}
\displaystyle{
\sum_{i=j-k}^{j}  \Bspline{i}{k}{\alpha}(x)
}
&=&
\displaystyle{
\sum_{i=j-k}^{j} w_{i}^{k}(x)  \Bspline{i}{k-1}{\alpha}(x)
}
+
\displaystyle{
\sum_{i=j-k}^{j} 
{\left[ 1 - w_{i+1}^{k} (x)\right]} \Bspline{i+1}{k-1}{\alpha}(x)
}\\
&=&
\displaystyle{
\sum_{i=j-k}^{j} w_{i}^{k}(x)  \Bspline{i}{k-1}{\alpha}(x)
}
+
\displaystyle{
\sum_{i=j-k+1}^{j+1} 
{\left[ 1 - w_{i}^{k}(x) \right]} \Bspline{i}{k-1}{\alpha}(x)
}\\
&=&
\displaystyle{
 w_{j-k}^{k}(x)  \Bspline{j-k}{k-1}{\alpha}(x)
}
+
\displaystyle{
\sum_{i=j-k+1}^{j} \Bspline{i}{k-1}{\alpha}(x)
}\\
&+&
\displaystyle{
{\left[ 1 - w_{j+1}^{k}(x) \right]} \Bspline{j+1}{k-1}{\alpha}(x)
}\\
&=&
\displaystyle{
\sum_{i=j-k+1}^{j} \Bspline{i}{k-1}{\alpha}(x)
}\\
\end{array}
$$
because
$$
\left\lbrace
\begin{array}{ll}
\Support{\Bspline{j-k}{k-1}{\alpha}} \cap [t_{j}\,,\, t_{j+1}) &=
  [t_{j-k}\, , \, t_{j}) \cap [t_{j}\,,\, t_{j+1}) =\emptyset
\\
\Support{\Bspline{j+1}{k-1}{\alpha}} \cap [t_{j}\,,\, t_{j+1}) &=
  [t_{j+1}\, , \, t_{j+k+1}) \cap [t_{j}\,,\, t_{j+1}) =\emptyset
\\
\end{array}
\right.
$$

\item
Let us show that for all $0\leq r \leq k-1$ we have
$$
\displaystyle{
\sum_{i=j-k+r}^{j} \Bspline{i}{k-r}{\alpha}(x)
=
\sum_{i=j-k+r+1}^{j} \Bspline{i}{k-r-1}{\alpha}(x)
}
$$
\begin{itemize}
\item
For $r=0$, it is verified.

\item Let $0<r\leq k-1$.
Suppose that the property is satisfied for all  $0\leq s <r$,
i.e.
$$
\displaystyle{
\sum_{i=j-k+s}^{j} \Bspline{i}{k-s}{\alpha}(x)
=
\sum_{i=j-k+s+1}^{j} \Bspline{i}{k-s-1}{\alpha}(x)
}
$$
Then, since
$$
\displaystyle{
\Bspline{i}{k-r}{\alpha}(x) =
w_{i}^{k-r}(x) \Bspline{i}{k-r-1}{\alpha}(x)+
{\left[ 1 - w_{i+1}^{k-r}(x) \right]}\Bspline{i+1}{k-r-1}{\alpha}(x)
}
$$
we have 

$$
\begin{array}{rcl}
\displaystyle{
\sum_{i=j-k+r}^{j} \Bspline{i}{k-r}{\alpha}(x)
}
&=&
\displaystyle{
\sum_{i=j-k+r}^{j} w_{i}^{k-r}(x) \Bspline{i}{k-r-1}{\alpha}(x)
}
\\
&+&
\displaystyle{
\sum_{i=j-k+r}^{j}{\left[ 1 - w_{i+1}^{k-r}(x) \right]}\Bspline{i+1}{k-r-1}{\alpha}(x)
}
\\
&=&
\displaystyle{
 w_{j-k+r}^{k-r}(x) \Bspline{j-k+r}{k-r-1}{\alpha}(x)
}
+
\displaystyle{
\sum_{i=j-k+r+1}^{j}  \Bspline{i}{k-r-1}{\alpha}(x)
}
\\
&+&
\displaystyle{
{\left[ 1 - w_{j+1}^{k-r}(x) \right]}\Bspline{j+1}{k-r-1}{\alpha}(x)
}
\\
&=&
\displaystyle{
\sum_{i=j-k+r+1}^{j}  \Bspline{i}{k-r-1}{\alpha}(x)
}
\\
\end{array}
$$
because
$$
\left\lbrace
\begin{array}{ll}
\Support{\Bspline{j-k+r}{k-r-1}{\alpha}} \cap  [t_{j}\,,\,t_{j+1}) &=
  [t_{j-k+r}\, , \, t_{j}) \cap [t_{j}\,,\,t_{j+1}) =\emptyset
\\
\Support{\Bspline{j+1}{k-r-1}{\alpha}} \cap [t_{j}\,,\,t_{j+1}) &=
  [t_{j+1}\, , \, t_{j+k-r+1}) \cap [t_{j}\,,\,t_{j+1}) =\emptyset
\\
\end{array}
\right.
$$

Therefore the result follows.

\end{itemize}

\item
By setting $r=k-1$ we obtain
$$
\displaystyle{
\sum_{i=j-k}^{j}  \Bspline{i}{k}{\alpha}(x)
}
=
\displaystyle{
\sum_{i=j}^{j}  \Bspline{i}{0}{\alpha}(x)
= \Bspline{j}{0}{\alpha}(x)=1
}
$$

\end{itemize}

\item (\emph{Symmetry})

Consider the symmetrical knot vector  $U=\suite{t_{i}}_{i=0}^{m}$, let 
$x\in[t_0\, , \, t_m]$, let us show that for all $k\geq 0$ and all $i\leq m-k-1$, 
we have 
$$
\displaystyle{
\Bspline{i}{k}{\alpha}(t_0+t_m -x)
=\Bspline{m-k-1-i}{k}{\alpha}(x)
}
$$

Let $T$ be the affine function on $\RR$ defined by
$\displaystyle{T(x)=t_0+t_m -x}$. $T$ is strictly decreasing.

\begin{itemize}
\item
For all $j_1<j_2$ such that $\displaystyle{ t_{j_1}< t_{j_2}}$
$$
\begin{array}{rcl}
\displaystyle{
x\in (t_{j_1} \, ,\, t_{j_2})
}
&\equivaut&
\displaystyle{
T(x) \in (T(t_{j_2}) \, ,\, T(t_{j_1}))
}
\\
&\equivaut&
\displaystyle{
T(x) \in (t_{m-j_2} \, ,\, t_{m-j_1})
} \textrm{ because } U \textrm{ is symmetric}
\\
\end{array}
$$

\item
We begin by checking for $k=0$, i.e. 
$$
\displaystyle{
\Bspline{i}{0}{\alpha}(T(x))
=\Bspline{m-1-i}{0}{1-\alpha}(x)
}
$$

$$
\begin{array}{rcl}
\displaystyle{
\Bspline{i}{0}{\alpha}(T(x)) \neq 0
}
&\implique&
\displaystyle{ t_{i} < T(x) <  t_{i+1}}
\\
&\equivaut&
\displaystyle{ t_{m-i-1}=T(t_{i+1}) < x < T( t_{i})=t_{m-i}}
\\
&\implique&
\displaystyle{
\Bspline{m-1-i}{0}{1-\alpha}(x) \neq 0
}
\\
\end{array}
$$
and conversely. The result follows as a consequence of the definition.
\item
Let $k\in  \NN^*$. We suppose that for all $j <k$ one has
$$
\displaystyle{
\Bspline{i}{j}{\alpha}(T(x))
=\Bspline{m-j-1-i}{j}{1-\alpha}(x)
}
$$

We first observe that  
$$
\displaystyle{
T(x) \in (t_{i} \, ,\, t_{i+k+1})
}
\equivaut
\displaystyle{
x \in (T( t_{i+k+1}) \, ,\,T(t_{i}))= (t_{m-i-k-1} \, ,\,t_{m-i})
}
$$
By definition:

$$
\begin{array}{rcl}
\displaystyle{
\Bspline{i}{k}{\alpha}(T(x))
}
&=&
\displaystyle{
\varphi_\alpha\triplet{T(x)}{t_{i}}{t_{i+k}} 
\Bspline{i}{k-1}{\alpha}(T(x))
}
\\
&+&
\displaystyle{
\left[ 1- \varphi_\alpha\triplet{T(x)}{t_{i+1}}{t_{i+k+1}} \right]
\Bspline{i+1}{k-1}{\alpha}(T(x))
}
\\
\end{array}
$$

By using corollary \ref{CorolParam1}

$$
\begin{array}{rcl}
\displaystyle{
\Bspline{i}{k}{\alpha}(T(x))
}
&=&
\displaystyle{
\varphi_\alpha\triplet{T(x)}{t_{i}}{t_{i+k}} 
\Bspline{i}{k-1}{\alpha}(T(x))
}
\\
&+&
\displaystyle{
\left[ 1- \varphi_\alpha\triplet{T(x)}{t_{i+1}}{t_{i+k+1}} \right]
\Bspline{i+1}{k-1}{\alpha}(T(x))
}
\\
&=&
\displaystyle{
\left[ 1- \varphi_{1-\alpha}\triplet{x}{T(t_{i+k})}{T(t_{i})}  \right]
\Bspline{i}{k-1}{\alpha}(T(x))
}
\\
&+&
\displaystyle{
\varphi_{1-\alpha}\triplet{x}{T(t_{i+k+1})}{T(t_{i+1})}
\Bspline{i+1}{k-1}{\alpha}(T(x))
}
\\
&=&
\displaystyle{
\varphi_{1-\alpha}\triplet{x}{t_{m-i-k-1}}{t_{m-i-1}}
\Bspline{i+1}{k-1}{\alpha}(T(x))
}
\\
&+&
\displaystyle{
\left[ 1- \varphi_{1-\alpha}\triplet{x}{t_{m-i-k}}{t_{m-i}}  \right]
\Bspline{i}{k-1}{\alpha}(T(x))
}
\\
\end{array}
$$

By using the recurrence hypothesis for $j=k-1$ we obtain:

$$
\begin{array}{rcl}
\displaystyle{
\Bspline{i}{k}{\alpha}(T(x))
}
&=&
\displaystyle{
\varphi_{1-\alpha}\triplet{x}{t_{m-i-k-1}}{t_{m-i-1}}
\Bspline{i+1}{k-1}{\alpha}(T(x))
}
\\
&+&
\displaystyle{
\left[ 1- \varphi_{1-\alpha}\triplet{x}{t_{m-i-k}}{t_{m-i}}  \right]
\Bspline{i}{k-1}{\alpha}(T(x))
}
\\
&=&
\displaystyle{
\varphi_{1-\alpha}\triplet{x}{t_{m-i-k-1})}{t_{m-i-1}}
\Bspline{m-k-i-1}{k-1}{1-\alpha}(x)
}
\\
&+&
\displaystyle{
\left[ 1- \varphi_{1-\alpha}\triplet{x}{t_{m-i-k}}{t_{m-i}}  \right]
\Bspline{m-k-i}{k-1}{1-\alpha}(x)
}
\\
&=&
\displaystyle{
\Bspline{m-k-i-1}{k}{1-\alpha}(x)
} \textrm{ by definition}
\\
\end{array}
$$
This completes the proof of the property.

\end{itemize}

\end{enumerate}

\begin{proposition}[Continuity property] \label{PropoContinu}

Let $m, k, n\in \NN^*$ such that \mbox{$n \geq k$} and $m=n+ k+1$.
Let $\displaystyle{U=\suite{t_i}_{i=0}^{m}}$ be a knot vector, 
let \mbox{$\alpha \in (-\infty \, , \, 0) \cup (1  \, , \,  \infty)$}.

Consider the rational B-spline basis of index $\alpha$,
with knot vector $U$ and of degree $k$,
$\suite{\Bspline{i}{k}{\alpha}}_{i=0}^{n}$.
The following properties hold: 
\begin{enumerate}
\item 
For all $i=0, \ldots , n$,
$\Bspline{i}{k}{\alpha}$  is a piecewise rational function.

\item 
For all $i=0, \ldots , n$,
$\Bspline{i}{k}{\alpha}$ is of class ${\cal C}^0$ if the knot vector $U$
does not have any interior nodes with multiplicity strictly greater than $k$.

\item
 If the knot vector $U$ is open we have

$$
\begin{array}{rcl}
\displaystyle{
\Bspline{0}{k}{\alpha}(t_{0})
}
&=&1
\\
\displaystyle{
\Bspline{i}{k}{\alpha}(t_{0})
}
&=& 0 \textrm{ for all } 0<i \leq n
\\
\displaystyle{
\Bspline{i}{k}{\alpha}(t_{m})
}
&\equiv&
\displaystyle{
\lim_{x \to t_{m}^-}\Bspline{i}{k}{\alpha}(x) =0
}
 \textrm{ for all } 0\leq i<n
\\
\displaystyle{
\Bspline{n}{k}{\alpha}(t_{m}) 
}
&\equiv&
\displaystyle{
\lim_{x \to t_{m}^-}\Bspline{n}{k}{\alpha}(x) =1
}
\\
\end{array}
$$

\end{enumerate}

\end{proposition}

\Preuve

Let $n,k \in \NN^*$ such that $n\geq k$, let  \mbox{$m=n+k+1$} and
$\displaystyle{U=\suite{t_{i}}_{i=0}^{m}}$ be a knot vector.
Let $t_i$ be an interior node with multiplicity $m_i$. 
Assume that \mbox{$1\leq m_i \leq k$}
\begin{enumerate}
\item We shall show simultaneously the two properties by recurrence on the degree $k$
\item 
We make use of the recurrence for $k\geq 1$.

\begin{itemize}
\item 
For $k=1$, we suppose a multiplicity $m_i=1$ for all
interior node  $t_{i}$.
$$
\begin{array}{rcl}
\Bspline{i}{1}{\alpha}(x)
&=&
\varphi_\alpha\triplet{x}{t_{i}}{t_{i+1}}\Bspline{i}{0}{\alpha}(x)
+
{\left[1-\varphi_\alpha\triplet{x}{t_{i+1}}{t_{i+2}}\right]}\Bspline{i+1}{0}{\alpha}(x)
\\
&=&
\left\lbrace
\begin{array}{rcl}
\varphi_\alpha\triplet{x}{t_{i}}{t_{i+1}}
&& 
\textrm{ if } x\in [t_{i}\,,\,t_{i+1})\neq \emptyset
\\
{1-\varphi_\alpha\triplet{x}{t_{i+1}}{t_{i+2}}}
&& 
\textrm{ if } x\in [t_{i}\,,\,t_{i+1})\neq \emptyset
\\
0
&& 
\textrm{ otherwise }
\\
\end{array}
\right.
\\
\end{array}
$$

Since $\displaystyle{x \in [t_{i}\,,\,t_{i+1}) \mapsto 
\varphi_\alpha\triplet{x}{t_{i}}{t_{i+1}}
}$
is homographic on  \mbox{$ [t_{i}\,,\,t_{i+1})\neq \emptyset$}
then $\Bspline{i}{1}{\alpha}$ is rational 
on  \mbox{$ [t_{i}\,,\,t_{i+1})\neq \emptyset$} and 
 \mbox{$ [t_{i+1}\,,\,t_{i+2})\neq \emptyset$} as well.
We then deduce that 
 $\Bspline{i}{1}{\alpha}$ is ${\cal C}^\infty$ on 
  \mbox{$ [t_{i}\,,\,t_{i+1})\neq \emptyset$} and also on 
 \mbox{$ [t_{i+1}\,,\,t_{i+2})\neq \emptyset$}.

Let show that $\Bspline{i}{1}{\alpha}$ is continuous at the nodes
$t_{i}$, $t_{i+1}$ et $t_{i+2}$
$$
\begin{array}{rcl}
\displaystyle{
\lim_{x \to t_{i}^-}\Bspline{i}{1}{\alpha}(x)
}
&=&
\displaystyle{
0 \textrm{ because } x \notin (t_{i}\,,\,t_{i+2})
}
\\
\displaystyle{
\lim_{x \to t_{i}^+}\Bspline{i}{1}{\alpha}(x)
}
&=& 
\displaystyle{
\lim_{x \to t_{i}^+} \varphi_\alpha \triplet{x}{t_{i}}{t_{i+1}}=0
\textrm{ if }  [t_{i}\,,\,t_{i+1})\neq \emptyset
}
\\
&=&
\displaystyle{
\Bspline{i}{1}{\alpha}( t_{i}) 
}
\\
\displaystyle{
\lim_{x \to t_{i+1}^-}\Bspline{i}{1}{\alpha}(x)
}
&=& 
\displaystyle{
\lim_{x \to t_{i+1}^-} \varphi_\alpha \triplet{x}{t_{i}}{t_{i+1}} =1
\textrm{ if }  [t_{i}\,,\,t_{i+1})\neq \emptyset
}
\\
\displaystyle{
\lim_{x \to t_{i+1}^+}\Bspline{i}{1}{\alpha}(x)
}
&=& 
\displaystyle{
\lim_{x \to t_{i+1}^+}{\left[1-\varphi_\alpha\triplet{x}{t_{i+1}}{t_{i+2}}\right]}=1
}
\\
&&
\displaystyle{
\textrm{ if }  [t_{i+1}\,,\,t_{i+2})\neq \emptyset
}
\\
&=& 
\displaystyle{
\Bspline{i}{1}{\alpha}( t_{i+1}) 
}
\\
\displaystyle{
\lim_{x \to t_{i+2}^-}\Bspline{i}{1}{\alpha}(x)
}
&=& 
\displaystyle{
\lim_{x \to t_{i+2}^-}{\left[1-\varphi_\alpha\triplet{x}{t_{i+1}}{t_{i+2}}\right]}
=0
}
\\
&&
\displaystyle{
\textrm{ if }  [t_{i+1}\,,\,t_{i+2})\neq \emptyset
}
\\
\displaystyle{
\lim_{x \to t_{i+2}^+}\Bspline{i}{1}{\alpha}(x)
}
&=& 
\displaystyle{
\Bspline{i}{1}{\alpha}( t_{i+2})=0
\textrm{ because } x\notin  (t_{i}\,,\,t_{i+2})\neq \emptyset
}
\\
\end{array}
$$

We conclude that $\Bspline{i}{1}{\alpha}$ is piecewise rational and of class ${\cal C}^0$.

\item For $k>1$  we suppose a multiplicity $1\leq m_i \leq k$ for all
interior node  $t_{i}$. 

Suppose that for all $1\leq j <k$ $\Bspline{i}{j}{\alpha}$ is 
piecewise rational and of class ${\cal C}^0$. Let us show that
$\Bspline{i}{k}{\alpha}$ is piecewise rational and of class ${\cal C}^0$ on
$\segment{t_{0}}{t_{m}}$.

By definition we know that 
$$
\begin{array}{rcl}
\displaystyle{
\Bspline{i}{k}{\alpha}
}
&=&
\displaystyle{
\varphi_\alpha\triplet{x}{t_{i}}{t_{i+k}}\Bspline{i}{k-1}{\alpha}(x)
}\\

&+&
\displaystyle{
{\left[1-\varphi_\alpha\triplet{x}{t_{i+1}}{t_{i+k+1}}\right]}\Bspline{i+1}{k-1}{\alpha}(x)
}\\
\end{array}
$$

Thus $\Bspline{i}{k}{\alpha}$ is piecewise rational as product and sum of
piecewise rational functions.
As the $\Bspline{i}{k-1}{\alpha}$ are  ${\cal C}^0$  
on $[t_{0}\,,\,t_{m})$ and if 
the multiplicity of interior nodes is at most $k$,
$$
\begin{array}{l}
\displaystyle{
x\mapsto \varphi_\alpha\triplet{x}{t_{i}}{t_{i+k}}
}\textrm{ is continuous on } 
[t_{0}\,,\,t_{k+i})\cup  (t_{k+i}\,,\,t_{m})
\\
\displaystyle{
x\mapsto \varphi_\alpha\triplet{x}{t_{i+1}}{t_{i+k+1}}
}\textrm{ is continuous on } 
[t_{0}\,,\,t_{k+i+1})\cup  (t_{k+i+1}\,,\,t_{m})
\\
\end{array}
$$
with
$$
\begin{array}{l}
\displaystyle{
\lim_{x\to t_{i+k}^- } \varphi_\alpha\triplet{x}{t_{i}}{t_{i+k}} =1
}
\\
\displaystyle{
\lim_{x\to t_{i+k}^+ } \varphi_\alpha\triplet{x}{t_{i}}{t_{i+k}} =0
}
\\
\displaystyle{
\lim_{x\to t_{i+k+1}^- } \varphi_\alpha\triplet{x}{t_{i+1}}{t_{i+k+1}}=1
}
\\
\displaystyle{
\lim_{x\to t_{i+k+1}^+ } \varphi_\alpha\triplet{x}{t_{i+1}}{t_{i+k+1}}=0
}
\\
\end{array}
$$

then $\Bspline{i}{k}{\alpha}$ is continuous on
$[t_{0}\,,\,t_{k+i})\cup  (t_{k+i}\,,\,t_{m})$ 
since
$$
\begin{array}{l}
\Support{\Bspline{i}{k-1}{\alpha}} \cap  (t_{k+i+1}\,,\,t_{m})=
\emptyset
\\
\Support{\Bspline{i+1}{k-1}{\alpha}} \cap  (t_{k+i+1}\,,\,t_{m})=
\emptyset
\\
\end{array}
$$

It is left with checking the continuity at $t_{k+i}$, which is obvious.

We can conclude that $\Bspline{i}{k}{\alpha}$ is of class ${\cal C}^0$ on 
$ [t_{0}\,,\,t_{m})$
\end{itemize}

\item 

For the endpoints values of the knot vector $U$, we have

$$
\begin{array}{l}
\displaystyle{
\Bspline{k}{0}{\alpha}(t_{0}) =\Bspline{k}{0}{\alpha}(t_{k})=1
}
\\
\displaystyle{
\lim_{x \to t_{m}^-} \Bspline{n}{0}{\alpha}(x) 
=\lim_{x \to t_{n+1}^-} \Bspline{n}{0}{\alpha}(x)=1
}
\\
\end{array}
$$

By using successively, for $r=0$ and $r=k-1$, the recurrence \ref{RecValExtremite1} 
of lemma \ref{LemmeValExtremite1} and the recurrence  \ref{RecValExtremite2} of lemma
 \ref{LemmeValExtremite2}, one can deduce that:

$$
\begin{array}{l}
\displaystyle{
\Bspline{0}{k}{\alpha}(t_{0}) =\Bspline{k}{0}{\alpha}(t_{0}) =
\Bspline{k}{0}{\alpha}(t_{k})=1
}
\\
\displaystyle{
\lim_{x \to t_{m}^-} \Bspline{n}{k}{\alpha}(x) 
=\lim_{x \to t_{m}^-} \Bspline{n}{0}{\alpha}(x) 
=\lim_{x \to t_{n+1}^-} \Bspline{n}{0}{\alpha}(x)=1
}
\\
\end{array}
$$

From the property of unit partition, we have
$$
\displaystyle{
\sum_{i=0}^{n} \Bspline{i}{k}{\alpha}(x)=1 \quad 
\forall x\in [t_{0}\,,\,t_{m})= [t_{k}\,,\,t_{n+1})
}
$$
Thus 
$$
\begin{array}{l}
\displaystyle{
\sum_{i=1}^{n} \Bspline{i}{k}{\alpha}(t_{0})=0
}
\\
\displaystyle{
\sum_{i=0}^{n-1} \left(\lim_{x\to t_{m}^- }\Bspline{i}{k}{\alpha}(x) \right)=
\lim_{x\to t_{m}^- }\sum_{i=0}^{n-1} \Bspline{i}{k}{\alpha}(x)=0
}
\\
\end{array}
$$
From the fact that the $\displaystyle{\Bspline{i}{k}{\alpha}}$ are positive,
we obtain
$$
\begin{array}{ll}
\displaystyle{
\Bspline{i}{k}{\alpha}(t_{0})=0 
} 
&\textrm{for all } i=1, \ldots , n
\\
\displaystyle{
\lim_{x\to t_{m}^- } \Bspline{i}{k}{\alpha}(x)=0
}
&\textrm{for all } i=0, \ldots , n-1
\\
\end{array}
$$

Each $\displaystyle{ \Bspline{i}{k}{\alpha}}$ admits a continuous extension
at $t_{m}$
\end{enumerate}

\begin{lemme}\label{LemmeValExtremite1}

Let $m, k, n\in \NN^*$ such that \mbox{$n \geq k$} and $m=n+ k+1$.
Let $\displaystyle{U=\suite{t_i}_{i=0}^{m}}$ be an open knot vector and 
  \mbox{$\alpha \in (-\infty \, , \, 0) \cup (1  \, , \,  \infty)$}.

Consider the rational B-spline basis of index $\alpha$ with knot vector $U$ and of degree $k$,
$\suite{\Bspline{i}{k}{\alpha}}_{i=0}^{n}$.
For all $0\leq r\leq k-1$ we have:
\begin{equation}\label{RecValExtremite1}
\begin{array}{l}
\displaystyle{
\Bspline{r}{k-r}{\alpha}(t_{0}) =\Bspline{r+1}{k-r-1}{\alpha}(t_{0})
}
\\
\displaystyle{
\Bspline{r+1}{k-r}{\alpha}(t_{0}) =\Bspline{r+2}{k-r-1}{\alpha}(t_{0})
}
\\
\end{array}
\end{equation}

\end{lemme}

\Preuve

\begin{itemize}
\item
For $r=0$, we have
$$
\begin{array}{rcl}
\displaystyle{
\Bspline{r}{k-r}{\alpha}(t_{0}) 
}
&=&
\displaystyle{
\Bspline{0}{k}{\alpha}(t_{0})
}
\\
&=&
\displaystyle{
\varphi_\alpha\triplet{t_{0}}{t_{0}}{t_{k}}\Bspline{0}{k-1}{\alpha}(t_{0})
}
\\
&+&
\displaystyle{
{\left[1-\varphi_\alpha\triplet{t_{0}}{t_{1}}{t_{k+1}}\right]}
\Bspline{1}{k-1}{\alpha}(t_{0})
}
\\
&=&
\displaystyle{
\Bspline{1}{k-1}{\alpha}(t_{0})=\Bspline{r+1}{k-r-1}{\alpha}(t_{0})
}
\\

\end{array}
$$

Besides
$$
\begin{array}{rcl}
\displaystyle{
\Bspline{r+1}{k-r}{\alpha}(t_{0}) 
}
&=&
\displaystyle{
\Bspline{1}{k}{\alpha}(t_{0})
}
\\
&=&
\displaystyle{
\varphi_\alpha\triplet{t_{0}}{t_{1}}{t_{k+1}}\Bspline{1}{k-1}{\alpha}(t_{0})
}
\\
&+&
\displaystyle{
{\left[1-\varphi_\alpha\triplet{t_{0}}{t_{2}}{t_{k+2}}\right]}
\Bspline{2}{k-1}{\alpha}(t_{0})
}
\\
&=&
\displaystyle{
\Bspline{2}{k-1}{\alpha}(t_{0})=\Bspline{r+2}{k-r-1}{\alpha}(t_{0})
}
\\
\end{array}
$$
because
$$
\begin{array}{l}
\displaystyle{
\varphi_\alpha\triplet{t_{0}}{t_{1}}{t_{k+1}}=
\varphi_\alpha\triplet{t_{0}}{t_{0}}{t_{k+1}}=0
}\\
\displaystyle{
\varphi_\alpha\triplet{t_{0}}{t_{2}}{t_{k+1}}=
\varphi_\alpha\triplet{t_{k}}{t_{k}}{t_{k+1}}=0
}\\
\end{array}
$$
since $U$ is open.

\item
Let $0<r<k$. 

We assume that for all $0\leq j< r$ we have
$$
\begin{array}{l}
\displaystyle{
\Bspline{j}{k-j}{\alpha}(t_{0}) 
=
\Bspline{j+1}{k-j-1}{\alpha}(t_{0}) 
}\\
\displaystyle{
\Bspline{j+1}{k-j}{\alpha}(t_{0}) 
=
\Bspline{j+2}{k-j-1}{\alpha}(t_{0}) 
}\\
\end{array}
$$

Then
$$
\begin{array}{rcl}
\displaystyle{
\Bspline{r}{k-r}{\alpha}(t_{0}) 
}
&=&
\displaystyle{
\varphi_\alpha\triplet{t_{0}}{t_{r}}{t_{k}}\Bspline{r}{k-r-1}{\alpha}(t_{0})
}
\\
&+&
\displaystyle{
{\left[1-\varphi_\alpha\triplet{t_{0}}{t_{r+1}}{t_{k+1}}\right]}
\Bspline{r+1}{k-r-1}{\alpha}(t_{0})
}
\\
&=&
\displaystyle{
\Bspline{r+1}{k-r-1}{\alpha}(t_{0})
}
\\

\end{array}
$$
and
$$
\begin{array}{rcl}
\displaystyle{
\Bspline{r+1}{k-r}{\alpha}(t_{0}) 
}
&=&
\displaystyle{
\varphi_\alpha\triplet{t_{0}}{t_{r+1}}{t_{k+1}}\Bspline{r+1}{k-r-1}{\alpha}(t_{0})
}
\\
&+&
\displaystyle{
{\left[1-\varphi_\alpha\triplet{t_{0}}{t_{r+2}}{t_{k+2}}\right]}
\Bspline{r+2}{k-r-1}{\alpha}(t_{0})
}
\\
&=&
\displaystyle{
\Bspline{r+2}{k-r-1}{\alpha}(t_{0})
}
\\

\end{array}
$$
because
$$
\begin{array}{l}
\displaystyle{
\varphi_\alpha\triplet{t_{0}}{t_{r+1}}{t_{k+1}}=
\varphi_\alpha\triplet{t_{0}}{t_{0}}{t_{k+1}}=0
}\\
\displaystyle{
\varphi_\alpha\triplet{t_{0}}{t_{r+2}}{t_{k+2}}=
\varphi_\alpha\triplet{t_{k}}{t_{k+1}}{t_{k+2}}=0
}\\
\end{array}
$$
since $U$ is open.

The result follows.

\end{itemize}

\begin{lemme}\label{LemmeValExtremite2}

Let $m, k, n\in \NN^*$ such that \mbox{$n \geq k$} and $m=n+ k+1$.
Let $\displaystyle{U=\suite{t_i}_{i=0}^{m}}$ be an open knot vector and 
\mbox{$\alpha \in (-\infty \, , \, 0) \cup (1  \, , \,  \infty)$}.

Consider the rational B-spline basis of index $\alpha$ with knot vector $U$ and of degree $k$,
$\suite{\Bspline{i}{k}{\alpha}}_{i=0}^{n}$.
For all $0\leq r\leq k-1$ we have
\begin{equation}\label{RecValExtremite2}
\begin{array}{l}
\displaystyle{
\lim_{x \to t_{m}^-} \Bspline{n}{k-r}{\alpha}(x) 
=\lim_{x \to t_{m}^-} \Bspline{n}{k-r-1}{\alpha}(x)
}
\\
\displaystyle{
\lim_{x \to t_{m}^-} \Bspline{n-1}{k-r}{\alpha}(x) 
=\lim_{x \to t_{m}^-} \Bspline{n-1}{k-r-1}{\alpha}(x)
} \textrm{ for } k\geq 2
\\
\end{array}
\end{equation}

\end{lemme}

\Preuve

\begin{itemize}
\item
For $r=0$, we have
$$
\begin{array}{rcl}
\displaystyle{
\lim_{x \to t_{m}^-}\Bspline{n}{k-r}{\alpha}(x) 
}
&=&
\displaystyle{
\lim_{x \to t_{m}^-}\Bspline{n}{k}{\alpha}(x)
}
\\
&=&
\displaystyle{
\lim_{x \to t_{m}^-}\varphi_\alpha\triplet{x}{t_{n}}{t_{n+k}}
\lim_{x \to t_{m}^-} \Bspline{n}{k-1}{\alpha}(x)
}
\\
&+&
\displaystyle{
\lim_{x \to t_{m}^-}{\left[1-\varphi_\alpha\triplet{x}{t_{n+1}}{t_{m}}\right]}
\lim_{x \to t_{m}^-}\Bspline{n+1}{k-1}{\alpha}(x)
}
\\
&=&
\displaystyle{
\lim_{x \to t_{m}^-}\varphi_\alpha\triplet{x}{t_{n}}{t_{m}}
\lim_{x \to t_{m}^-} \Bspline{n}{k-1}{\alpha}(x)
}
\\
&=&
\displaystyle{
\lim_{x \to t_{m}^-} \Bspline{n}{k-1}{\alpha}(x)
 = \lim_{x \to t_{m}^-} \Bspline{n}{k-r-1}{\alpha}(x)
}
\\
\end{array}
$$
since
$
\displaystyle{
\Support{\Bspline{n+1}{k-1}{\alpha}}
=[t_{n+1}\,,\,t_{m})= \emptyset
}
$

and
$$
\begin{array}{rcl}
\displaystyle{
\lim_{x \to t_{m}^-}\Bspline{n-1}{k-r}{\alpha}(x) 
}
&=&
\displaystyle{
\lim_{x \to t_{m}^-}\Bspline{n-1}{k}{\alpha}(x)
}
\\
&=&
\displaystyle{
\lim_{x \to t_{m}^-}\varphi_\alpha\triplet{x}{t_{n-1}}{t_{n+k-1}}
\lim_{x \to t_{m}^-} \Bspline{n-1}{k-1}{\alpha}(x)
}
\\
&+&
\displaystyle{
\lim_{x \to t_{m}^-}{\left[1-\varphi_\alpha\triplet{x}{t_{n}}{t_{n+k}}\right]}
\lim_{x \to t_{m}^-}\Bspline{n}{k-1}{\alpha}(x)
}
\\
&=&
\displaystyle{
\lim_{x \to t_{m}^-} \Bspline{n-1}{k-1}{\alpha}(x)
 = \lim_{x \to t_{m}^-} \Bspline{n-1}{k-r-1}{\alpha}(x)
}
\\
\end{array}
$$

since for $k\geq 2$ one has
$$
\begin{array}{l}
\displaystyle{
\lim_{x \to t_{m}^-}\varphi_\alpha\triplet{x}{t_{n-1}}{t_{n+k-1}}=
\lim_{x \to t_{m}^-}\varphi_\alpha\triplet{x}{t_{n-1}}{t_{m}}=1
}
\\
\displaystyle{
\lim_{x \to t_{m}^-}\varphi_\alpha\triplet{x}{t_{n}}{t_{n+k}}=
\lim_{x \to t_{m}^-}\varphi_\alpha\triplet{x}{t_{n}}{t_{m}}=1
}
\end{array}
$$.

\item
Let $0<r<k$. 

We suppose that for all $0\leq j\leq r$ we have
$
\displaystyle{
\lim_{x \to t_{m}^-}\Bspline{n}{j-r}{\alpha}(x) 
 = \lim_{x \to t_{m}^-} \Bspline{n}{j-r-1}{\alpha}(x)
}
$.
Then

$$
\begin{array}{rcl}
\displaystyle{
\lim_{x \to t_{m}^-}\Bspline{n}{k-r}{\alpha}(x) 
}
&=&
\displaystyle{
\lim_{x \to t_{m}^-}\varphi_\alpha\triplet{x}{t_{n}}{t_{n+k-r}}
\lim_{x \to t_{m}^-} \Bspline{n}{k-r-1}{\alpha}(x)
}
\\
&+&
\displaystyle{
\lim_{x \to t_{m}^-}{\left[1-\varphi_\alpha\triplet{x}{t_{n+1}}{t_{m-r}}\right]}
\lim_{x \to t_{m}^-}\Bspline{n+1}{k-r-1}{\alpha}(x)
}
\\
&=&
\displaystyle{
\lim_{x \to t_{m}^-}\varphi_\alpha\triplet{x}{t_{n}}{t_{m}}
\lim_{x \to t_{m}^-} \Bspline{n}{k-r-1}{\alpha}(x)
}
\\
&=&
\displaystyle{
 \lim_{x \to t_{m}^-} \Bspline{n}{k-r-1}{\alpha}(x)
}
\\
\end{array}
$$
because
$
\displaystyle{
\Support{\Bspline{n+1}{k-1}{\alpha}}
=[t_{n+1}\,,\,t_{m-r})
=[t_{n+1}\,,\,t_{m})= \emptyset
}
$

The result then follows.

On the other hand we assume that for all $0\leq j\leq r$ with $k\geq 2$, one has
$$
\displaystyle{
\lim_{x \to t_{m}^-}\Bspline{n-1}{j-r}{\alpha}(x) 
 = \lim_{x \to t_{m}^-} \Bspline{n-1}{j-r-1}{\alpha}(x)
}
$$

Then we get

$$
\begin{array}{rcl}
\displaystyle{
\lim_{x \to t_{m}^-}\Bspline{n-1}{k-r}{\alpha}(x) 
}
&=&
\displaystyle{
\lim_{x \to t_{m}^-}\varphi_\alpha\triplet{x}{t_{n-1}}{t_{n+k-r-1}}
\lim_{x \to t_{m}^-} \Bspline{n-1}{k-r-1}{\alpha}(x)
}
\\
&+&
\displaystyle{
\lim_{x \to t_{m}^-}{\left[1-\varphi_\alpha\triplet{x}{t_{n}}{t_{n+k-r}}\right]}
\lim_{x \to t_{m}^-}\Bspline{n}{k-r-1}{\alpha}(x)
}
\\
&=&
\displaystyle{
\lim_{x \to t_{m}^-} \Bspline{n-1}{k-r-1}{\alpha}(x)
}
\\
\end{array}
$$
because for $k\geq 2$ we have
$$
\begin{array}{l}
\displaystyle{
\lim_{x \to t_{m}^-}\varphi_\alpha\triplet{x}{t_{n-1}}{t_{n+k-r-1}}
=\lim_{x \to t_{m}^-}\varphi_\alpha\triplet{x}{t_{n-1}}{t_{m}}=1
}
\\
\displaystyle{
\lim_{x \to t_{m}^-}\varphi_\alpha\triplet{x}{t_{n}}{t_{n+k-r}}
=\lim_{x \to t_{m}^-}\varphi_\alpha\triplet{x}{t_{n}}{t_{m}}=1
}
\\
\end{array}
$$

\end{itemize}

\begin{lemme}\label{LemmeDerivExtremite1}

Let $m, k, n\in \NN^*$ such that \mbox{$n \geq k$} and $m=n+ k+1$.
Let $\displaystyle{U=\suite{t_i}_{i=0}^{m}}$ be an open knot vector 
and \mbox{$\alpha \in (-\infty \, , \, 0) \cup (1  \, , \,  \infty)$}.

Consider the rational B-spline basis $\suite{\Bspline{i}{k}{\alpha}}_{i=0}^{n}$ 
of index $\alpha$ with knot vector $U$
and of degree $k$.
For all $0\leq r\leq k-1$  and all $ i\geq 2$  we have:
\begin{equation}\label{RecDerivExtremite1}
\begin{array}{l}
\displaystyle{
\lim_{x \to t_{0}^+}\frac{d}{dx} \Bspline{r}{k-r}{\alpha}(x) 
=\lim_{x \to t_{0}^+}\frac{d}{dx} \Bspline{r+1}{k-r-1}{\alpha}(x)
-\lim_{x \to t_{0}^+} \frac{d}{dx} w_{r+1}^{k-r}(x)
}
\\
\displaystyle{
\lim_{x \to t_{0}^+}\frac{d}{dx} \Bspline{r+1}{k-r}{\alpha}(x) 
=\lim_{x \to t_{0}^+}\frac{d}{dx} \Bspline{r+2}{k-r-1}{\alpha}(x)
+\lim_{x \to t_{0}^+} \frac{d}{dx} w_{r+1}^{k-r}(x)
}
\\
\displaystyle{
\lim_{x \to t_{0}^+}\frac{d}{dx} \Bspline{i+r}{k-r}{\alpha}(x) 
=\lim_{x \to t_{0}^+}\frac{d}{dx} \Bspline{i+r+1}{k-r-1}{\alpha}(x)
}
\\
\end{array}
\end{equation}
with $\displaystyle{w_{i}^{j}(x)
=\varphi_\alpha\triplet{x}{t_{i}}{t_{i+j}}
}$

\end{lemme}

\Preuve
We proceed by recurrence on $r$
\begin{itemize}
\item
Let $r=0$ 
\begin{itemize}
\item
For $x\in (t_{0}\,,\,t_{k+1})$ we have
$$
\begin{array}{rcl}
\displaystyle{
\Bspline{r}{k-r}{\alpha}(x) 
}
&=&
\displaystyle{
\Bspline{0}{k}{\alpha}(x) 
}
\\
&=&
\displaystyle{
\varphi_\alpha\triplet{x}{t_{0}}{t_{k}}
\Bspline{0}{k-1}{\alpha}(x) 
}\\
&+&
\displaystyle{
\left(1-\varphi_\alpha\triplet{x}{t_{1}}{t_{k+1}}\right)
\Bspline{1}{k-1}{\alpha}(x) 
}
\\
&=&
\displaystyle{
\left(1-\varphi_\alpha\triplet{x}{t_{1}}{t_{k+1}}\right)
\Bspline{1}{k-1}{\alpha}(x) 
}
\\
\end{array}
$$
since 
$\Support{\varphi_\alpha\triplet{.}{t_{0}}{t_{k}}}=\emptyset$, $U$ is open.

Thus we have
$$
\begin{array}{rcl}
\displaystyle{
\frac{d}{dx}\Bspline{r}{k-r}{\alpha}(x) 
}
&=&
\displaystyle{
-\frac{d}{dx}\varphi_\alpha\triplet{x}{t_{1}}{t_{k+1}}
\Bspline{1}{k-1}{\alpha}(x) 
}\\
&+&
\displaystyle{
\left(1-\varphi_\alpha\triplet{x}{t_{1}}{t_{k+1}}\right)
\frac{d}{dx}\Bspline{1}{k-1}{\alpha}(x) 
}
\\
\end{array}
$$

We deduce that 
$$
\begin{array}{rcl}
\displaystyle{
\lim_{x\to t_{0}^+}\frac{d}{dx}\Bspline{r}{k-r}{\alpha}(x) 
}
&=&
\displaystyle{
-\lim_{x\to t_{0}^+}\frac{d}{dx}\varphi_\alpha\triplet{x}{t_{1}}{t_{k+1}}
\lim_{x\to t_{0}^+}\Bspline{1}{k-1}{\alpha}(x) 
}\\
&+&
\displaystyle{
\left(1-\lim_{x\to t_{0}^+}\varphi_\alpha\triplet{x}{t_{1}}{t_{k+1}}\right)
\lim_{x\to t_{0}^+}\frac{d}{dx}\Bspline{1}{k-1}{\alpha}(x) 
}
\\
&=&
\displaystyle{
-\lim_{x\to t_{0}^+}\frac{d}{dx}\varphi_\alpha\triplet{x}{t_{1}}{t_{k+1}}
}
+
\displaystyle{
\lim_{x\to t_{0}^+}\frac{d}{dx}\Bspline{1}{k-1}{\alpha}(x) 
}
\\
&=&
\displaystyle{
-\lim_{x\to t_{0}^+}\frac{d}{dx}\varphi_\alpha\triplet{x}{t_{r+1}}{t_{k+1}}
}
+
\displaystyle{
\lim_{x\to t_{0}^+}\frac{d}{dx}\Bspline{r+1}{k-r-1}{\alpha}(x) 
}
\\
&=&
\displaystyle{
-\lim_{x\to t_{0}^+}\frac{d}{dx}w_{r+1}^{k-r}
}
+
\displaystyle{
\lim_{x\to t_{0}^+}\frac{d}{dx}\Bspline{r+1}{k-r-1}{\alpha}(x) 
}
\\
\end{array}
$$
because $[t_{k}\,,\,t_{k+1})=[t_{0}\,,\,t_{k+1})$ 
and by using lemma ~\ref{LemmeValExtremite1} successively for $r=1$
and  $r=k-1$, we have
$
\displaystyle{
\lim_{x\to t_{0}^+}\Bspline{1}{k-1}{\alpha}(x) =
\lim_{x\to t_{0}^+}\Bspline{k}{0}{\alpha}(x) =1
}
$
and one should remark that
$
\displaystyle{
\lim_{x\to t_{0}^+}\varphi_\alpha\triplet{x}{t_{1}}{t_{k+1}}=
\lim_{x\to t_{0}^+}\varphi_\alpha\triplet{x}{t_{0}}{t_{k+1}}=0
}
$.

We have just proved that for $r=0$ we have
$$
\displaystyle{
\lim_{x\to t_{0}^+}\frac{d}{dx}\Bspline{r}{k-r}{\alpha}(x) 
}
=
\displaystyle{
\lim_{x\to t_{0}^+}\frac{d}{dx}\Bspline{r+1}{k-r-1}{\alpha}(x) 
-\lim_{x\to t_{0}^+}\frac{d}{dx}w_{r+1}^{k-r}
}
$$

\item
For $x\in (t_{0}\,,\,t_{k+1})$ and $k\geq 2$ we have
$$
\begin{array}{rcl}
\displaystyle{
\Bspline{r+1}{k-r}{\alpha}(x) 
}
&=&
\displaystyle{
\Bspline{1}{k}{\alpha}(x) 
}
\\
&=&
\displaystyle{
\varphi_\alpha\triplet{x}{t_{1}}{t_{k+1}}
\Bspline{1}{k-1}{\alpha}(x) 
}\\
&+&
\displaystyle{
\left(1-\varphi_\alpha\triplet{x}{t_{2}}{t_{k+2}}\right)
\Bspline{2}{k-1}{\alpha}(x) 
}
\\
&=&
\displaystyle{
\varphi_\alpha\triplet{x}{t_{0}}{t_{k+1}}
\Bspline{1}{k-1}{\alpha}(x) 
}\\
&+&
\displaystyle{
\left(1-\varphi_\alpha\triplet{x}{t_{0}}{t_{k+2}}\right)
\Bspline{2}{k-1}{\alpha}(x) 
}
\\
\end{array}
$$

We then have 
$$
\begin{array}{rcl}
\displaystyle{
\frac{d}{dx} \Bspline{r+1}{k-r}{\alpha}(x) 
}
&=&
\displaystyle{
\frac{d}{dx} \varphi_\alpha\triplet{x}{t_{1}}{t_{k+1}}
\Bspline{1}{k-1}{\alpha}(x) 
}\\
&+&
\displaystyle{
\varphi_\alpha\triplet{x}{t_{0}}{t_{k+1}}
\frac{d}{dx} \Bspline{1}{k-1}{\alpha}(x) 
}\\
&-&
\displaystyle{
\frac{d}{dx} \varphi_\alpha\triplet{x}{t_{2}}{t_{k+2}} 
\Bspline{2}{k-1}{\alpha}(x) 
}
\\
&+&
\displaystyle{
\left(1-\varphi_\alpha\triplet{x}{t_{0}}{t_{k+2}}\right)
\frac{d}{dx} \Bspline{2}{k-1}{\alpha}(x) 
}
\\
\end{array}
$$

Therefore
$$
\begin{array}{rcl}
\displaystyle{
\lim_{x\to t_{0}^+} \frac{d}{dx} \Bspline{r+1}{k-r}{\alpha}(x) 
}
&=&
\displaystyle{
\lim_{x\to t_{0}^+} \frac{d}{dx} \varphi_\alpha\triplet{x}{t_{1}}{t_{k+1}}
\lim_{x\to t_{0}^+} \Bspline{1}{k-1}{\alpha}(x) 
}\\
&+&
\displaystyle{
\lim_{x\to t_{0}^+} \varphi_\alpha\triplet{x}{t_{0}}{t_{k+1}}
\lim_{x\to t_{0}^+} \frac{d}{dx} \Bspline{1}{k-1}{\alpha}(x) 
}\\
&-&
\displaystyle{
\lim_{x\to t_{0}^+}\frac{d}{dx} \varphi_\alpha\triplet{x}{t_{2}}{t_{k+2}} 
\lim_{x\to t_{0}^+}\Bspline{2}{k-1}{\alpha}(x) 
}
\\
&+&
\displaystyle{
\left(1-\lim_{x\to t_{0}^+} \varphi_\alpha\triplet{x}{t_{0}}{t_{k+2}}\right)
\lim_{x\to t_{0}^+} \frac{d}{dx} \Bspline{2}{k-1}{\alpha}(x) 
}
\\
&=&
\displaystyle{
\lim_{x\to t_{0}^+} \frac{d}{dx} \varphi_\alpha\triplet{x}{t_{1}}{t_{k+1}}
}
+
\displaystyle{
\lim_{x\to t_{0}^+} \frac{d}{dx} \Bspline{2}{k-1}{\alpha}(x) 
}
\\
&=&
\displaystyle{
\lim_{x\to t_{0}^+} \frac{d}{dx} w_{1}^{k}(x)
}
+
\displaystyle{
\lim_{x\to t_{0}^+} \frac{d}{dx} \Bspline{2}{k-1}{\alpha}(x) 
}
\\
&=&
\displaystyle{
\lim_{x\to t_{0}^+} \frac{d}{dx} w_{r+1}^{k-r}(x)
}
+
\displaystyle{
\lim_{x\to t_{0}^+} \frac{d}{dx} \Bspline{r+2}{k-r-1}{\alpha}(x) 
}
\\
\end{array}
$$
since from lemma ~\ref{LemmeValExtremite1} we have
$$
\begin{array}{rcl}
\displaystyle{
\lim_{x\to t_{0}^+}  \Bspline{2}{k-1}{\alpha}(x) 
}
&=&
\displaystyle{
\lim_{x\to t_{0}^+}  \Bspline{k+1}{0}{\alpha}(x) =0
}
\\
\displaystyle{
\lim_{x\to t_{0}^+}  \Bspline{1}{k-1}{\alpha}(x) 
}
&=&
\displaystyle{
\lim_{x\to t_{0}^+}  \Bspline{k}{0}{\alpha}(x) =1
}
\\
\end{array}
$$

\item
Similarly
for $x\in (t_{0}\,,\,t_{k+1})$ and $i\geq 2$ we have
$$
\begin{array}{rcl}
\displaystyle{
\Bspline{i+r}{k-r}{\alpha}(x) 
}
&=&
\displaystyle{
\Bspline{i}{k}{\alpha}(x) 
}
\\
&=&
\displaystyle{
\varphi_\alpha\triplet{x}{t_{i}}{t_{i+k}}
\Bspline{i}{k-1}{\alpha}(x) 
}\\
&+&
\displaystyle{
\left(1-\varphi_\alpha\triplet{x}{t_{i+1}}{t_{k+i+1}}\right)
\Bspline{i+1}{k-1}{\alpha}(x) 
}
\\
\end{array}
$$

$$
\begin{array}{rcl}
\displaystyle{
 \frac{d}{dx} \Bspline{i+r}{k-r}{\alpha}(x) 
}
&=&
\displaystyle{
 \frac{d}{dx} \varphi_\alpha\triplet{x}{t_{i}}{t_{i+k}}
\Bspline{i}{k-1}{\alpha}(x) 
}\\
&+&
\displaystyle{
\varphi_\alpha\triplet{x}{t_{i}}{t_{i+k}}
 \frac{d}{dx} \Bspline{i}{k-1}{\alpha}(x) 
}\\
&-&
\displaystyle{
\frac{d}{dx} \varphi_\alpha\triplet{x}{t_{i+1}}{t_{k+i+1}}
\Bspline{i+1}{k-1}{\alpha}(x) 
}
\\
&+&
\displaystyle{
\left(1-\varphi_\alpha\triplet{x}{t_{i+1}}{t_{k+i+1}}\right)
 \frac{d}{dx} \Bspline{i+1}{k-1}{\alpha}(x) 
}
\\
\end{array}
$$

By passing to the limit

$$
\begin{array}{rcl}
\displaystyle{
\lim_{x\to t_{0}^+} \frac{d}{dx} \Bspline{i+r}{k-r}{\alpha}(x) 
}
&=&
\displaystyle{
\lim_{x\to t_{0}^+}  \frac{d}{dx} \varphi_\alpha\triplet{x}{t_{i}}{t_{i+k}}
\lim_{x\to t_{0}^+} \Bspline{i}{k-1}{\alpha}(x) 
}\\
&+&
\displaystyle{
\lim_{x\to t_{0}^+} \varphi_\alpha\triplet{x}{t_{i}}{t_{i+k}}
\lim_{x\to t_{0}^+} \frac{d}{dx} \Bspline{i}{k-1}{\alpha}(x) 
}\\
&-&
\displaystyle{
\lim_{x\to t_{0}^+} \frac{d}{dx} \varphi_\alpha\triplet{x}{t_{i+1}}{t_{k+i+1}}
\lim_{x\to t_{0}^+} \Bspline{i+1}{k-1}{\alpha}(x) 
}
\\
&+&
\displaystyle{
\lim_{x\to t_{0}^+} \left(1-\varphi_\alpha\triplet{x}{t_{i+1}}{t_{k+i+1}}\right)
\lim_{x\to t_{0}^+} \frac{d}{dx} \Bspline{i+1}{k-1}{\alpha}(x) 
}
\\
&=&
\displaystyle{
\lim_{x\to t_{0}^+} \frac{d}{dx} \Bspline{i+1}{k-1}{\alpha}(x) 
}
\\
&=&
\displaystyle{
\lim_{x\to t_{0}^+} \frac{d}{dx} \Bspline{i+r+1}{k-r-1}{\alpha}(x) 
}
\\
\end{array}
$$
since
$$
\begin{array}{rcl}
\displaystyle{
\lim_{x\to t_{0}^+} \Bspline{i}{k-1}{\alpha}(x) 
}
&=&
\displaystyle{
\lim_{x\to t_{0}^+} \Bspline{i+k-1}{0}{\alpha}(x) =0 \quad \forall i\geq 2
}
\\
\end{array}
$$

\end{itemize}

\item
Let $0 <r<k$. Suppose that for all $0\leq j < r$ we have
$$
\begin{array}{l}
\displaystyle{
\lim_{x\to t_{0}^+}\frac{d}{dx}\Bspline{j}{k-j}{\alpha}(x) 
}
=
\displaystyle{
\lim_{x\to t_{0}^+}\frac{d}{dx}\Bspline{j+1}{k-j-1}{\alpha}(x) 
-\lim_{x\to t_{0}^+}\frac{d}{dx}w_{j+1}^{k-j}
}
\\
\displaystyle{
\lim_{x\to t_{0}^+}\frac{d}{dx}\Bspline{j+1}{k-j}{\alpha}(x) 
}
=
\displaystyle{
\lim_{x\to t_{0}^+}\frac{d}{dx}\Bspline{j+2}{k-j-1}{\alpha}(x) 
+\lim_{x\to t_{0}^+}\frac{d}{dx}w_{j+1}^{k-j}
}
\\
\displaystyle{
\lim_{x\to t_{0}^+}\frac{d}{dx}\Bspline{i+j}{k-j}{\alpha}(x) 
}
=
\displaystyle{
\lim_{x\to t_{0}^+}\frac{d}{dx}\Bspline{i+j+1}{k-j-1}{\alpha}(x) 
}
\\
\end{array}
$$

\begin{itemize}
\item
Let us show that
$$
\displaystyle{
\lim_{x\to t_{0}^+} \frac{d}{dx}\Bspline{r}{k-r}{\alpha}(x) 
}
=
\displaystyle{
-\lim_{x\to t_{0}^+} \frac{d}{dx} w_{r+1}^{k-r}
}
+
\displaystyle{
\lim_{x\to t_{0}^+} \frac{d}{dx}\Bspline{r+1}{k-r-1}{\alpha}(x) 
}
$$

For $x\in (t_{0}\,,\,t_{k+1})$ we have
$$
\begin{array}{rcl}
\displaystyle{
\Bspline{r}{k-r}{\alpha}(x) 
}
&=&
\displaystyle{
\varphi_\alpha\triplet{x}{t_{r}}{t_{k}}
\Bspline{r}{k-r-1}{\alpha}(x) 
}\\
&+&
\displaystyle{
\left(1-\varphi_\alpha\triplet{x}{t_{r+1}}{t_{k+1}}\right)
\Bspline{r+1}{k-r-1}{\alpha}(x) 
}
\\
&=&
\displaystyle{
\left(1-\varphi_\alpha\triplet{x}{t_{r+1}}{t_{k+1}}\right)
\Bspline{r+1}{k-r-1}{\alpha}(x) 
}
\\
\end{array}
$$
because $\Support{\Bspline{r}{k-r-1}{\alpha}}=[t_{r}\,,\,t_{k})=\emptyset$
since $r<k$.

Hence 
$$
\begin{array}{rcl}
\displaystyle{
\frac{d}{dx}\Bspline{r}{k-r}{\alpha}(x) 
}
&=&
-
\displaystyle{
\frac{d}{dx}\varphi_\alpha\triplet{x}{t_{r+1}}{t_{k+1}} 
\Bspline{r+1}{k-r-1}{\alpha}(x) 
}
\\
&+&
\displaystyle{
\left(1-\varphi_\alpha\triplet{x}{t_{r+1}}{t_{k+1}}\right)
\frac{d}{dx}\Bspline{r+1}{k-r-1}{\alpha}(x) 
}
\\
&=&
-
\displaystyle{
\frac{d}{dx}\varphi_\alpha\triplet{x}{t_{r+1}}{t_{k+1}} 
\Bspline{r+1}{k-r-1}{\alpha}(x) 
}
\\
&+&
\displaystyle{
\left(1-\varphi_\alpha\triplet{x}{t_{0}}{t_{k+1}}\right)
\frac{d}{dx}\Bspline{r+1}{k-r-1}{\alpha}(x) 
}
\\
\end{array}
$$

By passing to the limit we have
$$
\begin{array}{rcl}
\displaystyle{
\lim_{x\to t_{0}^+} \frac{d}{dx}\Bspline{r}{k-r}{\alpha}(x) 
}
&=&
-
\displaystyle{
\lim_{x\to t_{0}^+} \frac{d}{dx}\varphi_\alpha\triplet{x}{t_{r+1}}{t_{k+1}} 
\lim_{x\to t_{0}^+} \Bspline{r+1}{k-r-1}{\alpha}(x) 
}
\\
&+&
\displaystyle{
\left(1-\lim_{x\to t_{0}^+} \varphi_\alpha\triplet{x}{t_{0}}{t_{k+1}}\right)
\lim_{x\to t_{0}^+}\frac{d}{dx}\Bspline{r+1}{k-r-1}{\alpha}(x) 
}
\\
&=&
-
\displaystyle{
\lim_{x\to t_{0}^+} \frac{d}{dx}\varphi_\alpha\triplet{x}{t_{r+1}}{t_{k+1}} 
}
+
\displaystyle{
\lim_{x\to t_{0}^+}\frac{d}{dx}\Bspline{r+1}{k-r-1}{\alpha}(x) 
}
\\
&=&
-
\displaystyle{
\lim_{x\to t_{0}^+} \frac{d}{dx} w_{r+1}^{k-r}
}
+
\displaystyle{
\lim_{x\to t_{0}^+}\frac{d}{dx}\Bspline{r+1}{k-r-1}{\alpha}(x) 
}
\\
\end{array}
$$
the expected result as 
$
\displaystyle{
\lim_{x\to t_{0}^+} \Bspline{r+1}{k-r-1}{\alpha}(x) 
=\lim_{x\to t_{0}^+} \Bspline{k}{0}{\alpha}(x) =1
}
$
by applying the lemma~\ref{LemmeValExtremite1}.

\item
Let us show that
$$
\displaystyle{
\lim_{x\to t_{0}^+} \frac{d}{dx}\Bspline{r+1}{k-r}{\alpha}(x) 
}
=
\displaystyle{
\lim_{x\to t_{0}^+} \frac{d}{dx} w_{r+1}^{k-r}
}
+
\displaystyle{
\lim_{x\to t_{0}^+} \frac{d}{dx}\Bspline{r+2}{k-r-1}{\alpha}(x) 
}
$$

For $x\in (t_{0}\,,\,t_{k+1})$ we have
$$
\begin{array}{rcl}
\displaystyle{
\Bspline{r+1}{k-r}{\alpha}(x) 
}
&=&
\displaystyle{
\varphi_\alpha\triplet{x}{t_{r+1}}{t_{k+1}}
\Bspline{r+1}{k-r-1}{\alpha}(x) 
}\\
&+&
\displaystyle{
\left(1-\varphi_\alpha\triplet{x}{t_{r+2}}{t_{k+2}}\right)
\Bspline{r+2}{k-r-1}{\alpha}(x) 
}
\\
\end{array}
$$
and
$$
\begin{array}{rcl}
\displaystyle{
\frac{d}{dx}\Bspline{r+1}{k-r}{\alpha}(x) 
}
&=&
\displaystyle{
\frac{d}{dx} \varphi_\alpha\triplet{x}{t_{r+1}}{t_{k+1}}
\Bspline{r+1}{k-r-1}{\alpha}(x) 
}\\
&+&
\displaystyle{
\varphi_\alpha\triplet{x}{t_{r+1}}{t_{k+1}}
\frac{d}{dx} \Bspline{r+1}{k-r-1}{\alpha}(x) 
}\\
&-&
\displaystyle{
\frac{d}{dx} \varphi_\alpha\triplet{x}{t_{r+2}}{t_{k+2}} 
\Bspline{r+2}{k-r-1}{\alpha}(x) 
}
\\
&+&
\displaystyle{
\left(1-\varphi_\alpha\triplet{x}{t_{r+2}}{t_{k+2}}\right)
\frac{d}{dx}\Bspline{r+2}{k-r-1}{\alpha}(x) 
}
\\
\end{array}
$$

By passing to the limit
$$
\begin{array}{rcl}
\displaystyle{
\lim_{x\to t_{0}^+} \frac{d}{dx}\Bspline{r+1}{k-r}{\alpha}(x) 
}
&=&
\displaystyle{
\lim_{x\to t_{0}^+} \frac{d}{dx} \varphi_\alpha\triplet{x}{t_{r+1}}{t_{k+1}}
\lim_{x\to t_{0}^+} \Bspline{r+1}{k-r-1}{\alpha}(x) 
}\\
&+&
\displaystyle{
\lim_{x\to t_{0}^+} \varphi_\alpha\triplet{x}{t_{r+1}}{t_{k+1}}
\lim_{x\to t_{0}^+} \frac{d}{dx} \Bspline{r+1}{k-r-1}{\alpha}(x) 
}\\
&-&
\displaystyle{
\lim_{x\to t_{0}^+} \frac{d}{dx} \varphi_\alpha\triplet{x}{t_{r+2}}{t_{k+2}} 
\lim_{x\to t_{0}^+} \Bspline{r+2}{k-r-1}{\alpha}(x) 
}
\\
&+&
\displaystyle{
\left(1-\lim_{x\to t_{0}^+} \varphi_\alpha\triplet{x}{t_{r+2}}{t_{k+2}}\right)
\lim_{x\to t_{0}^+} \frac{d}{dx}\Bspline{r+2}{k-r-1}{\alpha}(x) 
}
\\
\end{array}
$$

By using the lemma~\ref{LemmeValExtremite1} we obtain
$$
\begin{array}{rcl}
\displaystyle{
\lim_{x\to t_{0}^+} \frac{d}{dx}\Bspline{r+1}{k-r}{\alpha}(x) 
}
&=&
\displaystyle{
\lim_{x\to t_{0}^+} \frac{d}{dx} \varphi_\alpha\triplet{x}{t_{r+1}}{t_{k+1}}
}
\\
&+&
\displaystyle{
\lim_{x\to t_{0}^+} \frac{d}{dx}\Bspline{r+2}{k-r-1}{\alpha}(x) 
}
\\
&=&
\displaystyle{
\lim_{x\to t_{0}^+} \frac{d}{dx} w_{r+1}^{k-r}
}
+
\displaystyle{
\lim_{x\to t_{0}^+} \frac{d}{dx}\Bspline{r+2}{k-r-1}{\alpha}(x) 
}
\\
\end{array}
$$
the expected result.

\item
To finish, let us show that
$$
\displaystyle{
\lim_{x\to t_{0}^+}\frac{d}{dx}\Bspline{i+r}{k-r}{\alpha}(x) 
}
=
\displaystyle{
\lim_{x\to t_{0}^+}\frac{d}{dx}\Bspline{i+r+1}{k-r-1}{\alpha}(x) 
}
$$

For $x\in (t_{0}\,,\,t_{k+1})$ we have
$$
\begin{array}{rcl}
\displaystyle{
\Bspline{i+r}{k-r}{\alpha}(x) 
}
&=&
\displaystyle{
\varphi_\alpha\triplet{x}{t_{i+r}}{t_{k+i}}
\Bspline{i+r}{k-r-1}{\alpha}(x) 
}\\
&+&
\displaystyle{
\left(1-\varphi_\alpha\triplet{x}{t_{i+r+1}}{t_{k+i+1}}\right)
\Bspline{i+r+1}{k-r-1}{\alpha}(x) 
}
\\
\end{array}
$$

We infer
$$
\begin{array}{rcl}
\displaystyle{
\frac{d}{dx} \Bspline{i+r}{k-r}{\alpha}(x) 
}
&=&
\displaystyle{
\frac{d}{dx} \varphi_\alpha\triplet{x}{t_{i+r}}{t_{k+i}}
\Bspline{i+r}{k-r-1}{\alpha}(x) 
}\\
&+&
\displaystyle{
\varphi_\alpha\triplet{x}{t_{i+r}}{t_{k+i}}
\frac{d}{dx} \Bspline{i+r}{k-r-1}{\alpha}(x) 
}\\
&-&
\displaystyle{
\frac{d}{dx} \varphi_\alpha\triplet{x}{t_{i+r+1}}{t_{k+i+1}}
\Bspline{i+r+1}{k-r-1}{\alpha}(x) 
}
\\
&+&
\displaystyle{
\left(1-\varphi_\alpha\triplet{x}{t_{i+r+1}}{t_{k+i+1}}\right)
\frac{d}{dx} \Bspline{i+r+1}{k-r-1}{\alpha}(x) 
}
\\
\end{array}
$$

By passing to the limit, we obtain
$$
\begin{array}{rcl}
\displaystyle{
\lim_{x\to t_{0}^+} \frac{d}{dx} \Bspline{i+r}{k-r}{\alpha}(x) 
}
&=&
\displaystyle{
\lim_{x\to t_{0}^+}\frac{d}{dx} \varphi_\alpha\triplet{x}{t_{i+r}}{t_{k+i}}
\lim_{x\to t_{0}^+}\Bspline{i+r}{k-r-1}{\alpha}(x) 
}\\
&+&
\displaystyle{
\lim_{x\to t_{0}^+}\varphi_\alpha\triplet{x}{t_{i+r}}{t_{k+i}}
\lim_{x\to t_{0}^+}\frac{d}{dx} \Bspline{i+r}{k-r-1}{\alpha}(x) 
}\\
&-&
\displaystyle{
\lim_{x\to t_{0}^+}\frac{d}{dx} \varphi_\alpha\triplet{x}{t_{i+r+1}}{t_{k+i+1}}
\lim_{x\to t_{0}^+}\Bspline{i+r+1}{k-r-1}{\alpha}(x) 
}
\\
&+&
\displaystyle{
\left(1-\lim_{x\to t_{0}^+}\varphi_\alpha\triplet{x}{t_{i+r+1}}{t_{k+i+1}}\right)
\lim_{x\to t_{0}^+}\frac{d}{dx} \Bspline{i+r+1}{k-r-1}{\alpha}(x) 
}
\\
&=&
\displaystyle{
\lim_{x\to t_{0}^+}\frac{d}{dx} \Bspline{i+r+1}{k-r-1}{\alpha}(x) 
}
\\
\end{array}
$$
since for all $i\geq 2$
$$
\begin{array}{rcl}
\displaystyle{
\lim_{x\to t_{0}^+}  \Bspline{i+r}{k-r-1}{\alpha}(x) 
}
&=&
\displaystyle{
\lim_{x\to t_{0}^+}  \Bspline{i+k-1}{0}{\alpha}(x)=0 
}
\\
\end{array}
$$

\end{itemize}

\end{itemize}

\begin{lemme}\label{LemmeDerivExtremite2}

Let $m, k, n\in \NN^*$ such that \mbox{$n \geq k$} and $m=n+ k+1$.
Let $\displaystyle{U=\suite{t_i}_{i=0}^{m}}$ be an open knot vector 
and \mbox{$\alpha \in (-\infty \, , \, 0) \cup (1  \, , \,  \infty)$}.

Consider the rational B-spline basis $\suite{\Bspline{i}{k}{\alpha}}_{i=0}^{n}$ of index $\alpha$
with knot vector $U$ and of degree $k$.
For all $0\leq r\leq k-1$  we have:
\begin{equation}\label{RecDerivExtremite2}
\begin{array}{l}
\displaystyle{
\lim_{x \to t_{m}^-}\frac{d}{dx}  \Bspline{n}{k-r}{\alpha}(x) 
=\lim_{x \to t_{m}^-}\frac{d}{dx}  \Bspline{n}{k-r-1}{\alpha}(x)
+\lim_{x \to t_{m}^-}\frac{d}{dx}  w_{n}^{k-r}(x)
}
\\
\displaystyle{
\lim_{x \to t_{m}^-}\frac{d}{dx}  \Bspline{n-1}{k-r}{\alpha}(x) 
=\lim_{x \to t_{m}^-}\frac{d}{dx}  \Bspline{n-1}{k-r-1}{\alpha}(x)
-\lim_{x \to t_{m}^-}\frac{d}{dx}  w_{n}^{k-r}(x)
}
\\
\end{array}
\end{equation}
with $\displaystyle{w_{i}^{j}(x)
=\varphi_\alpha\triplet{x}{t_{i}}{t_{i+j}}
}$

\end{lemme}

\Preuve
The proof is similar to the one of lemma~\ref{LemmeDerivExtremite1}.
We proceed by recurrence on $r$.

\begin{itemize}
\item
Let $r=0$ 
\begin{itemize}
\item
For $x\in (t_{n}\,,\,t_{m})$ we have
$$
\begin{array}{rcl}
\displaystyle{
\Bspline{n}{k-r}{\alpha}(x) 
}
&=&
\displaystyle{
\Bspline{n}{k}{\alpha}(x) 
}
\\
&=&
\displaystyle{
\varphi_\alpha\triplet{x}{t_{n}}{t_{n+k}}
\Bspline{n}{k-1}{\alpha}(x) 
}\\
&+&
\displaystyle{
\left(1-\varphi_\alpha\triplet{x}{t_{n+1}}{t_{n+k+1}}\right)
\Bspline{n+1}{k-1}{\alpha}(x) 
}
\\
&=&
\displaystyle{
\varphi_\alpha\triplet{x}{t_{n}}{t_{m}}
\Bspline{n}{k-1}{\alpha}(x) 
}
\\
\end{array}
$$
since  
$\Support{\varphi_\alpha\triplet{.}{t_{n+1}}{t_{m}}}=\emptyset$, $U$ is
open.

Thus we have
$$
\begin{array}{rcl}
\displaystyle{
\frac{d}{dx}\Bspline{n}{k-r}{\alpha}(x) 
}
&=&
\displaystyle{
\frac{d}{dx}\varphi_\alpha\triplet{x}{t_{n}}{t_{m}}
\Bspline{n}{k-1}{\alpha}(x) 
}\\
&+&
\displaystyle{
\varphi_\alpha\triplet{x}{t_{n}}{t_{m}}
\frac{d}{dx}\Bspline{n}{k-1}{\alpha}(x) 
}
\\
\end{array}
$$

We deduce that
$$
\begin{array}{rcl}
\displaystyle{
\lim_{x\to t_{m}^-}\frac{d}{dx}\Bspline{n}{k-r}{\alpha}(x) 
}
&=&
\displaystyle{
\lim_{x\to t_{m}^-}\frac{d}{dx}\varphi_\alpha\triplet{x}{t_{n}}{t_{m}}
\lim_{x\to t_{m}^-}\Bspline{n}{k-1}{\alpha}(x) 
}\\
&+&
\displaystyle{
\lim_{x\to t_{m}^-}\varphi_\alpha\triplet{x}{t_{n}}{t_{m}}
\lim_{x\to t_{m}^-}\frac{d}{dx}\Bspline{n}{k-1}{\alpha}(x) 
}
\\
&=&
\displaystyle{
\lim_{x\to t_{m}^-}\frac{d}{dx}\varphi_\alpha\triplet{x}{t_{n}}{t_{m}}
}
+
\displaystyle{
\lim_{x\to t_{m}^-}\frac{d}{dx}\Bspline{n}{k-1}{\alpha}(x) 
}
\\
&=&
\displaystyle{
\lim_{x\to t_{m}^-}\frac{d}{dx}w_{n}^{k}
}
+
\displaystyle{
\lim_{x\to t_{m}^-}\frac{d}{dx}\Bspline{n}{k-1}{\alpha}(x) 
}
\\
&=&
\displaystyle{
\lim_{x\to t_{m}^-}\frac{d}{dx}w_{n}^{k-r}
}
+
\displaystyle{
\lim_{x\to t_{m}^-}\frac{d}{dx}\Bspline{n}{k-r-1}{\alpha}(x) 
}
\\
\end{array}
$$
because $[t_{n}\,,\,t_{m-1})=[t_{n}\,,\,t_{m})$ 
and by using lemma ~\ref{LemmeValExtremite2} successively for $r=1$
and $r=k-1$, we have
$
\displaystyle{
\lim_{x\to t_{m}^-}\Bspline{n}{k-1}{\alpha}(x) =
\lim_{x\to t_{m}^-}\Bspline{n}{0}{\alpha}(x) =
\lim_{x\to t_{n+1}^-}\Bspline{n}{0}{\alpha}(x) =1
}
$
and also the fact that
$
\displaystyle{
\lim_{x\to t_{m}^-}\varphi_\alpha\triplet{x}{t_{n}}{t_{m}}=1
}
$.

We have thus proved that for $r=0$ we have 
$$
\displaystyle{
\lim_{x\to t_{m}^-}\frac{d}{dx}\Bspline{n}{k-r}{\alpha}(x) 
}
=
\displaystyle{
\lim_{x\to t_{m}^-}\frac{d}{dx}w_{n}^{k-r}
+
\lim_{x\to t_{m}^-}\frac{d}{dx}\Bspline{n}{k-r-1}{\alpha}(x) 
}
$$

\item
For  $x\in (t_{n}\,,\,t_{m})$ and $k\geq 2$ 
we have
$$
\begin{array}{rcl}
\displaystyle{
\Bspline{n-1}{k-r}{\alpha}(x) 
}
&=&
\displaystyle{
\Bspline{n-1}{k}{\alpha}(x) 
}
\\
&=&
\displaystyle{
\varphi_\alpha\triplet{x}{t_{n-1}}{t_{n+k-1}}
\Bspline{n-1}{k-1}{\alpha}(x) 
}\\
&+&
\displaystyle{
\left(1-\varphi_\alpha\triplet{x}{t_{n}}{t_{n+k}}\right)
\Bspline{n}{k-1}{\alpha}(x) 
}
\\
&=&
\displaystyle{
\varphi_\alpha\triplet{x}{t_{n-1}}{t_{m}}
\Bspline{n-1}{k-1}{\alpha}(x) 
}\\
&+&
\displaystyle{
\left(1-\varphi_\alpha\triplet{x}{t_{n}}{t_{m}}\right)
\Bspline{n}{k-1}{\alpha}(x) 
}
\\
\end{array}
$$

We then have 
$$
\begin{array}{rcl}
\displaystyle{
\frac{d}{dx} \Bspline{n-1}{k-r}{\alpha}(x) 
}
&=&
\displaystyle{
\frac{d}{dx} \varphi_\alpha\triplet{x}{t_{n-1}}{t_{m}}
\Bspline{n-1}{k-1}{\alpha}(x) 
}\\
&+&
\displaystyle{
\varphi_\alpha\triplet{x}{t_{n-1}}{t_{m}}
\frac{d}{dx} \Bspline{n-1}{k-1}{\alpha}(x) 
}\\
&-&
\displaystyle{
\frac{d}{dx} \varphi_\alpha\triplet{x}{t_{n}}{t_{m}}
\Bspline{n}{k-1}{\alpha}(x) 
}
\\
&+&
\displaystyle{
\left(1-\varphi_\alpha\triplet{x}{t_{n}}{t_{m}}\right)
\frac{d}{dx} \Bspline{n}{k-1}{\alpha}(x) 
}
\\
\end{array}
$$

We thus deduce that 
$$
\begin{array}{rcl}
\displaystyle{
\lim_{x\to t_{m}^-} \frac{d}{dx} \Bspline{n-1}{k-r}{\alpha}(x) 
}
&=&
\displaystyle{
\lim_{x\to t_{m}^-} \frac{d}{dx} \varphi_\alpha\triplet{x}{t_{n-1}}{t_{m}}
\lim_{x\to t_{m}^-} \Bspline{n-1}{k-1}{\alpha}(x) 
}\\
&+&
\displaystyle{
\lim_{x\to t_{m}^-} \varphi_\alpha\triplet{x}{t_{n-1}}{t_{m}}
\lim_{x\to t_{m}^-} \frac{d}{dx} \Bspline{n-1}{k-1}{\alpha}(x) 
}\\
&-&
\displaystyle{
\lim_{x\to t_{m}^-} \frac{d}{dx} \varphi_\alpha\triplet{x}{t_{n}}{t_{m}}
\lim_{x\to t_{m}^-} \Bspline{n}{k-1}{\alpha}(x) 
}
\\
&+&
\displaystyle{
\left(1-\lim_{x\to t_{m}^-} \varphi_\alpha\triplet{x}{t_{n}}{t_{m}}\right)
\lim_{x\to t_{m}^-} \frac{d}{dx} \Bspline{n}{k-1}{\alpha}(x) 
}
\\
&=&
\displaystyle{
\lim_{x\to t_{m}^-} \frac{d}{dx} \Bspline{n-1}{k-1}{\alpha}(x) 
}
-
\displaystyle{
\lim_{x\to t_{m}^-} \frac{d}{dx} \varphi_\alpha\triplet{x}{t_{n}}{t_{m}}
}
\\
\end{array}
$$

because from lemma ~\ref{LemmeValExtremite2} we have
$$
\begin{array}{rcl}
\displaystyle{
\lim_{x\to t_{m}^-} \Bspline{n-1}{k-1}{\alpha}(x) 
}
&=&
\displaystyle{
\lim_{x\to t_{m}^-} \Bspline{n-1}{0}{\alpha}(x) =0
}
\\
\displaystyle{
\lim_{x\to t_{m}^-} \Bspline{n}{k-1}{\alpha}(x)  
}
&=&
\displaystyle{
\lim_{x\to t_{m}^-} \Bspline{n}{0}{\alpha}(x)  =1
}
\\
\end{array}
$$
and also the fact that 
$$
\displaystyle{
\lim_{x\to t_{m}^-} \varphi_\alpha\triplet{x}{t_{n-1}}{t_{m}} =
\lim_{x\to t_{m}^-} \varphi_\alpha\triplet{x}{t_{n}}{t_{m}} =1
}
$$

\end{itemize}

\item
Let $0 <r<k$. Suppose that for all $0\leq j < r$ we have
$$
\begin{array}{l}
\displaystyle{
\lim_{x \to t_{m}^-}\frac{d}{dx}  \Bspline{n}{k-j}{\alpha}(x) 
=\lim_{x \to t_{m}^-}\frac{d}{dx}  \Bspline{n}{k-j-1}{\alpha}(x)
+\lim_{x \to t_{m}^-}\frac{d}{dx}  w_{n}^{k-j}(x)
}
\\
\displaystyle{
\lim_{x \to t_{m}^-}\frac{d}{dx}  \Bspline{n-1}{k-j}{\alpha}(x) 
=\lim_{x \to t_{m}^-}\frac{d}{dx}  \Bspline{n-1}{k-j-1}{\alpha}(x)
-\lim_{x \to t_{m}^-}\frac{d}{dx}  w_{n}^{k-j}(x)
}
\\
\displaystyle{
\lim_{x \to t_{m}^-}\frac{d}{dx}  \Bspline{i}{k-j}{\alpha}(x) 
=\lim_{x \to t_{m}^-}\frac{d}{dx}  \Bspline{i}{k-j-1}{\alpha}(x)
},\, n-k\leq i \leq n-2,\, k\geq 2
\\
\end{array}
$$

\begin{itemize}
\item
Let us show that  
$$
\displaystyle{
\lim_{x \to t_{m}^-}\frac{d}{dx}  \Bspline{n}{k-r}{\alpha}(x) 
=\lim_{x \to t_{m}^-}\frac{d}{dx}  \Bspline{n}{k-r-1}{\alpha}(x)
+\lim_{x \to t_{m}^-}\frac{d}{dx}  w_{n}^{k-r}(x)
}
$$

For $x\in (t_{n}\,,\,t_{m})$ we have
$$
\begin{array}{rcl}
\displaystyle{
\Bspline{n}{k-r}{\alpha}(x) 
}
&=&
\displaystyle{
\varphi_\alpha\triplet{x}{t_{n}}{t_{n+k-r}}
\Bspline{n}{k-r-1}{\alpha}(x) 
}\\
&+&
\displaystyle{
\left(1-\varphi_\alpha\triplet{x}{t_{n+1}}{t_{n+k-r+1}}\right)
\Bspline{n+1}{k-r-1}{\alpha}(x) 
}
\\
&=&
\displaystyle{
\varphi_\alpha\triplet{x}{t_{n}}{t_{n+k-r}}
\Bspline{n}{k-r-1}{\alpha}(x) 
}\\
\end{array}
$$
because $\Support{\Bspline{n+1}{k-r-1}{\alpha}}=[t_{n+1}\,,\,t_{m-r})=\emptyset$
since $r<k$.

Then
$$
\begin{array}{rcl}
\displaystyle{
\frac{d}{dx}\Bspline{n}{k-r}{\alpha}(x) 
}
&=&
\displaystyle{
\frac{d}{dx} \varphi_\alpha\triplet{x}{t_{n}}{t_{n+k-r}}
\Bspline{n}{k-r-1}{\alpha}(x) 
}\\
&+&
\displaystyle{
\varphi_\alpha\triplet{x}{t_{n}}{t_{n+k-r}}
\frac{d}{dx} \Bspline{n}{k-r-1}{\alpha}(x) 
}\\
\end{array}
$$

Passing to the limit, we get
$$
\begin{array}{rcl}
\displaystyle{
\lim_{x \to t_{m}^-} \frac{d}{dx}\Bspline{n}{k-r}{\alpha}(x) 
}
&=&
\displaystyle{
\lim_{x \to t_{m}^-} \frac{d}{dx} \varphi_\alpha\triplet{x}{t_{n}}{t_{n+k-r}}
\lim_{x \to t_{m}^-} \Bspline{n}{k-r-1}{\alpha}(x) 
}\\
&+&
\displaystyle{
\lim_{x \to t_{m}^-} \varphi_\alpha\triplet{x}{t_{n}}{t_{n+k-r}}
\lim_{x \to t_{m}^-} \frac{d}{dx} \Bspline{n}{k-r-1}{\alpha}(x) 
}\\
&=&
\displaystyle{
\lim_{x \to t_{m}^-} \frac{d}{dx} \varphi_\alpha\triplet{x}{t_{n}}{t_{n+k-r}}
}\\
&+&
\displaystyle{
\lim_{x \to t_{m}^-} \frac{d}{dx} \Bspline{n}{k-r-1}{\alpha}(x) 
}\\
\end{array}
$$

the expected result because 
$$
\displaystyle{
\lim_{x \to t_{m}^-} \Bspline{n}{k-r-1}{\alpha}(x) 
=\lim_{x \to t_{m}^-} \Bspline{n}{0}{\alpha}(x) 
=\lim_{x \to t_{n+1}^-} \Bspline{n}{0}{\alpha}(x) =1
}
$$
and
$$
\displaystyle{
\lim_{x \to t_{m}^-} \varphi_\alpha\triplet{x}{t_{n}}{t_{n+k-r}}
=\lim_{x \to t_{m}^-} \varphi_\alpha\triplet{x}{t_{n}}{t_{m}}=1
}
$$
by applying lemma~\ref{LemmeValExtremite2}.

\item
Let us show that
$$
\displaystyle{
\lim_{x \to t_{m}^-}  \frac{d}{dx}\Bspline{n-1}{k-r}{\alpha}(x) 
}
=-
\displaystyle{
\lim_{x \to t_{m}^-}  \frac{d}{dx} w_{n}^{k-r}
}
+
\displaystyle{
\lim_{x \to t_{m}^-}  \frac{d}{dx}\Bspline{n-1}{k-r-1}{\alpha}(x) 
}
$$

For $x\in (t_{n}\,,\,t_{m})$ we have
$$
\begin{array}{rcl}
\displaystyle{
\Bspline{n-1}{k-r}{\alpha}(x) 
}
&=&
\displaystyle{
\varphi_\alpha\triplet{x}{t_{n-1}}{t_{n+k-r-1}}
\Bspline{n-1}{k-r-1}{\alpha}(x) 
}\\
&+&
\displaystyle{
\left(1-\varphi_\alpha\triplet{x}{t_{n}}{t_{n+k-r}}\right)
\Bspline{n}{k-r-1}{\alpha}(x) 
}
\\
\end{array}
$$
and
$$
\begin{array}{rcl}
\displaystyle{
\frac{d}{dx} \Bspline{n-1}{k-r}{\alpha}(x) 
}
&=&
\displaystyle{
\frac{d}{dx} \varphi_\alpha\triplet{x}{t_{n-1}}{t_{n+k-r-1}}
\Bspline{n-1}{k-r-1}{\alpha}(x) 
}\\
&+&
\displaystyle{
\varphi_\alpha\triplet{x}{t_{n-1}}{t_{n+k-r-1}}
\frac{d}{dx} \Bspline{n-1}{k-r-1}{\alpha}(x) 
}\\
&-&
\displaystyle{
\frac{d}{dx} \varphi_\alpha\triplet{x}{t_{n}}{t_{n+k-r}} 
\Bspline{n}{k-r-1}{\alpha}(x) 
}
\\
&+&
\displaystyle{
\left(1-\varphi_\alpha\triplet{x}{t_{n}}{t_{n+k-r}}\right)
\frac{d}{dx} \Bspline{n}{k-r-1}{\alpha}(x) 
}
\\
\end{array}
$$

Passing to the limit, we have
$$
\begin{array}{rcl}
\displaystyle{
\lim_{x \to t_{m}^-} \frac{d}{dx} \Bspline{n-1}{k-r}{\alpha}(x) 
}
&=&
\displaystyle{
\lim_{x \to t_{m}^-} \frac{d}{dx} \varphi_\alpha\triplet{x}{t_{n-1}}{t_{n+k-r-1}}
\lim_{x \to t_{m}^-} \Bspline{n-1}{k-r-1}{\alpha}(x) 
}\\
&+&
\displaystyle{
\lim_{x \to t_{m}^-} \varphi_\alpha\triplet{x}{t_{n-1}}{t_{n+k-r-1}}
\lim_{x \to t_{m}^-} \frac{d}{dx} \Bspline{n-1}{k-r-1}{\alpha}(x) 
}\\
&-&
\displaystyle{
\lim_{x \to t_{m}^-} \frac{d}{dx} \varphi_\alpha\triplet{x}{t_{n}}{t_{n+k-r}} 
\lim_{x \to t_{m}^-} \Bspline{n}{k-r-1}{\alpha}(x) 
}
\\
&+&
\displaystyle{
\lim_{x \to t_{m}^-} \left(1-\varphi_\alpha\triplet{x}{t_{n}}{t_{n+k-r}}\right)
\lim_{x \to t_{m}^-} \frac{d}{dx} \Bspline{n}{k-r-1}{\alpha}(x) 
}
\\
\end{array}
$$

By using lemma~\ref{LemmeValExtremite2} we obtain
$$
\begin{array}{rcl}
\displaystyle{
\lim_{x \to t_{m}^-} \frac{d}{dx} \Bspline{n-1}{k-r}{\alpha}(x) 
}
&=&
\displaystyle{
\lim_{x \to t_{m}^-} \varphi_\alpha\triplet{x}{t_{n-1}}{t_{n+k-r-1}}
\lim_{x \to t_{m}^-} \frac{d}{dx} \Bspline{n-1}{k-r-1}{\alpha}(x) 
}\\
&-&
\displaystyle{
\lim_{x \to t_{m}^-} \frac{d}{dx} \varphi_\alpha\triplet{x}{t_{n}}{t_{n+k-r}} 
\lim_{x \to t_{m}^-} \Bspline{n}{k-r-1}{\alpha}(x) 
}
\\
&=&
\displaystyle{
\lim_{x \to t_{m}^-} \frac{d}{dx} \Bspline{n-1}{k-r-1}{\alpha}(x) 
}
-
\displaystyle{
\lim_{x \to t_{m}^-} \frac{d}{dx} \varphi_\alpha\triplet{x}{t_{n}}{t_{n+k-r}} 
}
\\
\end{array}
$$
the desired result because 
$$
\begin{array}{rcl}
\displaystyle{
\lim_{x \to t_{m}^-}\Bspline{n-1}{k-r-1}{\alpha}(x) 
}
&=&
\displaystyle{
\lim_{x \to t_{m}^-}\Bspline{n-1}{0}{\alpha}(x) =0
}\\
\displaystyle{
\lim_{x \to t_{m}^-}\Bspline{n}{k-r-1}{\alpha}(x) 
}
&=&
\displaystyle{
\lim_{x \to t_{m}^-}\Bspline{n}{0}{\alpha}(x) =1
}\\
\end{array}
$$
and 
$$
\lim_{x \to t_{m}^-}  \varphi_\alpha\triplet{x}{t_{n}}{t_{n+k-r}} 
=\lim_{x \to t_{m}^-}  \varphi_\alpha\triplet{x}{t_{n}}{t_{m}} =1
$$

\end{itemize}

\end{itemize}


\begin{lemme}\label{LemmeDerivExtremite21}

Let $m, k, n\in \NN^*$ such that \mbox{$n \geq k$} and $m=n+ k+1$.
Let $\displaystyle{U=\suite{t_i}_{i=0}^{m}}$ be an open knot vector, 
let \mbox{$\alpha \in (-\infty \, , \, 0) \cup (1  \, , \,  \infty)$}.

Consider the rational B-spline basis $\suite{\Bspline{i}{k}{\alpha}}_{i=0}^{n}$ of index $\alpha$,
$U$ as a knot vector and of degree $k$.
For all $i \leq n-2,\, k\geq 2$  we have:
\begin{equation}\label{RecDerivExtremite21}
\displaystyle{
\lim_{x \to t_{m}^-}\frac{d}{dx}  \Bspline{i}{k}{\alpha}(x) 
=0
}
\end{equation}
with $\displaystyle{w_{i}^{j}(x)
=\varphi_\alpha\triplet{x}{t_{i}}{t_{i+j}}
}$

\end{lemme}

\Preuve
We want to show that
$
\displaystyle{
\lim_{x \to t_{m}^-}\frac{d}{dx}  \Bspline{i}{k}{\alpha}(x) 
=0
}
$ 
for all $i\leq n-2$.

Let us first remark that
$$
\begin{array}{rcl}
\Support{\Bspline{i}{k}{\alpha}}\cap [t_{n}\,,\,t_{n+1}) \neq \emptyset
&\equivaut&
 [t_{i}\,,\,t_{i+k+1})\cap [t_{n}\,,\,t_{n+1}) \neq \emptyset
\\
&\equivaut&
 [t_{n}\,,\,t_{n+1})\subset  [t_{i}\,,\,t_{i+k+1}) 
\\
&\equivaut&
 t_{i} \leq t_{n} < t_{n+1} \leq t_{i+k+1}
\\
&\equivaut&
n-k \leq i \leq n
\\
\end{array}
$$

\begin{itemize}
\item Suppose $i \leq n-k-1$. Then we have  
$
\Support{\Bspline{i}{k}{\alpha}}\cap [t_{n}\,,\,t_{n+1}) = \emptyset
$.
Thus for all
$
x \in [t_{n}\,,\,t_{n+1})
$
we have 
$
\displaystyle{
\Bspline{i}{k}{\alpha} (x) = 0
}
$
and then
$
\displaystyle{
\frac{d}{dx}\Bspline{i}{k}{\alpha} (x) = 0
}
$. 

We deduce that 
$
\displaystyle{
\lim_{x\to t_{m}^-} \frac{d}{dx}\Bspline{i}{k}{\alpha} (x) = 0
}
$. 

\item Suppose $i = n-k$ with $k\geq 2$. 

Let $x \in [t_{n}\,,\,t_{n+1})$.  We have
$$
\begin{array}{rcl}
\displaystyle{
\Bspline{n-k}{k}{\alpha} (x)
}
&=&
\displaystyle{
 \varphi_\alpha\triplet{x}{t_{n-k}}{t_{n}}
 \Bspline{n-k}{k-1}{\alpha} (x)
}
\\
&+&
\displaystyle{
\left(1- \varphi_\alpha\triplet{x}{t_{n-k+1}}{t_{n+1}} \right) 
\Bspline{n-k+1}{k-1}{\alpha} (x)
}
\\
&=&
\displaystyle{
\left(1- \varphi_\alpha\triplet{x}{t_{n-k+1}}{t_{n+1}} \right) 
\Bspline{n-k+1}{k-1}{\alpha} (x)
}
\\
\end{array}
$$
because
$
\Support{\Bspline{n-k}{k-1}{\alpha}}
\cap [t_{n}\,,\,t_{n+1}) = \emptyset
$.

Thus we have
$$
\begin{array}{rcl}
\displaystyle{
\frac{d}{dx} \Bspline{n-k}{k}{\alpha} (x)
}
&=&
\displaystyle{
- \frac{d}{dx} \varphi_\alpha\triplet{x}{t_{n-k+1}}{t_{n+1}} 
\Bspline{n-k+1}{k-1}{\alpha} (x)
}\\
&+&
\displaystyle{
\left(1- \varphi_\alpha\triplet{x}{t_{n-k+1}}{t_{n+1}} \right)
 \frac{d}{dx} \Bspline{n-k+1}{k-1}{\alpha} (x)
}
\\
\end{array}
$$

Passing to the limit, we have
$$
\begin{array}{rcl}
\displaystyle{
\lim_{x\to t_{m}^-} \frac{d}{dx} \Bspline{n-k}{k}{\alpha} (x)
}
&=&
\displaystyle{
-\lim_{x\to t_{m}^-}  \frac{d}{dx} \varphi_\alpha\triplet{x}{t_{n-k+1}}{t_{n+1}} 
\lim_{x\to t_{m}^-} \Bspline{n-k+1}{k-1}{\alpha} (x)
}\\
&+&
\displaystyle{
\left(1- \lim_{x\to t_{m}^-} \varphi_\alpha\triplet{x}{t_{n-k+1}}{t_{n+1}} \right)
\lim_{x\to t_{m}^-}  \frac{d}{dx} \Bspline{n-k+1}{k-1}{\alpha} (x)
}
\\
&=&
\displaystyle{
-\lim_{x\to t_{m}^-}  \frac{d}{dx} \varphi_\alpha\triplet{x}{t_{n-k+1}}{t_{n+1}} 
\lim_{x\to t_{m}^-} \Bspline{n-k+1}{k-1}{\alpha} (x)
}=0\\
\end{array}
$$
because
$$
\begin{array}{rcl}
\displaystyle{
\lim_{x\to t_{m}^-}  \varphi_\alpha\triplet{x}{t_{n-k+1}}{t_{n+1}} 
}
&=& 1\\
\displaystyle{
\lim_{x\to t_{m}^-} \Bspline{n-k+1}{k-1}{\alpha} (x)
}
&=& 0\\
\end{array}
$$

\item Suppose $ n-k <i \leq n-2$ with $k\geq 2$. 
Then we have $t_{i}\leq t_{i+1} \leq t_{n} < t_{n+1} \leq t_{i+k} \leq t_{i+k+1} $.

Let $x \in [t_{n}\,,\,t_{n+1})$.  
We have
$$
\begin{array}{rcl}
\displaystyle{
\Bspline{i}{k}{\alpha}(x) 
}
&=&
\displaystyle{
\varphi_\alpha\triplet{x}{t_{i}}{t_{i+k}}
\Bspline{i}{k-1}{\alpha}(x) 
}\\
&+&
\displaystyle{
\left(1-\varphi_\alpha\triplet{x}{t_{i+1}}{t_{k+i+1}}\right)
\Bspline{i+1}{k-1}{\alpha}(x) 
}
\\
\end{array}
$$
Hence

$$
\begin{array}{rcl}
\displaystyle{
 \frac{d}{dx} \Bspline{i}{k}{\alpha}(x) 
}
&=&
\displaystyle{
 \frac{d}{dx} \varphi_\alpha\triplet{x}{t_{i}}{t_{i+k}}
\Bspline{i}{k-1}{\alpha}(x) 
}\\
&+&
\displaystyle{
\varphi_\alpha\triplet{x}{t_{i}}{t_{i+k}}
 \frac{d}{dx} \Bspline{i}{k-1}{\alpha}(x) 
}\\
&-&
\displaystyle{
\frac{d}{dx} \varphi_\alpha\triplet{x}{t_{i+1}}{t_{k+i+1}}
\Bspline{i+1}{k-1}{\alpha}(x) 
}
\\
&+&
\displaystyle{
\left(1-\varphi_\alpha\triplet{x}{t_{i+1}}{t_{k+i+1}}\right)
 \frac{d}{dx} \Bspline{i+1}{k-1}{\alpha}(x) 
}
\\
\end{array}
$$

Passing to the limit, we have
$$
\begin{array}{rcl}
\displaystyle{
 \lim_{x\to t_{m}^-} \frac{d}{dx} \Bspline{i}{k}{\alpha}(x) 
}
&=&
\displaystyle{
 \lim_{x\to t_{m}^-} \frac{d}{dx} \varphi_\alpha\triplet{x}{t_{i}}{t_{i+k}}
\lim_{x\to t_{m}^-} \Bspline{i}{k-1}{\alpha}(x) 
}\\
&+&
\displaystyle{
\lim_{x\to t_{m}^-} \varphi_\alpha\triplet{x}{t_{i}}{t_{i+k}}
 \lim_{x\to t_{m}^-} \frac{d}{dx} \Bspline{i}{k-1}{\alpha}(x) 
}\\
&-&
\displaystyle{
\lim_{x\to t_{m}^-} \frac{d}{dx} \varphi_\alpha\triplet{x}{t_{i+1}}{t_{k+i+1}}
\lim_{x\to t_{m}^-} \Bspline{i+1}{k-1}{\alpha}(x) 
}
\\
&+&
\displaystyle{
\left(1-\lim_{x\to t_{m}^-} \varphi_\alpha\triplet{x}{t_{i+1}}{t_{k+i+1}}\right)
 \lim_{x\to t_{m}^-} \frac{d}{dx} \Bspline{i+1}{k-1}{\alpha}(x) 
}
\\
\end{array}
$$

Thus we obtain
$$
\displaystyle{
 \lim_{x\to t_{m}^-} \frac{d}{dx} \Bspline{i}{k}{\alpha}(x) 
}
=
\displaystyle{
 \lim_{x\to t_{m}^-} \frac{d}{dx} \Bspline{i}{k-1}{\alpha}(x) 
}
$$
since
$$
\begin{array}{rcl}
\displaystyle{
\lim_{x\to t_{m}^-} \Bspline{i}{k-1}{\alpha}(x) 
}
&=&
\displaystyle{
 0 \quad \forall i\leq n-2
}
\\
\end{array}
$$
and for all $n-k\leq i \leq n-2$
$$
\begin{array}{rcl}
\displaystyle{
\lim_{x\to t_{m}^-} \varphi_\alpha\triplet{x}{t_{i}}{t_{i+k}}
}
&=&
1
\\
\displaystyle{
\lim_{x\to t_{m}^-} \varphi_\alpha\triplet{x}{t_{i+1}}{t_{i+k+1}}
}
&=&
1
\\
\end{array}
$$

As it is true for $r=0$  then let us show by  recurrence
that for all  $0 < r \leq k-1$ we have
$$
\displaystyle{
\lim_{x \to t_{m}^-}\frac{d}{dx}  \Bspline{i}{k-r}{\alpha}(x) 
=\lim_{x \to t_{m}^-}\frac{d}{dx}  \Bspline{i}{k-r-1}{\alpha}(x)
}
$$

\end{itemize}

\begin{proposition}[Regularity property]\label{PropoRegular}

Let $m, k, n\in \NN^*$ such that \mbox{$n \geq k$} and  \mbox{$m=n+ k+1$}.
Let $\displaystyle{U=\suite{t_i}_{i=0}^{m}}$ be a knot vector, 
let  \mbox{$\alpha \in (-\infty \, , \, 0) \cup (1  \, , \,  \infty)$}.

Consider the rational B-spline $\suite{\Bspline{i}{k}{\alpha}}_{i=0}^{n}$
of index $\alpha$, $U$ as knot vector and of degree $k$.
We have the following properties:
\begin{enumerate}
\item
For all $i=0, \ldots , n$,
$\Bspline{i}{k}{\alpha}$ is of class ${\cal C}^\infty$ 
on all $(t_{j}\,,\,t_{j+1})$ if  \mbox{$t_{j}<t_{j+1}$}.

\item
For all $i=0, \ldots , n$,
$\Bspline{i}{k}{\alpha}$ is left and right differentiable at all
 $t_{j}$ for all $j$.

\item
If $U$ is an open knot vector then we have  
\begin{enumerate}
\item
$$
\begin{array}{rcl}
\displaystyle{
\lim_{x \to t_{0}^+}\frac{d}{dx}\Bspline{0}{k}{\alpha}(x)
}
&=&
-\displaystyle{
\lim_{x \to t_{0}^+}\frac{d}{dx}\Bspline{1}{k}{\alpha}(x)
}
\\
&=&
-\displaystyle{
\frac{\alpha k}{(\alpha -1){\left(t_{k+1}-t_{0} \right)}}
}
\\
\displaystyle{
\lim_{x \to t_{0}^+}\frac{d}{dx}\Bspline{i}{k}{\alpha}(x)
}
&=& 0 \textrm{ for all } 2\leq i \leq n
\\
\end{array}
$$

\item
$$
\begin{array}{rcl}
\displaystyle{
\lim_{x \to t_{m}^-}\frac{d}{dx}\Bspline{n}{k}{\alpha}(x)
}
&=&
-\displaystyle{
\lim_{x \to t_{m}^-}\frac{d}{dx}\Bspline{n-1}{k}{\alpha}(x)
}
\\
&=&
\displaystyle{
\frac{(\alpha -1)k}{\alpha{\left(t_{m}-t_{n} \right)}}
}
\\
\displaystyle{
\lim_{x \to t_{m}^-}\frac{d}{dx}\Bspline{i}{k}{\alpha}(x)
}
&=& 0 \textrm{ for all } 0\leq i \leq n-2
\\
\end{array}
$$

\end{enumerate}

\end{enumerate}
By definition,  for all $0\leq i \leq n$,
$$
\begin{array}{rcl}
\displaystyle{
\frac{d}{dx}\Bspline{i}{k}{\alpha}(t_{0})
}
&=&
\displaystyle{
\lim_{x \to t_{0}^+}\frac{d}{dx}\Bspline{i}{k}{\alpha}(x)
}\\
\displaystyle{
\frac{d}{dx}\Bspline{i}{k}{\alpha}(t_{m})
}
&=&
\displaystyle{
\lim_{x \to t_{m}^-}\frac{d}{dx}\Bspline{i}{k}{\alpha}(x)
}\\
\end{array}
$$.

\end{proposition}

\Preuve

\begin{enumerate}
\item  ${\cal C}^\infty$ regularity except on the nodes is a consequence 
of the fact that
$\displaystyle{\Bspline{i}{k}{\alpha}}$ is picewise rational function,
as stated in proposition~ \ref{PropoContinu} 
on continuity property.

\item 
The basis functions 
$\displaystyle{ \Bspline{i}{k}{\alpha}}$ are of  ${\cal C}^0$
on $\segment{t_{0}}{t_{m}}$ and  ${\cal C}^1$ on
 $\displaystyle{\bigcup_{i=0}^{m-1} (t_{i}\,,\,t_{i+1})}$. 
 It is sufficient to prove that for all
  $i=0, \ldots, n$ and all $j=0, \ldots, m-1$ such that 
  \mbox{$\displaystyle{t_{j}<t_{j+1}}$}, we have  
 $
 \displaystyle{
 \lim_{x\to t_{j}^+} \frac{d}{dx} \Bspline{i}{k}{\alpha}(x) \in \RR
 }
 $
and
 $
 \displaystyle{
 \lim_{x\to t_{j+1}^-} \frac{d}{dx}   \Bspline{i}{k}{\alpha}(x) \in \RR
 }
 $.

We will proceed by recurrence on $k$.

\begin{itemize}
 \item 
 Let $k=1$. Assume a multiplicity$m_i=1$ for all
interior node  $t_{i}$. Thus
$$
\Bspline{i}{1}{\alpha}(x)
=
\left\lbrace
\begin{array}{rcl}
\varphi_\alpha\triplet{x}{t_{i}}{t_{i+1}}
&& 
\textrm{ if } x\in [t_{i}\,,\,t_{i+1})	\neq \emptyset
\\
{1-\varphi_\alpha\triplet{x}{t_{i+1}}{t_{i+2}}}
&& 
\textrm{ if } x\in [t_{i+1}\,,\,t_{i+2})\neq \emptyset
\\
0
&& 
\textrm{ otherwise }
\\
\end{array}
\right.
$$

One deduces that
$$
\frac{d}{dx} \Bspline{i}{1}{\alpha}(x)
=
\left\lbrace
\begin{array}{rcl}
\frac{d}{dx} \varphi_\alpha\triplet{x}{t_{i}}{t_{i+1}}
&& 
\textrm{ if } x\in (t_{i}\,,\,t_{i+1})\neq \emptyset
\\
-\frac{d}{dx} \varphi_\alpha\triplet{x}{t_{i+1}}{t_{i+2}}
&& 
\textrm{ if } x\in (t_{i+1}\,,\,t_{i+2})\neq \emptyset
\\
0
&& 
\textrm{ otherwise }
\\
\end{array}
\right.
$$

From this we obtain:
$$
\begin{array}{rcl}
\displaystyle{
 \lim_{x\to t_{i}^-}\frac{d}{dx} \Bspline{i}{1}{\alpha}(x)
}
&=&
0
\\
\displaystyle{
\lim_{x\to t_{i}^+}\frac{d}{dx} \Bspline{i}{1}{\alpha}(x)
}
&=&
\displaystyle{
\lim_{x\to t_{i}^+} \frac{d}{dx} \varphi_\alpha\triplet{x}{t_{i}}{t_{i+1}}
}
\\
&=&
\displaystyle{
\frac{\alpha}{(\alpha-1)\left(t_{i+1}-t_{i} \right)} \in \RR
}
\\
\displaystyle{
\lim_{x\to t_{i+1}^-}\frac{d}{dx} \Bspline{i}{1}{\alpha}(x)
}
&=&
\displaystyle{
\lim_{x\to t_{i+1}^-} \frac{d}{dx} \varphi_\alpha\triplet{x}{t_{i}}{t_{i+1}}
}
\\
&=&
\displaystyle{
\frac{\alpha-1}{\alpha\left(t_{i+1}-t_{i} \right)} \in \RR
}
\\
\displaystyle{
\lim_{x\to t_{i+1}^+}\frac{d}{dx} \Bspline{i}{1}{\alpha}(x)
}
&=&
\displaystyle{
-\lim_{x\to t_{i+1}^+} \frac{d}{dx} \varphi_\alpha\triplet{x}{t_{i+1}}{t_{i+2}}
}
\\
&=&
\displaystyle{
-\frac{\alpha}{(\alpha-1)\left(t_{i+2}-t_{i+1} \right)} \in \RR
}
\\
\displaystyle{
\lim_{x\to t_{i+2}^-}\frac{d}{dx} \Bspline{i}{1}{\alpha}(x)
}
&=&
\displaystyle{
-\lim_{x\to t_{i+2}^-} \frac{d}{dx} \varphi_\alpha\triplet{x}{t_{i+1}}{t_{i+2}}
}
\\
&=&
\displaystyle{
-\frac{\alpha-1}{\alpha\left(t_{i+2}-t_{i+1} \right)} \in \RR
}
\\
\displaystyle{
\lim_{x\to t_{i+2}^+}\frac{d}{dx} \Bspline{i}{1}{\alpha}(x)
}
&=&
0
\\
\end{array}
$$

We can conclude that $\Bspline{i}{1}{\alpha}$ is left and right differentiable at any point 
if $U$ only admits interior points of multiplicity $1$.
\item 
Let $k> 1$ and suppose that for all \mbox{$1\leq s \leq k-1$}
and all  \mbox{$i=0, \ldots, m-s-1$}
$\Bspline{i}{s}{\alpha}$ is left and right differentiable at all node 
of multiplicity at most $s$.

As for all $x\in \RR$
$$
\begin{array}{rcl}
\displaystyle{
\Bspline{i}{k}{\alpha}(x)
}
&=&
\displaystyle{
 \varphi_\alpha\triplet{x}{t_{i}}{t_{i+k}}
\Bspline{i}{k-1}{\alpha}(x)
}
\\
&+&
\displaystyle{
\left(1- \varphi_\alpha\triplet{x}{t_{i+1}}{t_{i+k+1}}\right)
\Bspline{i+1}{k-1}{\alpha}(x)
}\\
\end{array}
$$
then  if for all  $i$ $\Bspline{i}{k-1}{\alpha}$ is left and right differentiable
at a certain node $t_{j}$, 
$\Bspline{i}{k}{\alpha}$ is also left differentiable at $t_{j}$
as product and sum of left differentiable functions at $t_{j}$ because
from remark ~\ref{RemRegularParam}, all
$\displaystyle{\varphi_\alpha\triplet{.}{t_{i}}{t_{i+k}}}$
is left and right differentiable at any point of $\RR$

It is also the case for the right differentiability.
\end{itemize}

\item 
\begin{enumerate}
\item
Using lemma~\ref{LemmeDerivExtremite1} one can prove that:
\begin{itemize}
\item
on one hand,
$$
\begin{array}{rcl}
\displaystyle{
\lim_{x \to t_{0}^+}\frac{d}{dx}\Bspline{0}{k}{\alpha}(x)
}
&=&
\displaystyle{
\lim_{x \to t_{0}^+}\frac{d}{dx}\Bspline{k}{0}{\alpha}(x)
- \sum_{i=0}^{k-1} \lim_{x \to t_{0}^+}\frac{d}{dx}w_{i+1}^{k-i}(x)
}
\\
&=&
\displaystyle{
- \sum_{i=0}^{k-1}\lim_{x \to t_{0}^+}\frac{d}{dx}
\varphi_\alpha\triplet{x}{t_{i+1}}{t_{k+1}}
}
\\
&=&
\displaystyle{
- \sum_{i=0}^{k-1}\lim_{x \to t_{0}^+}\frac{d}{dx}
\varphi_\alpha\triplet{x}{t_{0}}{t_{k+1}}
}
\\
&=&
\displaystyle{
- k\frac{\alpha}{(\alpha-1) \left(t_{k+1} -t_{0} \right)}
}
\\
\end{array}
$$

\item
on other hand
$$
\begin{array}{rcl}
\displaystyle{
\lim_{x \to t_{0}^+}\frac{d}{dx}\Bspline{1}{k}{\alpha}(x)
}
&=&
\displaystyle{
\lim_{x \to t_{0}^+}\frac{d}{dx}\Bspline{k+1}{0}{\alpha}(x)
+ \sum_{i=0}^{k-1} \lim_{x \to t_{0}^+}\frac{d}{dx}w_{i+1}^{k-i}(x)
}
\\
&=&
\displaystyle{
 \sum_{i=0}^{k-1}\lim_{x \to t_{0}^+}\frac{d}{dx}
\varphi_\alpha\triplet{x}{t_{i+1}}{t_{k+1}}
}
\\
&=&
\displaystyle{
 \sum_{i=0}^{k-1}\lim_{x \to t_{0}^+}\frac{d}{dx}
\varphi_\alpha\triplet{x}{t_{0}}{t_{k+1}}
}
\\
\end{array}
$$

\item
and finally for  $i\geq 2$ we obtain
$$
\displaystyle{
\lim_{x \to t_{0}^+}\frac{d}{dx}\Bspline{i}{k}{\alpha}(x)
= \lim_{x \to t_{0}^+}\frac{d}{dx}\Bspline{i+k}{0}{\alpha}(x) =0
}
$$
because
$
\displaystyle{
\Support{\Bspline{i+k}{0}{\alpha}}\cap [t_{0}\,,\,t_{k+1})
=\emptyset
}
$
\end{itemize}

\item
Similarly by using lemma~\ref{LemmeDerivExtremite2} one shows that:
\begin{itemize}
\item
from one hand,
$$
\begin{array}{rcl}
\displaystyle{
\lim_{x \to t_{m}^-}\frac{d}{dx}\Bspline{n}{k}{\alpha}(x)
}
&=&
\displaystyle{
\lim_{x \to t_{m}^-}\frac{d}{dx}\Bspline{n}{0}{\alpha}(x)
+ \sum_{i=0}^{k-1} \lim_{x \to t_{m}^-}\frac{d}{dx}w_{n}^{k-i}(x)
}
\\
&=&
\displaystyle{
\sum_{i=0}^{k-1}\lim_{x \to t_{m}^-}\frac{d}{dx}
\varphi_\alpha\triplet{x}{t_{n}}{t_{n+k-i}}
}
\\
&=&
\displaystyle{
 \sum_{i=0}^{k-1}\lim_{x \to t_{m}^-}\frac{d}{dx}
\varphi_\alpha\triplet{x}{t_{n}}{t_{m}}
}
\\
&=&
\displaystyle{
 k\frac{\alpha-1}{\alpha \left(t_{m} -t_{n} \right)}
}
\\
\end{array}
$$

\item
On another hand, we have 
$$
\begin{array}{rcl}
\displaystyle{
\lim_{x \to t_{m}^-}\frac{d}{dx}\Bspline{n-1}{k}{\alpha}(x)
}
&=&
\displaystyle{
\lim_{x \to t_{m}^-}\frac{d}{dx}\Bspline{n-1}{0}{\alpha}(x)
- \sum_{i=0}^{k-1} \lim_{x \to t_{m}^-}\frac{d}{dx}w_{n}^{k-i}(x)
}
\\
&=&
\displaystyle{
-\sum_{i=0}^{k-1}\lim_{x \to t_{m}^-}\frac{d}{dx}
\varphi_\alpha\triplet{x}{t_{n}}{t_{n+k-i}}
}
\\
&=&
\displaystyle{
- \sum_{i=1}^{k}\lim_{x \to t_{m}^-}\frac{d}{dx}
\varphi_\alpha\triplet{x}{t_{n}}{t_{m}}
}
\\
&=&
\displaystyle{
- k\frac{\alpha-1}{\alpha \left(t_{m} -t_{n} \right)}
}
\\
\end{array}
$$

\item
Finally for $i\leq n-2$ by directly applying lemma \ref{LemmeDerivExtremite21}
we have:
$$
\displaystyle{
\lim_{x \to t_{m}^-}\frac{d}{dx}\Bspline{i}{k}{\alpha}(x)
}
=
\displaystyle{
0
}
$$
\end{itemize}

\end{enumerate}
\end{enumerate}

\begin{remarque}
As shown by the illustrations of appendix, for $k\geq 1$ the functions
$\suite{\Bspline{i}{k}{\alpha}}_{i=0}^{n}$ are not of class ${\cal C}^1$, even when
the nodes are of multiplicity $1$, this perfectly contradicts the classical results
~\cite{piegl1997} page 57.
\end{remarque}

\begin{conjecture}[Existence property and unicity of a maximum]
\label{ConjecMaximum}

Let $m, k, n\in \NN^*$ such that \mbox{$n \geq k$} and $m=n+ k+1$.
Let $\displaystyle{U=\suite{t_i}_{i=0}^{m}}$ be a knot vectors, 
let  \mbox{$\alpha \in (-\infty \, , \, 0) \cup (1  \, , \,  \infty)$}.

Any element of the rational B-spline $\suite{\Bspline{i}{k}{\alpha}}_{i=0}^{n}$ of index $\alpha$ with 
knot vector $U$ and of degree $k$ admits one and only one maximum.

\end{conjecture}

\begin{remarque}
We admit for any useful purpose  this conjecture which is widely illustrated by 
numerical experience and cited in classical review~\cite{rogers2001}
to the page 58 and ~\cite{piegl1997} to the page 45.
\end{remarque}

\begin{proposition}[Linear independence property]

Let $m, k, n\in \NN^*$ such that  \mbox{$n \geq k$} and $m=n+ k+1$.
Let $\displaystyle{U=\suite{t_i}_{i=0}^{m}}$ be an open knot vector
with interior nodes of multiplicity at most $k$, 
let \mbox{$\alpha \in (-\infty \, , \, 0) \cup (1  \, , \,  \infty)$}.

The rational B-spline basis $\suite{\Bspline{i}{k}{\alpha}}_{i=0}^{n}$ of index $\alpha$
with knot vector $U$ and of degree $k$ is a free system in the vector space 
${\cal C}^0(\segment{t_{0}}{t_{m}})$ of continuous functions on $\segment{t_{0}}{t_{m}}$.
\end{proposition}

\Preuve
To show that the B-spline basis $\suite{\Bspline{i}{k}{\alpha}}_{i=0}^{n}$
is linear independent, we will proceed by recurrence on the degree $k$.

\begin{itemize}
\item
Let $k=1$ we search
$\displaystyle{
\suite{\lambda_{i}}_{i=0}^{m-k-1}\subset \RR
}$
such that
$\displaystyle{
\sum_{i=0}^{m-k-1} \lambda_{i} \Bspline{i}{k}{\alpha} =0
}$

Let $\displaystyle{x \in \segment{t_{0}}{t_{m}}}$ by setting
$
\displaystyle{
w_{i}^{r}(x)=\varphi_\alpha\triplet{x}{t_{i}}{t_{i+r}}
}$

$$
\begin{array}{rcl}
0
&=&
\displaystyle{
\sum_{i=0}^{m-k-1} \lambda_{i} \Bspline{i}{k}{\alpha}(x)
=\sum_{i=0}^{m-2} \lambda_{i} \Bspline{i}{1}{\alpha}(x)
}
\\
&=&
\displaystyle{
\sum_{i=0}^{m-2} \lambda_{i}
w_{i}^{1}(x) \Bspline{i}{0}{\alpha}(x)
}
\\
&+&
\displaystyle{
\sum_{i=0}^{m-2} \lambda_{i}
{\left(1 - w_{i+1}^{1}(x)\right)} \Bspline{i+1}{0}{\alpha}(x)
}
\\
&=&
\displaystyle{
\lambda_{0} w_{0}^{1}(x) \Bspline{0}{0}{\alpha}(x)
+\lambda_{m-2} {\left(1 - w_{m-1}^{1}(x)\right)} \Bspline{m-1}{0}{\alpha}(x)
}
\\
&+&
\displaystyle{
\sum_{i=1}^{m-2}
\left[
\lambda_{i} w_{i}^{1}(x) 
+\lambda_{i-1} {\left(1 - w_{i}^{1}(x)\right)}
\right]
\Bspline{i}{0}{\alpha}(x)
}
\\
&=&
\displaystyle{
\sum_{i=1}^{m-2}
\left[
\lambda_{i} w_{i}^{1}(x) 
+\lambda_{i-1} {\left(1 - w_{i}^{1}(x)\right)}
\right]
\Bspline{i}{0}{\alpha}(x)
}
\\
\end{array}
$$
since $U$ is open and 
$$
\begin{array}{l}
\Support{w_{0}^{1}}=[t_{0}\,,\,t_{1})=\emptyset \\
\Support{w_{m-1}^{1}}=[t_{m-1}\,,\,t_{m})=\emptyset \\
\end{array}
$$

As the interior nodes of $U$ are of multiplicity at most $k=1$ then
for all $1\leq j \leq m-2$ 
$\displaystyle{[t_{j}\,,\,t_{j+1})\neq \emptyset}$.

Thus for all  $1\leq j \leq m-2$  and all
$\displaystyle{ 
x \in [t_{j}\,,\,t_{j+1})
}$ 
we have

$$
\begin{array}{rcl}
0
&=&
\displaystyle{
\sum_{i=1}^{m-2}
\left[
\lambda_{i} w_{i}^{1}(x) 
+\lambda_{i-1} {\left(1 - w_{i}^{1}(x)\right)}
\right]
\Bspline{i}{0}{\alpha}(x)
}
\\
&=&
\displaystyle{
\lambda_{j} w_{j}^{1}(x) 
+\lambda_{j-1} {\left(1 - w_{j}^{1}(x)\right)}
}
\\
\end{array}
$$

Moreover we have
$\displaystyle{
0=\sum_{i=0}^{m-2}
\lambda_{i} 
\Bspline{i}{1}{\alpha}(t_{0})
=\lambda_{0} 
}$

All in all we get this linear system:
$$
\left\lbrace
\begin{array}{rcl}
\lambda_{0} &=&0\\
\displaystyle{
\lambda_{j-1} {\left(1 - w_{j}^{1}(x_{j})\right)}
+
\lambda_{j} w_{j}^{1}(x_{j}) 
}&=&0 \textrm{ for } j=1, \ldots, m-2 
\\
&&\textrm{ and } 
x_{j}\in \osegment{t_{j}}{t_{j+1}}
\\
\end{array}
\right.
$$
where $\displaystyle{w_{j}^{1}(x_{j} >0}$ and $\displaystyle{1 - w_{j}^{1}(x_{j} >0}$ 
for all $1\leq j \leq m-2$. Since the system is lower-triangular with null diagonal terms and homogeneous
then we have $\displaystyle{\lambda_j=0}$ for all
$j=0, \ldots, m-2$. We conclude that 
$\displaystyle{\suite{\Bspline{i}{1}{\alpha}}_{i=0}^{m-2}}$ is a free system.

\item
let $k>1$ and suppose that for all $1\leq p\leq k-1$
$\displaystyle{\suite{\Bspline{i}{p}{\alpha}}_{i=0}^{m-p-1}}$ is a free system.
Let show that $\displaystyle{\suite{\Bspline{i}{k}{\alpha}}_{i=0}^{m-k-1}}$
is a free system.

$$
\begin{array}{rcl}
0
&=&
\displaystyle{
\sum_{i=0}^{m-k-1} \lambda_{i} \Bspline{i}{k}{\alpha}(x)
}
\\
&=&
\displaystyle{
\sum_{i=0}^{m-k-1} \lambda_{i}
w_{i}^{k}(x) \Bspline{i}{k-1}{\alpha}(x)
}
\\
&+&
\displaystyle{
\sum_{i=0}^{m-k-1} \lambda_{i}
{\left(1 - w_{i+1}^{k}(x)\right)} \Bspline{i+1}{k-1}{\alpha}(x)
}
\\
&=&
\displaystyle{
\lambda_{0} w_{0}^{k}(x) \Bspline{0}{k-1}{\alpha}(x)
+\lambda_{m-k-1} {\left(1 - w_{m-k}^{k}(x)\right)} \Bspline{m-k}{k-1}{\alpha}(x)
}
\\
&+&
\displaystyle{
\sum_{i=1}^{m-k-1}
\left[
\lambda_{i} w_{i}^{k}(x) 
+\lambda_{i-1} {\left(1 - w_{i}^{k}(x)\right)}
\right]
\Bspline{i}{k-1}{\alpha}(x)
}
\\
&=&
\displaystyle{
\sum_{i=1}^{m-k-1}
\left[
\lambda_{i} w_{i}^{k}(x) 
+\lambda_{i-1} {\left(1 - w_{i}^{k}(x)\right)}
\right]
\Bspline{i}{k-1}{\alpha}(x)
}
\\
\end{array}
$$
since $U$ is open and
$$
\begin{array}{l}
\Support{w_{0}^{k}}=[t_{0}\,,\,t_{k})=\emptyset \\
\Support{w_{m-k}^{k}}=[t_{m-k}\,,\,t_{m})=\emptyset \\
\end{array}
$$

As by hypothesis 
$\displaystyle{\suite{\Bspline{i}{k-1}{\alpha}}_{i=0}^{m-k}}$ is a free system
and the multiplicity of a node of $U$ is at most $k$,
then for all $1\leq j \leq m-k-1$ and all
$x_{j} \in (t_{j}\,,\,t_{j+k})\neq \emptyset$ we have
$
\displaystyle{
\lambda_{j} w_{j}^{k}(x_{j}) 
+\lambda_{j-1} {\left(1 - w_{j}^{k}(x_{j})\right)}
=0
}
$
with $\displaystyle{ w_{j}^{k}(x_{j})>0}$ and $\displaystyle{1 - w_{j}^{k}(x_{j})>0}$.

Moreover we have
$\displaystyle{
0=\sum_{i=0}^{m-k-1}
\lambda_{i} 
\Bspline{i}{k}{\alpha}(t_{0})
=\lambda_{0} 
}$

We then obtain the following linear system:
$$
\left\lbrace
\begin{array}{rcl}
\lambda_{0} &=&0\\
\displaystyle{
\lambda_{j-1} {\left(1 - w_{j}^{k}(x_{j})\right)}
+
\lambda_{j} w_{j}^{k}(x_{j}) 
}&=&0 \textrm{ for } j=1, \ldots, m-k-1 
\\
&&\textrm{ and } 
x_{j}\in \osegment{t_{j}}{t_{j+1}}
\\
\end{array}
\right.
$$

This lower-triangular system with positive diagonal terms admits a unique solution
 $\displaystyle{\lambda_j =0}$ for all $0\leq j \leq m-k-1$.
Hence $\displaystyle{\suite{\Bspline{i}{k}{\alpha}}_{i=0}^{m-k-1}}$ is free.

\end{itemize}

\subsection{Case of an open knot vector with no interior node}

\begin{proposition}

Let $a, b\in \RR$ such that $a<b$.
Let $m, k, n\in \NN^*$ such that  \mbox{$n = k$} and $m=2k+1$.
Let $\displaystyle{U_{k}=\suite{t_{i}^{k}}_{i=0}^{2k+1}}$ 
be the open knot vector such that 
 $t_{k}^{k}=a$ and $t_{k+1}^{k}=b$
let  \mbox{$\alpha \in (-\infty \, , \, 0) \cup (1  \, , \,  \infty)$}.

Let $\suite{\Bezier{i}{k}{\alpha}}_{i=0}^{k}$
be the rational B-spline basis of index $\alpha$
with knot vectors $U_{k}$ and of degree $k$,
let $\suite{\Bezier{i}{k-1}{\alpha}}_{i=0}^{k-1}$
be the rational B-spline basis of index $\alpha$
with knot vectors $U_{k-1}$ and of degree $k-1$. 

For all $x\in [a\, , \, b]$ and by setting 
$\displaystyle{w(x) = \varphi_\alpha\triplet{x}{a}{b}}$
we have the following:

\begin{enumerate}
\item\emph{Recurrence relation}
\begin{equation}\label{RecurrenceBezier}
\displaystyle{
\Bezier{i}{k}{\alpha}(x)=
w(x)\Bezier{i-1}{k-1}{\alpha}(x) 
+
{\left(1-w(x) \right)}\Bezier{i}{k-1}{\alpha}(x)
}
\end{equation}

\item\emph{Explicit formula}
$$
\displaystyle{
\Bezier{i}{k}{\alpha}(x)=
C_{k}^{i}{\left(w(x) \right)}^{i}{\left(1-w(x) \right)}^{k-i}
}
$$

\end{enumerate}

\end{proposition}

By definition $\suite{\Bezier{i}{k}{\alpha}}_{i=0}^{k}$ will be called
 Bernstein basis of index $\alpha$ and of degree $k$ on the parametrization space $[a\, , \, b]$.

\Preuve

\begin{enumerate}
\item
\emph{Recurrence relation} 

Consider the open knot vectors: \\ 
$\displaystyle{U_{k}=\suite{t_{i}^{k}}_{i=0}^{2k+1}}$ and
$\displaystyle{U_{k-1}=\suite{t_{i}^{k-1}}_{i=0}^{2k-1}}$ 
satisfy
$$
\begin{array}{lcr}
t_{k}^{k}=a
&\textrm{ and }&
t_{k+1}^{k}=b
\\
t_{k-1}^{k-1}=a
&\textrm{ and }&
t_{k}^{k-1}=b
\\
\end{array}
$$

Let $\displaystyle{g_{k} : i\in  \ZZ \mapsto g_{k}(i) =i-1 \in \ZZ }$.
Based on this bijection, we have 
$$
\displaystyle{
t_{i}^{k}=t_{g_{k}(i)}^{k-1} \quad \forall i=0, \ldots, 2k+1
}
$$ 
by imposing
$
\displaystyle{t_{0}^{k}=t_{-1}^{k-1}=t_{0}^{k-1}}
$
and
$
\displaystyle{t_{2k+1}^{k}=t_{2k}^{k-1}}.
$

Thus $U_{k}$ is seen as a natural extension of $U_{k-1}$.

Consider the family 
$\displaystyle{\suite{\Bspline{i}{j}{\alpha}}_{i=0}^{2k-j}}$ of B-spline basis
of index $\alpha$ with knot vector  $U_{k}$ and of degree $j$ with $0\leq j \leq k$.

Let $\displaystyle{\suite{\Bezier{i}{k}{\alpha}}_{i=0}^{k}}$ be the B-spline basis
of index $\alpha$ with knot vector $U_{k}$  and of degree $k$.

Let $\displaystyle{\suite{\Bezier{i}{k-1}{\alpha}}_{i=0}^{k-1}}$ be the B-spline basis
of index $\alpha$ with knot vector  $U_{k-1}$  and of degree $k-1$.

From the definition, for all $i=0, \ldots, k$ and all $x\in \segment{a}{b}$ we have
$$
\begin{array}{rcl}
\displaystyle{\Bezier{i}{k}{\alpha}(x)}
&=&
\displaystyle{\Bspline{i}{k}{\alpha}(x)}
\\
&=&
\displaystyle{w_{i}^{k}(x) \Bspline{i}{k-1}{\alpha}(x)}
+
\displaystyle{\left(1- w_{i+1}^{k}(x)\right) \Bspline{i+1}{k-1}{\alpha}(x)}
\\
\end{array}
$$

$\displaystyle{\Bspline{i}{k-1}{\alpha}}$ is of degree $k-1$ respect to the knot vector
  $U_{k}$ which is an extension of the knot vector $U_{k-1}$.
 
  Relative to the knot vector  $U_{k-1}$ by imposing
 $$
 \displaystyle{\Bezier{-1}{k-1}{\alpha}}=
 \displaystyle{\Bezier{k}{k-1}{\alpha}}\equiv 0
 $$
we have for all $i=0, \ldots, k+1$
 $$
 \displaystyle{\Bspline{i}{k-1}{\alpha}}=
 \displaystyle{\Bezier{g_{k}(i)}{k-1}{\alpha}}=
 \displaystyle{\Bezier{i-1}{k-1}{\alpha}}
 $$

 Thus we have
 $$
 \displaystyle{\Bezier{i}{k}{\alpha}(x)}
=
\displaystyle{w_{i}^{k}(x) \Bezier{i-1}{k-1}{\alpha}(x)}
+
\displaystyle{\left(1- w_{i+1}^{k}(x)\right) \Bezier{i}{k-1}{\alpha}(x)}
 $$
 
As
$$
\begin{array}{rcl}
\displaystyle{w_{i}^{k}(x)}
&=&
\displaystyle{\varphi_\alpha\triplet{x}{t_{i}}{t_{i+k}}}
\\
&=&
\left\lbrace
\begin{array}{rcl}
\displaystyle{\varphi_\alpha\triplet{x}{a}{b}}
&& \textrm{ if } 1\leq i\leq k
\\
0 && \textrm{ otherwise }
\\
\end{array}
\right.
\\
\end{array}
$$
we can set
$\displaystyle{w(x)=\varphi_\alpha\triplet{x}{a}{b}}$
and obtain for all $k\in \NN^*$ and
all $0\leq i \leq k$, the recurrence relation 
 $$
 \displaystyle{\Bezier{i}{k}{\alpha}(x)}
=
\displaystyle{w(x) \Bezier{i-1}{k-1}{\alpha}(x)}
+
\displaystyle{\left(1- w(x)\right) \Bezier{i}{k-1}{\alpha}(x)}
 $$

\item\emph{Explicit formula}

We will now show that the recurrence relation \ref{RecurrenceBezier}
leads to 
$$
\left\lbrace
\begin{array}{rcl}
 \displaystyle{\Bezier{0}{k}{\alpha}(x)}
 &=&
 \displaystyle{\left(1- w(x)\right)^{k}}
 \\
 \displaystyle{\Bezier{k}{k}{\alpha}(x)}
 &=&
 \displaystyle{\left( w(x)\right)^{k} }
 \\
 \displaystyle{\Bezier{i}{k}{\alpha}(x)}
 &=&
 C_{k}^{i}
 \displaystyle{\left( w(x)\right)^{i} \left(1- w(x)\right)^{k-i} }
 \textrm{ for } 1\leq i \leq k-1
 \\
\end{array}
\right.
$$
\begin{itemize}
\item
For all $k\in \NN^*$, if $i=0$ then the equation \ref{RecurrenceBezier} becomes
$$
\displaystyle{\Bezier{0}{k}{\alpha}(x)}=
\displaystyle{ \left(1- w(x)\right)\Bezier{0}{k-1}{\alpha}(x)}
$$

The sequence $\displaystyle{\suite{\Bezier{0}{k}{\alpha}(x)}_{k\geq 0}}$
is geometric with common ratio $1- w(x)$. We deduce that
$$
\displaystyle{\Bezier{0}{k}{\alpha}(x)}=
\displaystyle{ \left(1- w(x)\right)^{k}\Bezier{0}{0}{\alpha}(x)}
=\displaystyle{ \left(1- w(x)\right)^{k}}
$$
since
$
\displaystyle{ \Bezier{0}{0}{\alpha}(x)=\Bspline{0}{0}{\alpha}(x)=1}
$
for all $x \in [a\,,\,b)$.

We remark that for all $x \in (a\,,\,b)$
$
\displaystyle{ \Bezier{0}{k}{\alpha}(x)}
=\displaystyle{C_{k}^{0} \left(w(x)\right)^{0} \left(1- w(x)\right)^{k}}
$
since $\displaystyle{C_{k}^{0}=1}$,  $w(x)> 0$ and  $1-w(x)> 0$
\item
For all $k\in \NN^*$, if $i =k$ then the equation \ref{RecurrenceBezier} gives
$$
\displaystyle{\Bezier{k}{k}{\alpha}(x)}=
\displaystyle{ \left(w(x)\right)\Bezier{k-1}{k-1}{\alpha}(x)}
$$

The sequence $\displaystyle{\suite{\Bezier{k}{k}{\alpha}(x)}_{k\geq 0}}$
is geometric with common ratio $w(x)$. We deduce that 
$$
\displaystyle{\Bezier{k}{k}{\alpha}(x)}=
\displaystyle{ \left( w(x)\right)^{k}\Bezier{0}{0}{\alpha}(x)}
=\displaystyle{ \left( w(x)\right)^{k}}
$$

As previously we observe that for $x \in (a\,,\,b)$
$
\displaystyle{ \Bezier{k}{k}{\alpha}(x)}
=\displaystyle{C_{k}^{k} \left(w(x)\right)^{k} \left(1- w(x)\right)^{0}}
$
because $\displaystyle{C_{k}^{k}=1}$

\item
For all $k\in \NN^*$, if $1\leq i <k$ then the equation \ref{RecurrenceBezier} gives
$$
\displaystyle{\Bezier{i}{k}{\alpha}(x)}=
\displaystyle{ \left(w(x)\right)\Bezier{i-1}{k-1}{\alpha}(x)}
+
\displaystyle{ \left(1-w(x)\right)\Bezier{i}{k-1}{\alpha}(x)}
$$
Let us prove by recurrence on $k$ that 
$
\displaystyle{ \Bezier{i}{k}{\alpha}(x)}
=\displaystyle{C_{k}^{i} \left(w(x)\right)^{i} \left(1- w(x)\right)^{k-i}}
$

\begin{itemize}
\item
The relation is true for $k=1$. 

\item Let $k>1$.
Suppose that for all $1\leq j <k$, one has for all $0\leq i\leq j$
$
\displaystyle{ \Bezier{i}{j}{\alpha}(x)}
=\displaystyle{C_{j}^{i} \left(w(x)\right)^{i} \left(1- w(x)\right)^{j-i}}
$.

For all $1\leq i \leq k-1$, we have
$$
\begin{array}{rcl}
\displaystyle{\Bezier{i}{k}{\alpha}(x)}
&=&
\displaystyle{ \left(w(x)\right)\Bezier{i-1}{k-1}{\alpha}(x)}
+
\displaystyle{ \left(1-w(x)\right)\Bezier{i}{k-1}{\alpha}(x)}
\\
&=&
\displaystyle{ \left(w(x)\right)}
\displaystyle{C_{k-1}^{i-1} \left(w(x)\right)^{i-1} \left(1- w(x)\right)^{k-i}}
\\
&+&
\displaystyle{ \left(1-w(x)\right)}
\displaystyle{C_{k-1}^{i} \left(w(x)\right)^{i} \left(1- w(x)\right)^{k-1-i}}
\\
&=&
\displaystyle{C_{k-1}^{i-1} \left(w(x)\right)^{i} \left(1- w(x)\right)^{k-i}}
\\
&+&
\displaystyle{C_{k-1}^{i} \left(w(x)\right)^{i} \left(1- w(x)\right)^{k-i}}
\\
&=&
\displaystyle{\left[C_{k-1}^{i-1}+C_{k-1}^{i}\right] 
\left(w(x)\right)^{i} \left(1- w(x)\right)^{k-i}}
\\
&=&
\displaystyle{C_{k}^{i}\left(w(x)\right)^{i} \left(1- w(x)\right)^{k-i}}
\\
\end{array}
$$
because $\displaystyle{C_{k}^{i}=C_{k-1}^{i-1}+C_{k-1}^{i}}$.

\end{itemize}

\end{itemize}

\end{enumerate}

\section{Properties of the new class of \mbox{B-spline} curves\label{SecNewCourbBspline} }

Let $m, k, n\in \NN^*$ such that \mbox{$n \geq k$} and $m=n+ k+1$.
Let $\displaystyle{U=\suite{t_i}_{i=0}^{m}}$ be an open knot vector, 
let \mbox{$\alpha \in (-\infty \, , \, 0) \cup (1  \, , \,  \infty)$}.

Consider the rational B-spline basis $\suite{\Bspline{i}{k}{\alpha}}_{i=0}^{n}$
of index $\alpha$ with knot vector $U$ 
and of degree $k$,

Consider the B-spline curve $G_\alpha$ of index $\alpha$,of knot vector $U$,
 of control points $\displaystyle{\suite{d_i}_{i=0}^{n}  \subset \RR^d}$ 
and defined for all 
$x \in [t_{0}\, , \, t_{m}]$ by
$$
\displaystyle{G_\alpha(x) = \sum_{i=0}^{n}d_i \Bspline{i}{k}{\alpha}(x) }
$$

\subsection{Geometric properties}
The curves of this new class verify the classical properties of B-spline curve.
They also show some exotic properties namely related to the symmetry. These
properties are given in the following propositions.

\begin{proposition}
\label{PropoGeometrie1}
We have the following properties:

\begin{enumerate}
\item \emph{Local control property:} 

Let $j\in \NN$ such that $0\leq j \leq n$. Any variation of the control point
$d_{j}$ does influence $\displaystyle{G_\alpha (x)}$ only for  
$x \in [t_{j}\,,\,t_{j+k+1})$

\item \emph{Second local control property:} 

Let $j\in \NN$ such that $k\leq j \leq n$  and $t_{j} <t_{j+1} $.
For all $x \in [t_{j} \,,\,t_{j+1} )$, we have 
$$
\displaystyle{
G_\alpha (x)
=
\sum_{i=j-k}^{j} d_{i}\Bspline{i}{k}{\alpha}(x)
}
$$
This computation uses only the  $k+1$ control points 
$\displaystyle{\suite{d_{i}}_{i=j-k}^{j}}$.

\item \emph{Convex hull property:}

$G_\alpha$ is in convex hull of its control points
 $\displaystyle{\suite{d_i}_{i=0}^{n}}$.

In other words, for all $x \in [a\, , \, b]$, there exists
$\suite{\lambda_i}_{i=0}^{n} \subset \RR_+$ such that
$
G_\alpha(x)=
\displaystyle{\sum_{i=0}^{n}{\lambda_i d_i}}
$ 
with
$\displaystyle{\sum_{i=0}^{n}{\lambda_i}=1}$

\item \emph{Invariance by affine transformation property:}

For any affine transformation $T$ in $\RR^d$, we have
$$
T\left(G_\alpha(x)\right) =
\displaystyle{
\sum_{i=0}^{n}T(d_i)\Bspline{i}{k}{\alpha}(x)
}
$$

\end{enumerate}

\end{proposition}

\Preuve

\begin{enumerate}
\item \emph{Local control property:} 

Consider the control polygons
$\displaystyle{\Pi=\suite{d_{i}}_{i=0}^{n} \subset \RR^d}$ and
$\displaystyle{\hat{\Pi}=\suite{\hat{d}_{i}}_{i=0}^{n} \subset \RR^d}$.
Suppose that for a fixed $0\leq j \leq n$ we have
$$
\left\lbrace
\begin{array}{rcl}
\hat{d}_{i}&=&d_{i} \textrm{ if } i \neq j\\
\hat{d}_{j}&\neq& d_{j}\\
\end{array}
\right.
$$

Let $G_\alpha$ and  $\hat{G}_\alpha$ be the B-spline curves of index $\alpha$ of degree
$k$ and of control polygons $\Pi$ and  $\hat{\Pi}$ respectively.

For $x \in \segment{t_{0}}{t_{m}}$ we have 
$$
\left\lbrace
\begin{array}{rcl}
\displaystyle{G_\alpha (x)}
&=&
\displaystyle{\sum_{i=0}^{n}d_{i}\Bspline{i}{k}{\alpha} (x)}
\\
\displaystyle{\hat{G}_\alpha (x)}
&=&
\displaystyle{\sum_{i=0}^{n}\hat{d}_{i}\Bspline{i}{k}{\alpha} (x)}
\\
\end{array}
\right.
$$

The variation $\displaystyle{\Delta d_{j}= d_{j} -\hat{d}_{j}}$ 
of the control point $d_{j}$ induces a variation at $x$
of the curve $G_\alpha$ denoted by 
$\displaystyle{\Delta G_\alpha (x)=G_\alpha (x) -\hat{G}_\alpha (x)}$.

One has
$$
\displaystyle{\Delta G_\alpha (x)=G_\alpha (x) -\hat{G}_\alpha (x)}
=
\displaystyle{\left( d_{j} -\hat{d}_{j}\right)\Bspline{j}{k}{\alpha}(x)}
=
\displaystyle{\Delta d_{j}\Bspline{j}{k}{\alpha}(x)}
$$

Thus
$$
\displaystyle{\Delta G_\alpha (x)\neq 0}
\equivaut
\displaystyle{\Bspline{j}{k}{\alpha}(x) \neq 0}
\equivaut
\displaystyle{x \in (t_{j}\,,\,t_{j+k+1})}
$$

The effect of the variation $\displaystyle{\Delta d_{j}}$ 
can then only be viewed on the computation of $\displaystyle{G_\alpha (x)}$ for 
$\displaystyle{x \in (t_{j}\,,\,t_{j+k+1})}$.

\item \emph{Second local control property:} 

Let $j\in \NN$. Since $\displaystyle{U=\suite{t_{i}}_{i=0}^{m}}$ is open,
$$
t_{j}< t_{j+1}
\implique
j\geq k
\textrm{ and }
j\leq n=m-k-1
\equivaut 
 k \leq j\leq n=m-k-1
$$

Let then $k\leq j \leq n$ such that $t_{j}<t_{j+1}$ and
$\displaystyle{x\in \segment{t_{j}}{t_{j+1}}}$.

A control point $d_{s}$ influences the computation of
$\displaystyle{G_\alpha(x)=\sum_{i=0}^{n}d_{i}\Bspline{i}{k}{\alpha}(x)}$
if and only if $\displaystyle{\Bspline{s}{k}{\alpha}(x) \neq 0}$
$$
\begin{array}{rcl}
\displaystyle{\Bspline{s}{k}{\alpha}(x) \neq 0}
&\equivaut&
\displaystyle{
\Support{\Bspline{s}{k}{\alpha}}
\cap [t_{j}\,,\,t_{j+1}) \neq \emptyset
}
\\
&\equivaut&
\displaystyle{
 \emptyset \neq [t_{j}\,,\,t_{j+1}) \subset
[t_{s}\,,\,t_{s+k+1})
}
\\
&\equivaut&
\displaystyle{
t_{s} \leq t_{j}<t_{j+1}\leq t_{s+k+1}
}
\\
&\equivaut&
\displaystyle{
s \leq j<j+1 \leq  s+k+1 
}
\\
&\equivaut&
\displaystyle{
j-k \leq   s \leq j
}
\\
\end{array}
$$

We deduce that 
$$
\displaystyle{G_\alpha(x)
=\sum_{i=0}^{n}d_{i}\Bspline{i}{k}{\alpha}(x)
=\sum_{i=j-k}^{j}d_{i}\Bspline{i}{k}{\alpha}(x)
}
$$

This computation does use only the  $k+1$ control points
$\displaystyle{\suite{d_{i}}_{i=j-k}^{j}}$.

This result gives another point of view of local control.

\item \emph{Convex hull property:}

Let $x\in \segment{t_{0}}{t_{m}}$
$$
\begin{array}{rcl}
\displaystyle{G_\alpha (x)}
&=&
\displaystyle{\sum_{i=0}^{n} d_{i}  \Bspline{i}{n}{\alpha}(x)}\\
&=&
\displaystyle{\sum_{i=0}^{n}\lambda_i d_{i} }\\
\textrm{where}&&\\
\displaystyle{\lambda_i}
&=&
\displaystyle{ \Bspline{i}{n}{\alpha}(x)\in \RR_+  \, \forall i}\\
\end{array}
$$

But from unit partition property,  one gets 
$
\displaystyle{\sum_{i=0}^{n}\lambda_i  
=\sum_{i=0}^{n}  \Bspline{i}{n}{\alpha}(x) =1 }
$. 
$G_\alpha(x)$ is in the convex hull of control polygon 
$\displaystyle{\suite{d_i}_{i=0}^{n}}$

\item \emph{Invariance by affine transformation property:}

Let $T$ be an affine transformation in $\RR^d$. There exists a square matrix $M$
of order $d$ and a point $C\in \RR^d$ such that for all  $X\in \RR^d$, 
$\displaystyle{T(X)=M\,X+C}$.
Let $x\in  \segment{t_{0}}{t_{m}}$. 
Since $\displaystyle{G_\alpha (x) \in \RR^d}$ then we have
$$
\begin{array}{rcl}
\displaystyle{T{\left( G_\alpha (x)\right)}}
&=&
\displaystyle{T{\left( \sum_{i=0}^{n} d_{i}  \Bspline{i}{n}{\alpha}(x)\right)}}\\
&=&
\displaystyle{M{\left( \sum_{i=0}^{n} d_{i}  \Bspline{i}{n}{\alpha}(x)\right)}+C}\\
&=&
\displaystyle{\sum_{i=0}^{n}M{\left(  d_{i}  \Bspline{i}{n}{\alpha}(x)\right)}}+
\displaystyle{{\left( \sum_{i=0}^{n}\Bspline{i}{n}{\alpha}(x)\right)}C}\\
&=&
\displaystyle{\sum_{i=0}^{n}{\left( M d_{i}  \Bspline{i}{n}{\alpha}(x)\right)}}+
\displaystyle{ \sum_{i=0}^{n}{\left(C\Bspline{i}{n}{\alpha}(x)\right)}}\\
&=&
\displaystyle{ \sum_{i=0}^{n}{\left(M d_{i} +C\right)}\Bspline{i}{n}{\alpha}(x)}=
\displaystyle{ \sum_{i=0}^{n}T{\left(d_{i} \right)}\Bspline{i}{n}{\alpha}(x)}\\
\end{array}
$$
what is expected.

\end{enumerate}

\begin{proposition}
\label{PropoExtremites}
The following properties hold:
\begin{enumerate}
\item \emph{Interpolation property of extreme points:} 

The curve $G_\alpha$ 
interpolates the extreme points of is control polygon, that is 
$G_\alpha(t_{0})=d_0$ and $G_\alpha(t_{m})=d_n$

\item \emph{Tangent property at extreme points:}

The curve $G_\alpha$ is tangent to its control polygon
at extreme points. More precisely, we have
$$
\left\lbrace
\begin{array}{l}
\displaystyle{
\frac{d G_\alpha}{dx}(t_{0})=
\frac{k \alpha}{(\alpha-1)(t_{k+1}-t_{0})}{\left( d_1- d_0\right)}
}
\\
\\
\displaystyle{
\frac{d G_\alpha}{dx}(t_{m})=
\frac{k (\alpha-1)}{\alpha(t_{m}-t_{n})}{\left( d_{n} - d_{n-1}\right)}
}
\end{array}
\right.
$$ 
\end{enumerate}

\end{proposition}

\Preuve

We draw attention on the fact that once the knot vector $U=\suite{t_{i}}_{i=0}^{m}$
has no interior node of multiplicity greater than $k$, the associated basis 
$\suite{\Bspline{i}{k}{\alpha}}_{i=0}^{n}$ is of class ${\cal C}^0$. We have a curve
$\displaystyle{
G_\alpha =\sum_{i=0}^{n} d_{i}\Bspline{i}{k}{\alpha}
}$
which is ${\cal C}^0$ on $\displaystyle{\segment{t_{0}}{t_{m}}}$
for all control polygon
$\displaystyle{\Pi =\suite{d_{i}}_{i=0}^{n} \subset \RR^d}$.

\begin{enumerate}
\item \emph{Interpolation property of extreme points:} 

By using proposition \ref{PropoContinu} we have
$$
\displaystyle{
G_\alpha(t_{0}) =\sum_{i=0}^{n} d_{i}\Bspline{i}{k}{\alpha}(t_{0})
=d_{0}\Bspline{0}{k}{\alpha}(t_{0})=d_{0}
}
$$
and
$$
\displaystyle{
G_\alpha(t_{m}) =\sum_{i=0}^{n} d_{i}\Bspline{i}{k}{\alpha}(t_{m})
=d_{n}\Bspline{n}{k}{\alpha}(t_{m})=d_{n}
}
$$

\item \emph{Tangent property at extreme points:}

By making use of proposition \ref{PropoRegular}
we obtain
$$
\begin{array}{rcl}
\displaystyle{
\frac{d}{dx} G_\alpha(t_{0}) 
}
&=&
\displaystyle{
\sum_{i=0}^{n} d_{i}
\frac{d}{dx}\Bspline{i}{k}{\alpha}(t_{0})
}
\\
&=&
\displaystyle{
d_{0}\frac{d}{dx}\Bspline{0}{k}{\alpha}(t_{0})
+d_{1}\frac{d}{dx}\Bspline{1}{k}{\alpha}(t_{0})
}
\\
&=&
\displaystyle{
\left( d_{1}-d_{0}\right)\frac{d}{dx}\Bspline{1}{k}{\alpha}(t_{0})
}
\\
&=&
\displaystyle{
\left( d_{1}-d_{0}\right)\frac{k\alpha}{(\alpha -1) \left(t_{k+1}-t_{0} \right)}
}
\\
\end{array}
$$
and
$$
\begin{array}{rcl}
\displaystyle{
\frac{d}{dx} G_\alpha(t_{m}) 
}
&=&
\displaystyle{
\sum_{i=0}^{n} d_{i}
\frac{d}{dx} \Bspline{i}{k}{\alpha}(t_{m})
}
\\
&=&
\displaystyle{
d_{n-1}\frac{d}{dx}\Bspline{n-1}{k}{\alpha}(t_{m})
+d_{n}\frac{d}{dx}\Bspline{n}{k}{\alpha}(t_{m})
}
\\
&=&
\displaystyle{
\left(d_{n}-d_{n-1} \right)\frac{d}{dx} \Bspline{n}{k}{\alpha}(t_{m})
}
\\
&=&
\displaystyle{
\left(d_{n}-d_{n-1} \right)\frac{k(\alpha -1)}{\alpha \left(t_{m}-t_{n} \right)} 
}
\\
\end{array}
$$

\end{enumerate}

\begin{proposition}[Symmetry property]
\label{PropoSymetrie}
If the knot vector $U=\suite{t_i}_{i=0}^{n}$ is symmetric
and the control polygon $\Pi=\suite{d_i}_{i=0}^{n}$ is also symmetric
with respect to the perpendicular bisector ${\cal D}$ of segment $\couple{d_{0}}{d_{n}}$
then the curves of degree $k$ :  $G_\alpha$ and $G_{1-\alpha}$ of the same
 knot vector
$U$ and of the same control polygon $\Pi$ are symmetric with respect to 
the line ${\cal D}$

\end{proposition}

\Preuve

Let $\displaystyle{U=\suite{t_{i}}_{i=0}^{m}}$ be symmetric.

We suppose that $\RR^d$ is endowed with orthonormed coordinate system
$\displaystyle{{\cal R}=\quadruplet{O}{\vec{e}_{1}}{\ldots}{\vec{e}_{d}}}$.

Let $\displaystyle{\Pi=\suite{d_{i}}_{i=0}^{n}\subset \RR^d}$ be a symmetric control polygon  
with respect to
the perpendicular bisector ${\cal D}$ of segment $\couple{d_{0}}{d_{n}}$.

Then for all $0\leq i \leq n$,
${\cal D}$ is the perpendicular bisector of $\couple{d_{i}}{d_{n-i}}$; there exists a unique
$M_i \in {\cal D}$ such that 
$\displaystyle{\overrightarrow{M_{i}\!d_{i}}=-\overrightarrow{M_{i}\!d_{n-i}}}$
and ${\cal D}$  orthogonal to  $\couple{d_{i}}{d_{n-i}}$.
Without loss of generality, suppose that 
$\lbrace O \rbrace ={\cal D}\cap  \couple{d_{0}}{d_{n}}$,
${\cal D}$ is the line $\couple{O}{\vec{e}_{d}}$ and ${\cal R}$ the canonical coordinate system.
Hence for all $0\leq i \leq n$, there exists 
$\hat{d}_{i} \in \RR^{d-1}$ and $z_{i}\in \RR$ both unique such that
$$
\left\lbrace
\begin{array}{rcl}
d_{i}&=&\couple{\hat{d}_{i}}{z_{i}}\equiv \hat{d}_{i}+z_{i}\vec{e}_{d} \\
d_{n-i}&=&\couple{-\hat{d}_{i}}{z_{i}}\equiv -\hat{d}_{i}+z_{i}\vec{e}_{d}\\
\end{array}
\right.
$$

Consider the B-spline curves $\displaystyle{G_{\alpha}}$ and  $\displaystyle{G_{1-\alpha}}$
of degree $k$, of knot vector $U$ which is symmetric and 
of symmetric control polygon $\Pi$.

For all $x \in \segment{t_{0}}{t_{m}}$, we have
$$
\begin{array}{rcl}
\displaystyle{G_{\alpha}}(x)
&=&
\displaystyle{
\sum_{i=0}^{n} d_{i}\Bspline{i}{k}{\alpha}(x)
}\\
&=&
\displaystyle{
\sum_{i=0}^{n}{\left( \hat{d}_{i}+z_{i}\vec{e}_{d}\right)} 
\Bspline{i}{k}{\alpha}(x)
}\\
&=&
\displaystyle{
\sum_{i=0}^{n} \hat{d}_{i}\Bspline{i}{k}{\alpha}(x)
+
{\left(\sum_{i=0}^{n} z_{i}\Bspline{i}{k}{\alpha}(x)\right)} \vec{e}_{d}
}\\
\end{array}
$$

Also

$$
\begin{array}{rcl}
\displaystyle{G_{1-\alpha}}(t_{0}+t_{m}-x)
&=&
\displaystyle{
\sum_{i=0}^{n} d_{i}\Bspline{i}{k}{1-\alpha}(t_{0}+t_{m}-x)
}\\
&=&
\displaystyle{
\sum_{i=0}^{n} d_{i}\Bspline{n-i}{k}{\alpha}(x)
}\\
&=&
\displaystyle{
\sum_{i=0}^{n}{\left( \hat{d}_{i}+z_{i}\vec{e}_{d}\right)} 
\Bspline{n-i}{k}{\alpha}(x)
}\\
&=&
\displaystyle{
\sum_{i=0}^{n} \hat{d}_{i}\Bspline{n-i}{k}{\alpha}(x)
+
{\left(\sum_{i=0}^{n} z_{i}\Bspline{n-i}{k}{\alpha}(x)\right)} \vec{e}_{d}
}\\
&=&
\displaystyle{
-\sum_{i=0}^{n} \hat{d}_{n-i}\Bspline{n-i}{k}{\alpha}(x)
+
{\left(\sum_{i=0}^{n} z_{n-i}\Bspline{n-i}{k}{\alpha}(x)\right)} \vec{e}_{d}
}\\
&=&
\displaystyle{
-\sum_{i=0}^{n} \hat{d}_{i}\Bspline{i}{k}{\alpha}(x)
+
{\left(\sum_{i=0}^{n} z_{i}\Bspline{i}{k}{\alpha}(x)\right)} \vec{e}_{d}
}\\
\end{array}
$$

We deduce that
$$
\begin{array}{rcl}
\displaystyle{
\frac{1}{2}
\left[
G_{\alpha}(x) +G_{1-\alpha}(t_{0}+t_{m}-x)
\right]
}
&=&
\displaystyle{
{\left(\sum_{i=0}^{n} z_{i}\Bspline{i}{k}{\alpha}(x)\right)} \vec{e}_{d}
\in {\cal D}
}\\
&&\\
\displaystyle{
\frac{1}{2}
\left[
G_{\alpha}(x) -G_{1-\alpha}(t_{0}+t_{m}-x)
\right] . \vec{e}_{d}
}
&=&
\displaystyle{
\sum_{i=0}^{n} {\left(\hat{d}_{i} .  \vec{e}_{d}\right)}
\Bspline{i}{k}{\alpha}(x)=0
}\\
\end{array}
$$

Thus ${\cal D}$ is the perpendicular bisector of segment 
$\displaystyle{\segment{G_{\alpha}(x) }{G_{1-\alpha}(t_{0}+t_{m}-x)}}$, we can then conclude that both
$G_{\alpha}$ and  $G_{1-\alpha}$ are symmetric
with respect to ${\cal D}$.

\subsection{Algorithms of computation of B-spline curve}
These algorithms show that it is possible to compute a point of B-spline curve or all of them
without making use of the explicit construction of the associated B-spline basis. The fundamental 
algorithm is of deBoor and can be defined as follows:

\begin{proposition}[ deBoor algorithm] \label{algodeBoor}
Let  $m, k, n\in \NN^*$ such that \mbox{$n \geq k$} and $m=n+ k+1$.
Let $U=\suite{t_i}_{i=0}^{m}$  be a knot vector.
Let  $\Pi=\suite{d_i}_{i=0}^{n}\subset \RR^d$ be a control polygon.

For all $j=k, \ldots , m-k-1$ such that $t_{j}<t_{j+1}$ 
and for all $x\in [t_{j}\,,\,t_{j+1})$
$$
\displaystyle{G_\alpha(x) = 
\sum_{i=j-k+r}^{j}d_{i}^{r}(x) \Bspline{i}{k-r}{\alpha}(x) }
$$
with
$$
\left\lbrace
\begin{array}{ll}
d_{i}^{0}(x)=d_i & \forall i=0, \ldots , n\\
\\
\displaystyle{
d_{i}^{r+1}(x)=w_{i}^{k-r}(x) d_{i+1}^{r}(x) 
+{\left(1-w_{i}^{k-r}(x)\right)} d_{i}^{r}(x)}
&\forall r=0, \ldots , k-1\\
&\forall i=j-k+r, \ldots , j\\
\end{array}
\right.
$$
where
$\displaystyle{w_{i}^{k-r}(x) = \varphi_\alpha\triplet{x}{t_{i}}{t_{i+k-r}}}$

Moreover we have 
$\displaystyle{G_\alpha(x) = d_{j}^{k}(x)  }$

\end{proposition}

\Preuve
Let $j=k, \ldots , m-k-1$ such that $t_{j}<t_{j+1}$ 
and  $x\in [t_{j}\,,\,t_{j+1})$.
Since for all $i$
$$
\displaystyle{
\Bspline{i}{k}{\alpha}(x) = 
w_{i}^{k}(x) \Bspline{i}{k-1}{\alpha}(x) 
+\left(1- w_{i+1}^{k}(x) \right) \Bspline{i+1}{k-1}{\alpha}(x) 
}
$$
then
$$
\begin{array}{rcl}
\displaystyle{
G_\alpha(x)
}
&=&
\displaystyle{
\sum_{i=j-k}^{j}d_{i} \Bspline{i}{k}{\alpha}(x) 
}
\\
&=&
\displaystyle{
\sum_{i=j-k}^{j}d_{i}
w_{i}^{k}(x) \Bspline{i}{k-1}{\alpha}(x) 
}
\\
&+&
\displaystyle{
\sum_{i=j-k}^{j}d_{i}
\left(1- w_{i+1}^{k}(x) \right) \Bspline{i+1}{k-1}{\alpha}(x) 
}
\\
&=&
\displaystyle{
\sum_{i=j-k}^{j}d_{i}
w_{i}^{k}(x) \Bspline{i}{k-1}{\alpha}(x) 
}
\\
&+&
\displaystyle{
\sum_{i=j-k+1}^{j+1}d_{i-1}
\left(1- w_{i}^{k}(x) \right) \Bspline{i}{k-1}{\alpha}(x) 
}
\\
&=&
\displaystyle{
d_{j-k}
w_{j-k}^{k}(x) \Bspline{j-k}{k-1}{\alpha}(x) 
+d_{j}
\left(1- w_{j+1}^{k}(x) \right) \Bspline{j+1}{k-1}{\alpha}(x) 
}
\\
&+&
\displaystyle{
\sum_{i=j-k+1}^{j}
\left[
d_{i-1}
\left(1- w_{i}^{k}(x) \right) 
+d_{i} w_{i}^{k}(x)
\right]
\Bspline{i}{k-1}{\alpha}(x) 
}
\\
\end{array}
$$

$$
\begin{array}{rcl}
\displaystyle{
G_\alpha(x)
}
&=&
\displaystyle{
d_{j-k}
w_{j-k}^{k}(x) \Bspline{j-k}{k-1}{\alpha}(x) 
+d_{j}
\left(1- w_{j+1}^{k}(x) \right) \Bspline{j+1}{k-1}{\alpha}(x) 
}
\\
&+&
\displaystyle{
\sum_{i=j-k+1}^{j}
\left[
d_{i-1}
\left(1- w_{i}^{k}(x) \right) 
+d_{i} w_{i}^{k}(x)
\right]
\Bspline{i}{k-1}{\alpha}(x) 
}
\\
&=&
\displaystyle{
\sum_{i=j-k+1}^{j}
\left[
d_{i-1}
\left(1- w_{i}^{k}(x) \right) 
+d_{i} w_{i}^{k}(x)
\right]
\Bspline{i}{k-1}{\alpha}(x) 
}
\\
&=&
\displaystyle{
\sum_{i=j-k+1}^{j}
d_{i}^{1}(x) 
\Bspline{i}{k-1}{\alpha}(x) 
}
\\
\end{array}
$$
with for all $j-k-1 \leq i \leq j$
$$
\begin{array}{rcl}
\displaystyle{
d_{i}^{1}(x)
}
&=&
\displaystyle{
d_{i-1}
\left(1- w_{i}^{k}(x) \right) 
+d_{i} w_{i}^{k}(x)
}\\
&=&
\displaystyle{
d_{i-1}^{0}(x)
\left(1- w_{i}^{k}(x) \right) 
+d_{i}^{0}(x) w_{i}^{k}(x)
}\\
\end{array}
$$
by setting
$\displaystyle{d_{i}^{0}(x)=d_{i}}$ for all $i$;
since
$$
\begin{array}{l}
\displaystyle{
\Support{\Bspline{j-k}{k-1}{\alpha}}\cap  [t_{j}\,,\,t_{j+1})=\emptyset
}
\\
\displaystyle{
\Support{\Bspline{j+1}{k-1}{\alpha}}\cap  [t_{j}\,,\,t_{j+1})=\emptyset
}
\\
\end{array}
$$

We have established
$$
\displaystyle{
G_\alpha(x)
}
=
\displaystyle{
\sum_{i=j-k}^{j}
d_{i}^{0}(x) 
\Bspline{i}{k}{\alpha}(x) 
}
=
\displaystyle{
\sum_{i=j-k+1}^{j}
d_{i}^{1}(x) 
\Bspline{i}{k-1}{\alpha}(x) 
}
$$

Let us show by recurrence that for all $0\leq r \leq k$ we have
$$
\displaystyle{
G_\alpha(x)
}
=
\displaystyle{
\sum_{i=j-k+r}^{j}
d_{i}^{r}(x) 
\Bspline{i}{k-r}{\alpha}(x) 
}
$$
with for all $r\leq k$
$$
\begin{array}{rcl}
\displaystyle{
d_{i}^{r}(x)
}
&=&
\displaystyle{
d_{i-1}^{r-1}(x)
\left(1- w_{i}^{k-r+1}(x) \right) 
+d_{i}^{r-1}(x) w_{i}^{k-r+1}(x)
}\\
\end{array}
$$

We assume that for all $1\leq r <k$ we have 
$$
\displaystyle{
G_\alpha(x)
}
=
\displaystyle{
\sum_{i=j-k+r}^{j}
d_{i}^{r}(x) 
\Bspline{i}{k-r}{\alpha}(x) 
}
$$
with
$$
\begin{array}{rcl}
\displaystyle{
d_{i}^{r}(x)
}
&=&
\displaystyle{
d_{i-1}^{r-1}(x)
\left(1- w_{i}^{k-r+1}(x) \right) 
+d_{i}^{r-1}(x) w_{i}^{k-r+1}(x)
}\\
\end{array}
$$

Then
$$
\begin{array}{rcl}
\displaystyle{
G_\alpha(x)
}
&=&
\displaystyle{
\sum_{i=j-k+r}^{j}
d_{i}^{r}(x) 
\Bspline{i}{k-r}{\alpha}(x) 
}\\
&=&
\displaystyle{
\sum_{i=j-k+r}^{j}
d_{i}^{r}(x) 
w_{i}^{k-r}
\Bspline{i}{k-r-1}{\alpha}(x) 
}\\
&+&
\displaystyle{
\sum_{i=j-k+r}^{j}
d_{i}^{r}(x)
\left(1 - w_{i+1}^{k-r}\right)
\Bspline{i+1}{k-r-1}{\alpha}(x) 
}\\
&=&
\displaystyle{
\sum_{i=j-k+r}^{j}
d_{i}^{r}(x) 
w_{i}^{k-r}
\Bspline{i}{k-r-1}{\alpha}(x) 
}\\
&+&
\displaystyle{
\sum_{i=j-k+r+1}^{j+1}
d_{i-1}^{r}(x)
\left(1 - w_{i}^{k-r}\right)
\Bspline{i}{k-r-1}{\alpha}(x) 
}\\
&=&
\displaystyle{
d_{j-k+r}^{r}(x) 
w_{j-k+r}^{k-r}
\Bspline{j-k+r}{k-r-1}{\alpha}(x) 
+
\left(1 - w_{j+1}^{k-r}\right)
d_{j}^{r}(x) 
w_{j+1}^{k-r}
\Bspline{j+1}{k-r-1}{\alpha}(x) 
}\\
&+&
\displaystyle{
\sum_{i=j-k+r+1}^{j}
\left[
d_{i-1}^{r}(x)
+
\left(1 - w_{i}^{k-r}\right)
d_{i}^{r}(x) 
w_{i}^{k-r}
\right]
\Bspline{i}{k-r-1}{\alpha}(x) 
}\\
&=&
\displaystyle{
\sum_{i=j-k+r+1}^{j}
\left[
d_{i-1}^{r}(x)
+
\left(1 - w_{i}^{k-r}\right)
d_{i}^{r}(x) 
w_{i}^{k-r}
\right]
\Bspline{i}{k-r-1}{\alpha}(x) 
}\\
&=&
\displaystyle{
\sum_{i=j-k+r+1}^{j}
d_{i}^{r+1}(x) 
\Bspline{i}{k-r-1}{\alpha}(x) 
}\\
\end{array}
$$
with
$$
d_{i}^{r+1}(x) 
=
d_{i-1}^{r}(x)
+
\left(1 - w_{i}^{k-r}\right)
d_{i}^{r}(x) 
w_{i}^{k-r}
$$
since
$$
\begin{array}{l}
\Support{\Bspline{j-k+r}{k-r-1}{\alpha}}\cap [t_{j}\,,\,t_{j+1})=\emptyset\\
\Support{\Bspline{j+1}{k-r-1}{\alpha}}\cap [t_{j}\,,\,t_{j+1})=\emptyset\\
\end{array}
$$
We have thus proved that for all  
 $0\leq r \leq k$ we have
$$
\displaystyle{
G_\alpha(x)
}
=
\displaystyle{
\sum_{i=j-k+r}^{j}
d_{i}^{r}(x) 
\Bspline{i}{k-r}{\alpha}(x) 
}
$$
with for all $r\leq k$
$$
\begin{array}{rcl}
\displaystyle{
d_{i}^{r}(x)
}
&=&
\displaystyle{
d_{i-1}^{r-1}(x)
\left(1- w_{i}^{k-r+1}(x) \right) 
+d_{i}^{r-1}(x) w_{i}^{k-r+1}(x)
}\\
\end{array}
$$

For $r=k$, we have for all $x \in  [t_{j}\,,\,t_{j+1})$
$$
\displaystyle{
G_\alpha(x)
}
=
\displaystyle{
\sum_{i=j}^{j}
d_{i}^{k}(x)
\Bspline{i}{0}{\alpha}(x) 
=d_{j}^{k}(x)
\Bspline{j}{0}{\alpha}(x) 
=d_{j}^{k}(x)
}
$$

This completes the proof.

\section{Some illustrations of properties of the new class of rational
B-spline curves}
In this section, we will present a set of practical cases which depicts the 
established properties in previous sections. Here the aim is just to give some illustration view 
without being concerned with the issue of algorithm optimization. To this end, 
we have adopted \textbf{Scilab} scripts and sometimes \textbf{Maxima} scripts particularly for
the formal expressions of B-spline basis listed in appendix.

We will first present the basis and then the B-spline curves.

\subsection{The new class of rational B-spline basis}

We emphasize on illustrations of first properties of the new class of B-spline basis.

We know that the B-spline basis are grouped in two categories regarding the fact that 
they are spanned by a periodic knot vector or not and in each category, the knot vector
may be uniform or not. We shall go through all of these variations.

\subsubsection{Case of periodic knot vectors}

We plan two illustrations. The first one explores the influence
of the uniformity of knot vector while the second one explores the non-uniformity.

\begin{castest}
We present here B-spline basis of degree $0$ to $3$ for the uniform
periodic knot vector 
$\displaystyle{U_0=\left(0,1,2,3,4,5,6 \right)}$ with
$\displaystyle{\alpha \in\left\lbrace -1,2,5,\infty \right\rbrace}$

\begin{figure}[h!]
\begin{center}
\includegraphics[width=11cm]{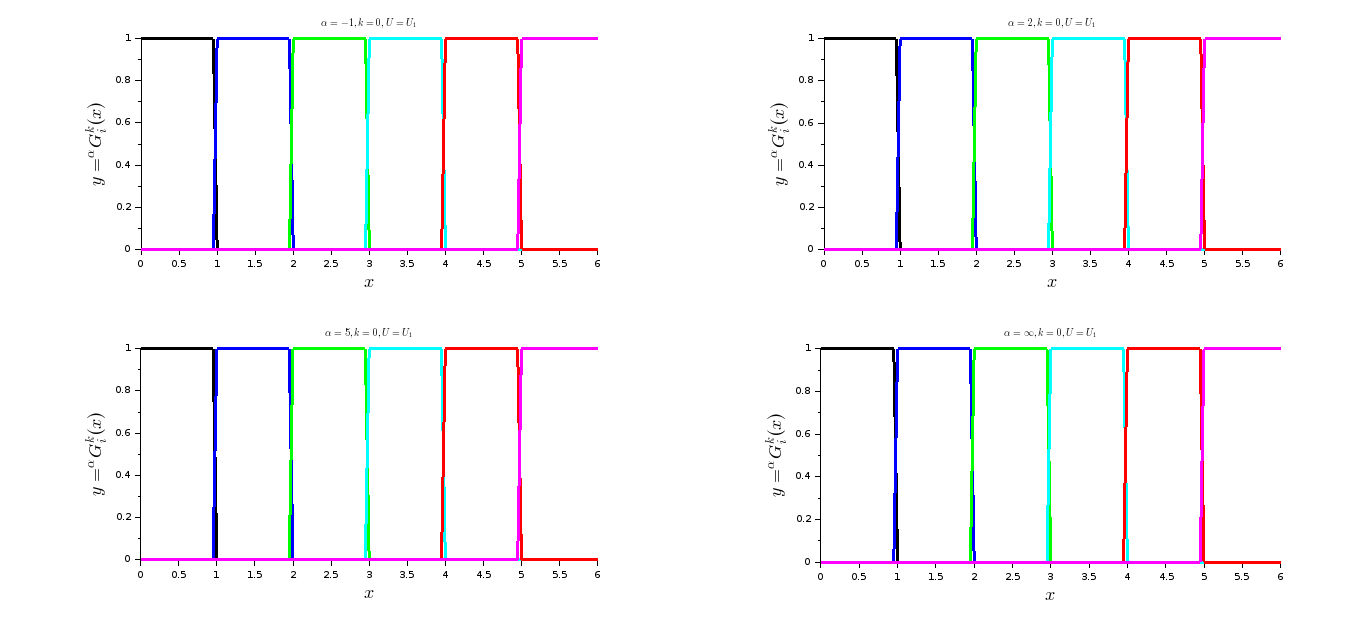}
\caption{The B-spline basis $\displaystyle{\Bspline{i}{0}{\alpha}}$ 
of knot vector $U_0$}
\label{figBaseBspline1}
\end{center}
\end{figure}

\begin{figure}[h!]
\begin{center}
\includegraphics[width=11cm]{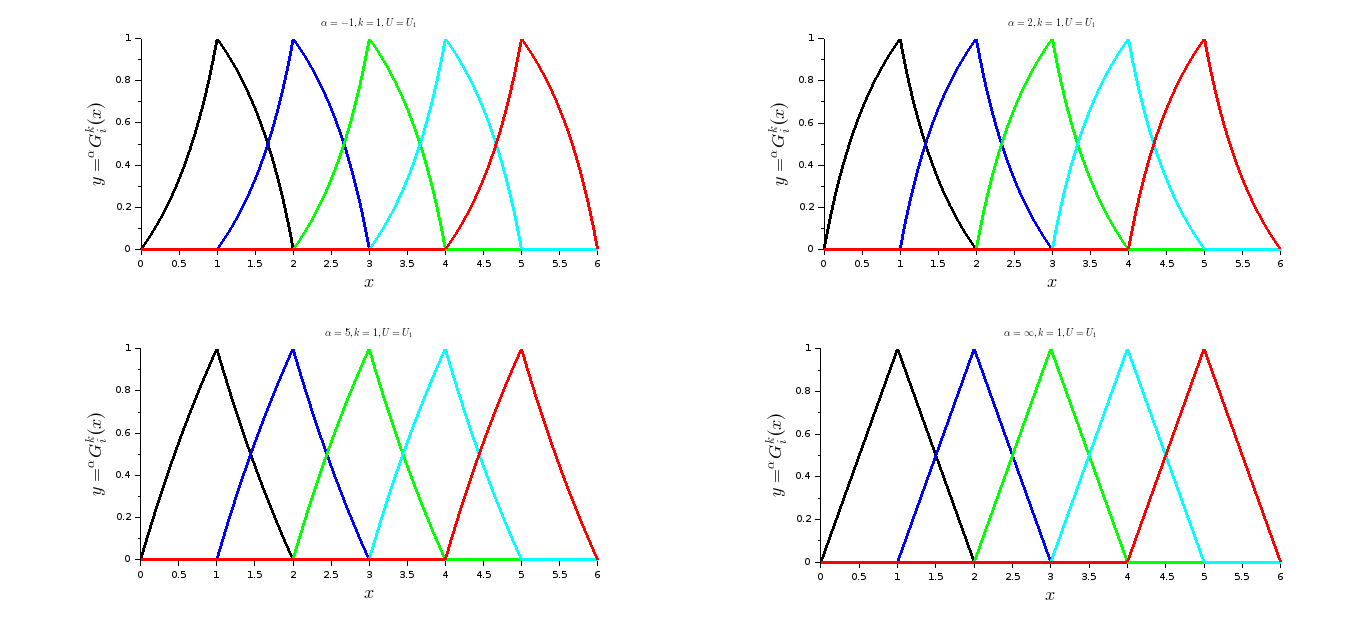}
\caption{The B-spline basis $\displaystyle{\Bspline{i}{1}{\alpha}}$
of knot vector $U_0$}
\label{figBaseBspline2}
\end{center}
\end{figure}

\begin{figure}[h!]
\begin{center}
\includegraphics[width=11cm]{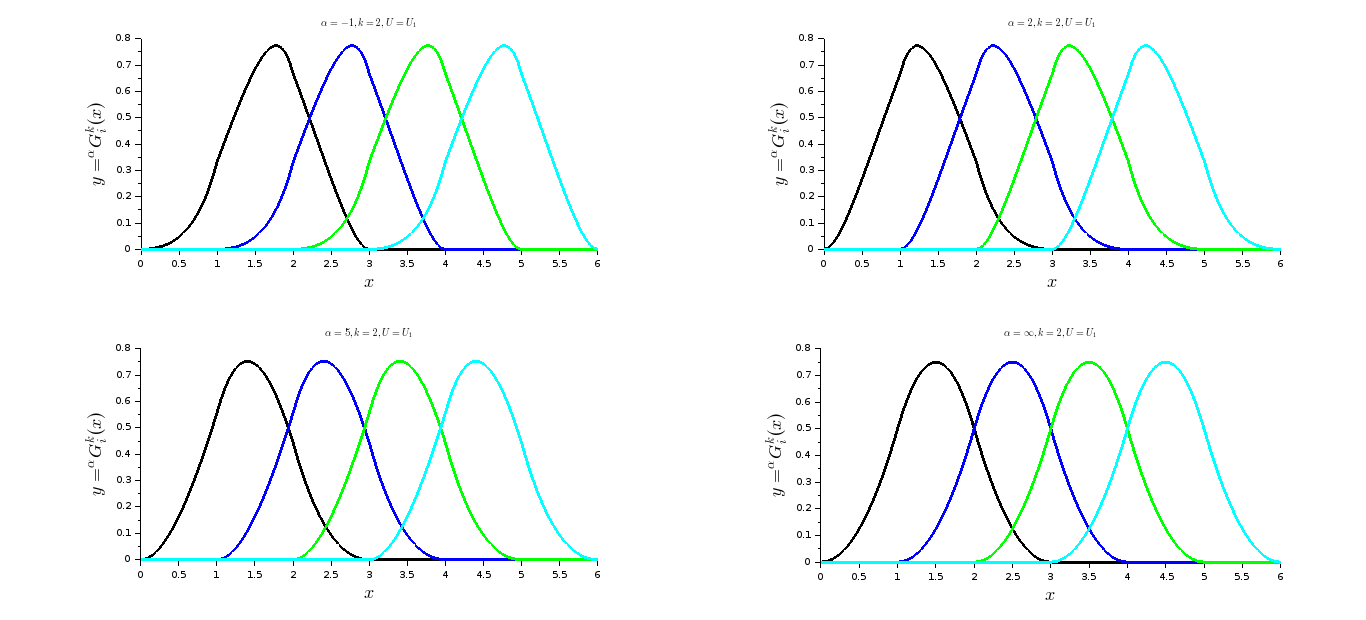}
\caption{The B-spline basis $\displaystyle{\Bspline{i}{2}{\alpha}}$
of knot vector $U_0$}
\label{figBaseBspline3}
\end{center}
\end{figure}

\begin{figure}[h!]
\begin{center}
\includegraphics[width=11cm]{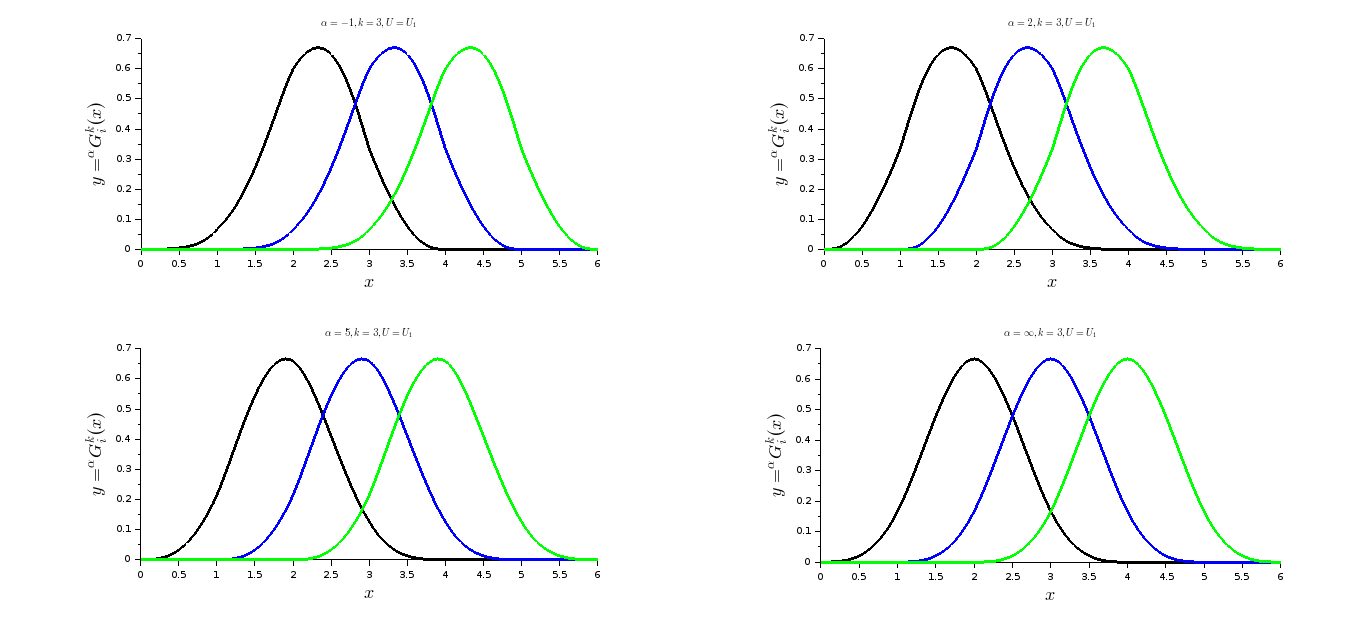}
\caption{The B-spline basis $\displaystyle{\Bspline{i}{3}{\alpha}}$
of knot vector $U_0$}
\label{figBaseBspline4}
\end{center}
\end{figure}

\end{castest}

From the analysis of figures
\ref{figBaseBspline1}  to \ref{figBaseBspline4}, 
we deduce that since $U_0$ is a uniform periodic knot vector, an element of the basis 
$\displaystyle{\suite{\Bspline{i}{k}{\alpha}}_{i=0}^{m-k-1}}$ is obtained by simple translation of 
$\displaystyle{\Bspline{0}{k}{\alpha}}$ that is    
$\displaystyle{
\Bspline{i}{k}{\alpha}(x)=
\Bspline{0}{k}{\alpha}(t_{0}-t_{i}+x)
}$.

We observe that
$\displaystyle{\Support{\Bspline{i}{k}{\alpha}} =\segment{t_{i}}{t_{i+k+1}}}$
and also the effect of parameter $\alpha$ is crucial at the neighborhood of $0^{-}$ and $1^{+}$.
The figure \ref{figBaseBspline4} seems to show that $\alpha$ does not have any influence
 on $\displaystyle{\suite{\Bspline{i}{3}{\alpha}}_{i=0}^{m-4}}$ which corresponds to a context 
of knot vector with no interior nodes. 

\begin{castest}
We present the influence of the non-uniformity of a 
periodic knot vector by restricting ourselves on B-spline basis
of degree $2$ in the following cases:

\begin{tabular}{l}
$\displaystyle{U_1=\left(0,1,2,3,3,5,6 \right)}$\\
$\displaystyle{U_2=\left(0,1,1,2,4,5,6 \right)}$\\
$\displaystyle{U_3=\left(0,1,1.5,2,3.5,5,6 \right)}$\\
\end{tabular}


\begin{figure}[h!]
\begin{center}
\includegraphics[width=11cm]{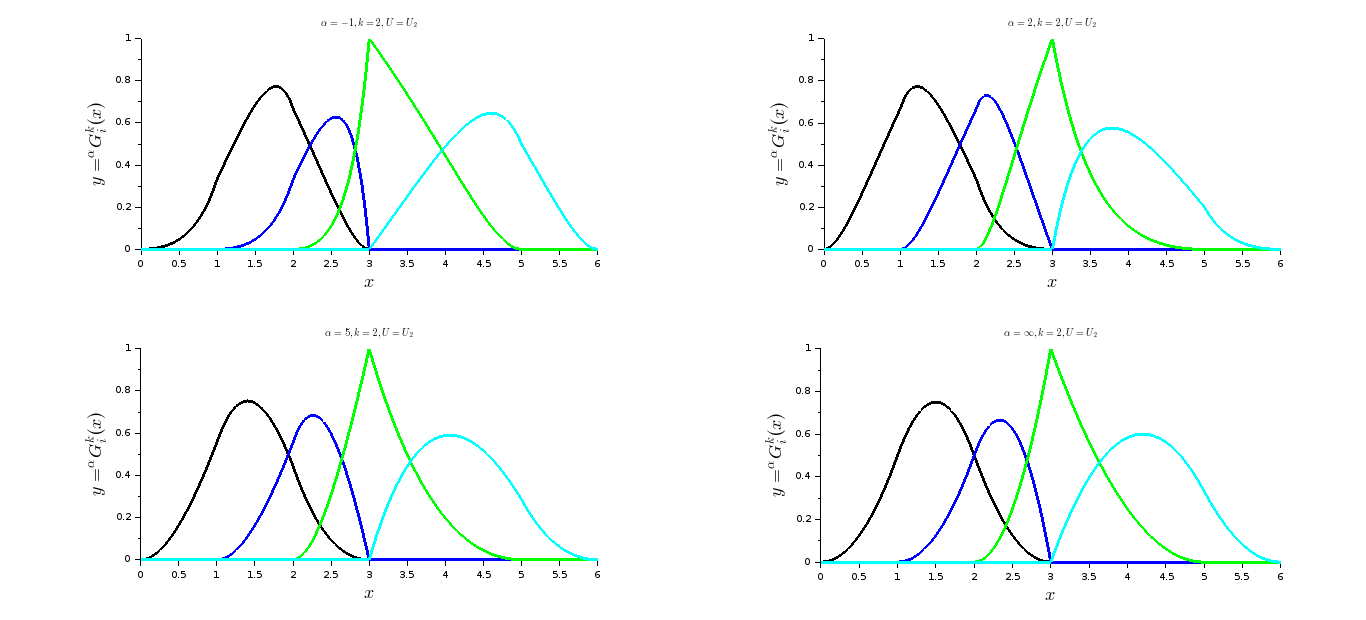}
\caption{The B-spline basis  $\displaystyle{\Bspline{i}{2}{\alpha}}$
of knot vector $U_1$}
\label{figBaseBspline6}
\end{center}
\end{figure}

\begin{figure}[h!]
\begin{center}
\includegraphics[width=11cm]{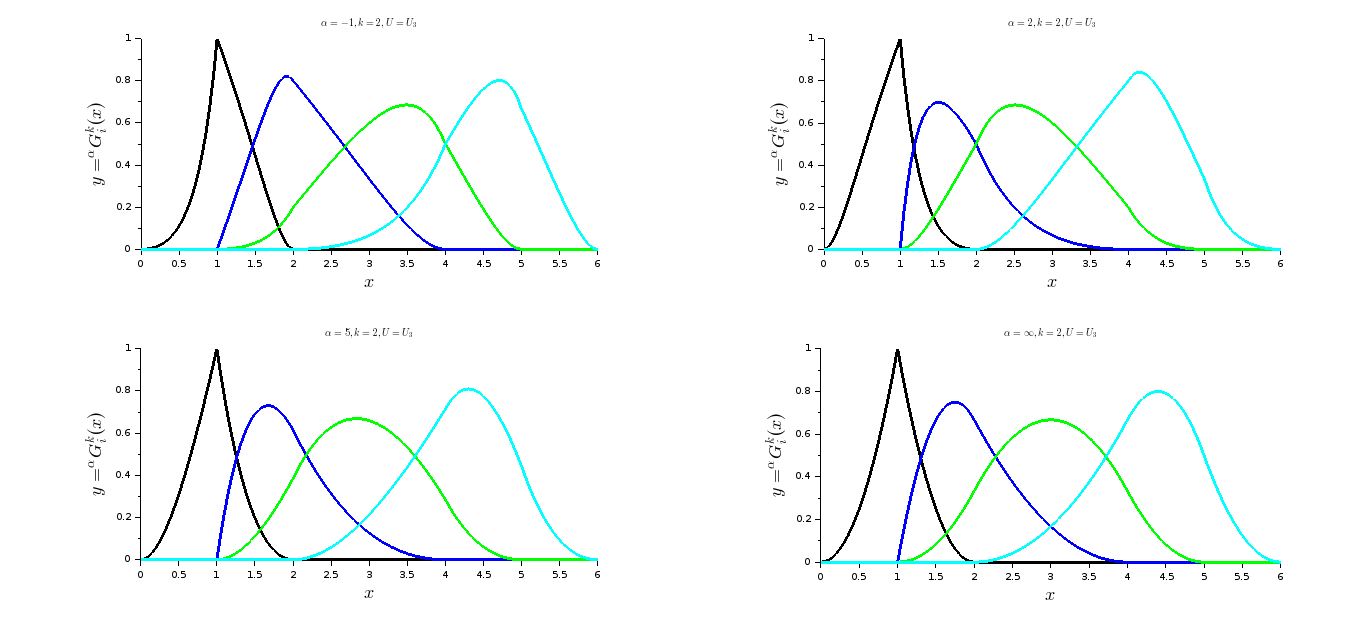}
\caption{The B-spline basis $\displaystyle{\Bspline{i}{2}{\alpha}}$
of knot vector $U_2$}
\label{figBaseBspline7}
\end{center}
\end{figure}

\begin{figure}[h!]
\begin{center}
\includegraphics[width=11cm]{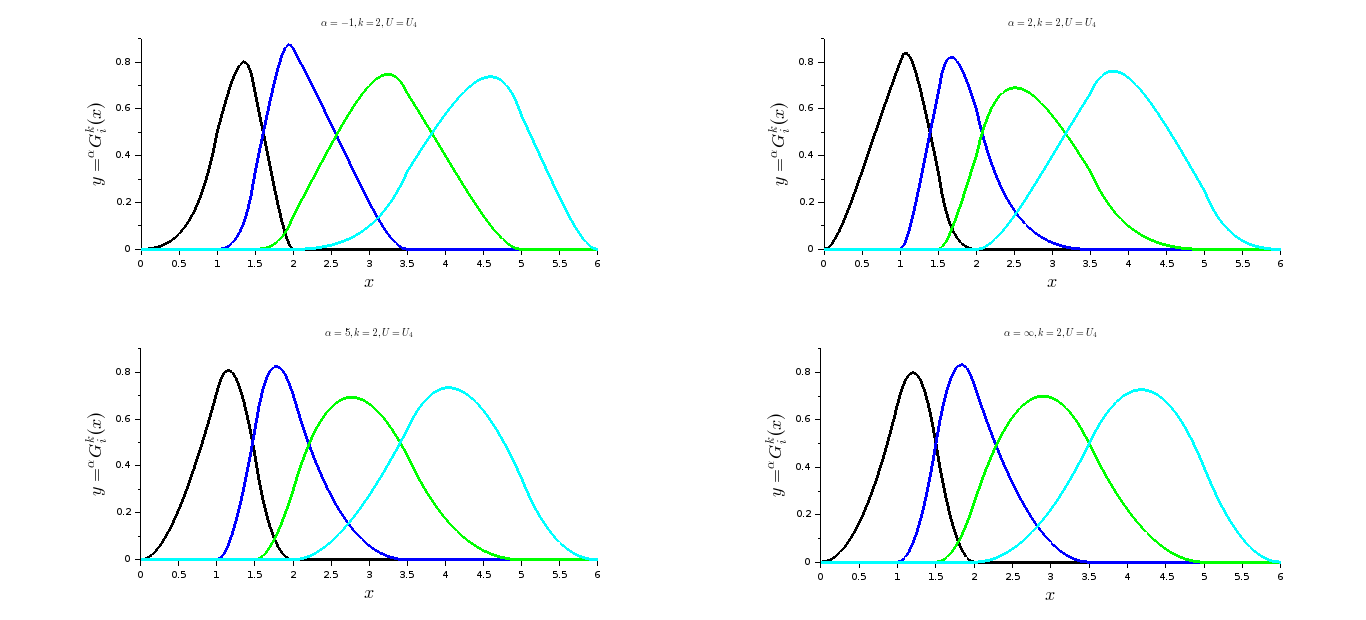}
\caption{The B-spline basis $\displaystyle{\Bspline{i}{2}{\alpha}}$
of knot vector $U_3$}
\label{figBaseBspline8}
\end{center}
\end{figure}

\end{castest}

The non-uniformity may come from the presence of a multiple node, it is the case of knot vectors
$U_1$ and  $U_2$. It may be also due to the step of variable between nodes as in $U_3$.

The figures \ref{figBaseBspline6} to  \ref{figBaseBspline8} show that
in all the cases we have
$\displaystyle{\Support{\Bspline{i}{2}{\alpha}} =\segment{t_{i}}{t_{i+3}}}$
and the effect of the parameter $\alpha$ remains important at the neighborhood of  
 $0^{-}$ and $1^{+}$. We observe a large diversity among the elements of the basis concerning 
the regularity.

The two illustrations of this subsection seem to confirm 
the conjecture~\ref{ConjecMaximum} related to the existence of a unique maximum for $\displaystyle{\Bspline{i}{k}{\alpha}}$ when $k>0$.

\subsubsection{Case of open knot vectors}

This subsection is also based on two test cases which give light on the basis of degree $2$ generated by open knot vectors for $\displaystyle{\alpha \in \lbrace -1, 2, 5, \infty  \rbrace}$.

The first test case  dealts with five knot vectors  having two multiple interior nodes or not.

In the second test case we also have five knot vectors but having three interior nodes
where the multiplicity may reach $3$.

\begin{castest}
We explore the case of B-spline basis of degree $2$ associated with an open knot vector
in the following cases:

\begin{tabular}{l}
$\displaystyle{U_4=\left(0,0,0,1,2,3,3,3 \right)}$\\
$\displaystyle{U_5=\left(0,0,0,0.4,2.6,3,3,3 \right)}$\\
$\displaystyle{U_6=\left(0,0,0,1.8,2.2,3,3,3 \right)}$\\
$\displaystyle{U_7=\left(0,0,0,1,1,3,3,3 \right)}$\\
$\displaystyle{U_8=\left(0,0,0,2,2,3,3,3 \right)}$\\
\end{tabular}

\begin{figure}[h!]
\begin{center}
\includegraphics[width=11cm]{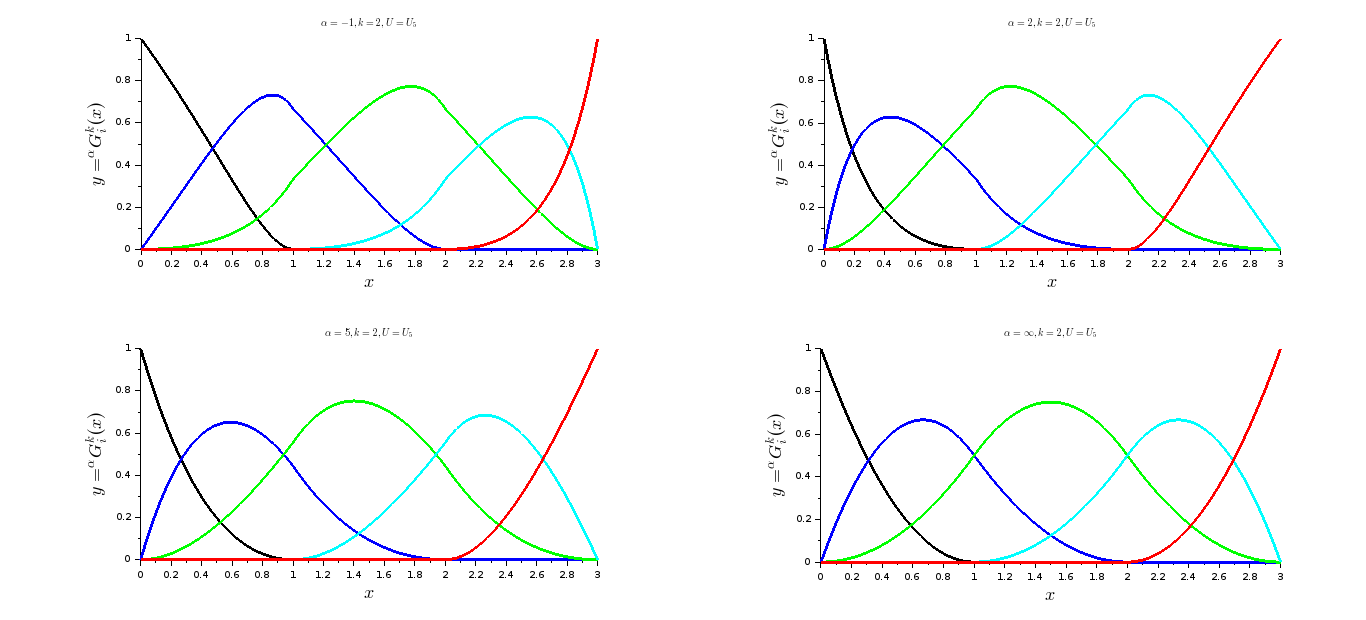}
\caption{The B-spline basis $\displaystyle{\Bspline{i}{2}{\alpha}}$
of knot vector $U_4$}
\label{figBaseBspline9}
\end{center}
\end{figure}

\begin{figure}[h!]
\begin{center}
\includegraphics[width=11cm]{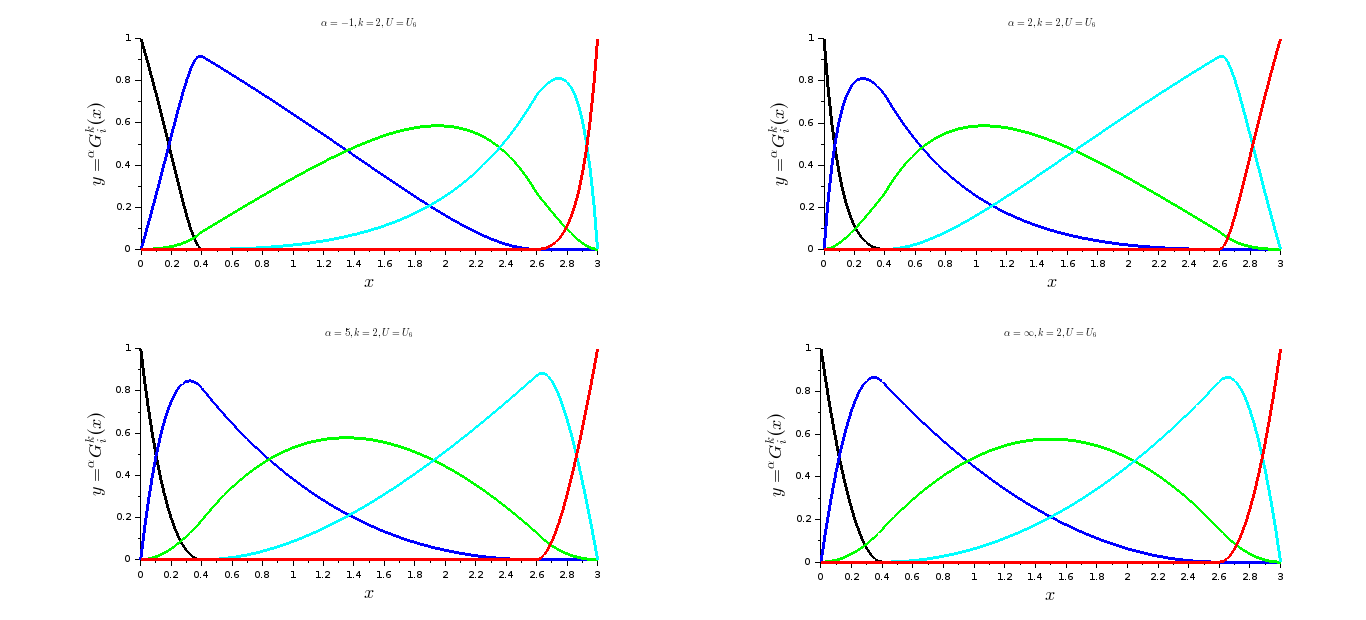}
\caption{The B-spline basis $\displaystyle{\Bspline{i}{2}{\alpha}}$
of knot vector $U_5$}
\label{figBaseBspline10}
\end{center}
\end{figure}

\begin{figure}[h!]
\begin{center}
\includegraphics[width=11cm]{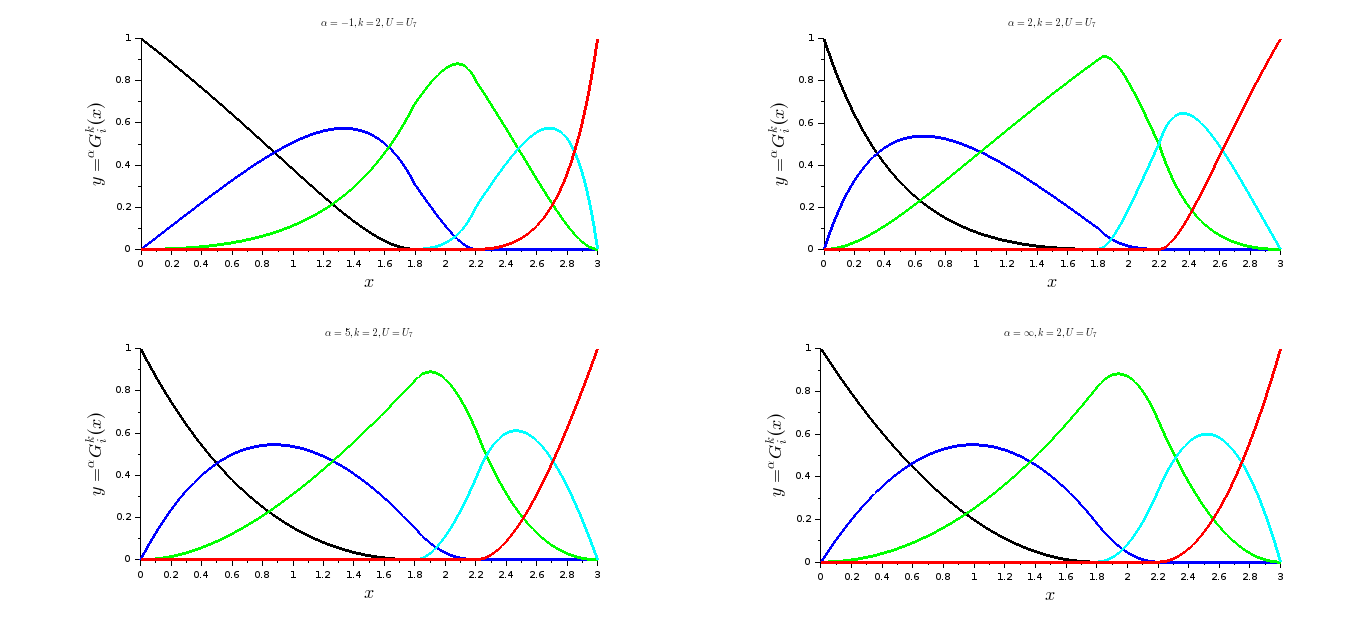}
\caption{The B-spline basis $\displaystyle{\Bspline{i}{2}{\alpha}}$
of knot vector $U_6$}
\label{figBaseBspline11}
\end{center}
\end{figure}

\begin{figure}[h!]
\begin{center}
\includegraphics[width=11cm]{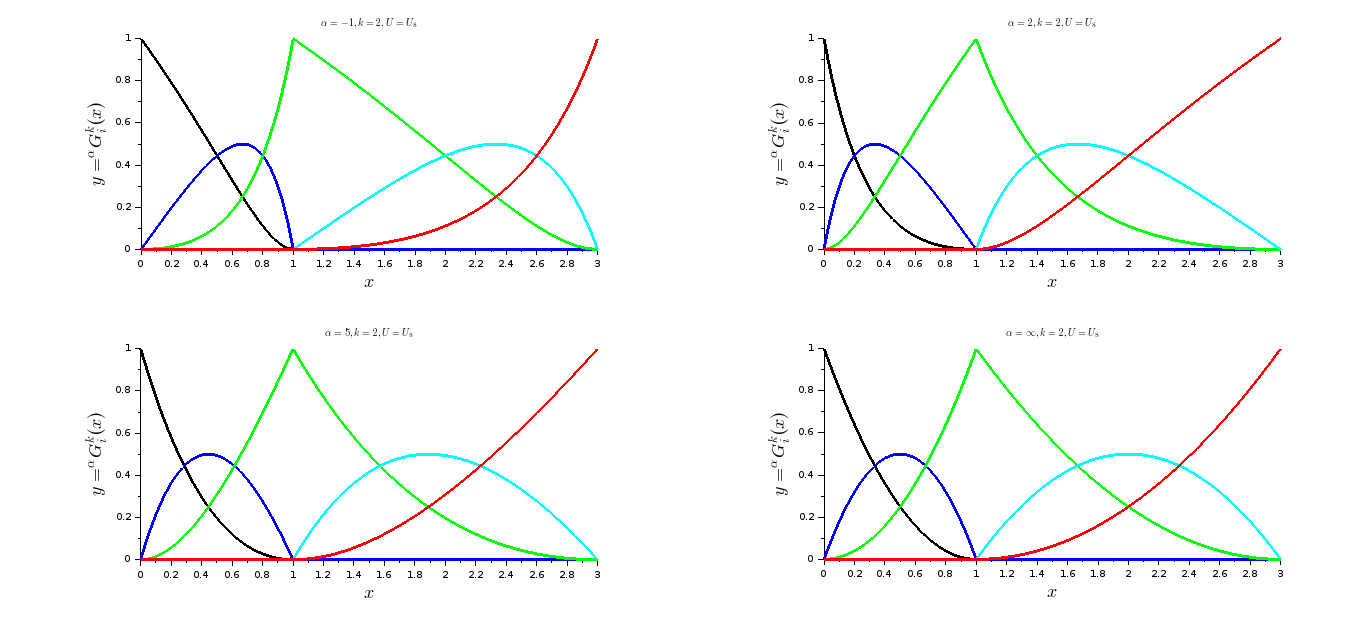}
\caption{The B-spline basis $\displaystyle{\Bspline{i}{2}{\alpha}}$
of knot vector $U_7$}
\label{figBaseBspline12}
\end{center}
\end{figure}

\begin{figure}[h!]
\begin{center}
\includegraphics[width=11cm]{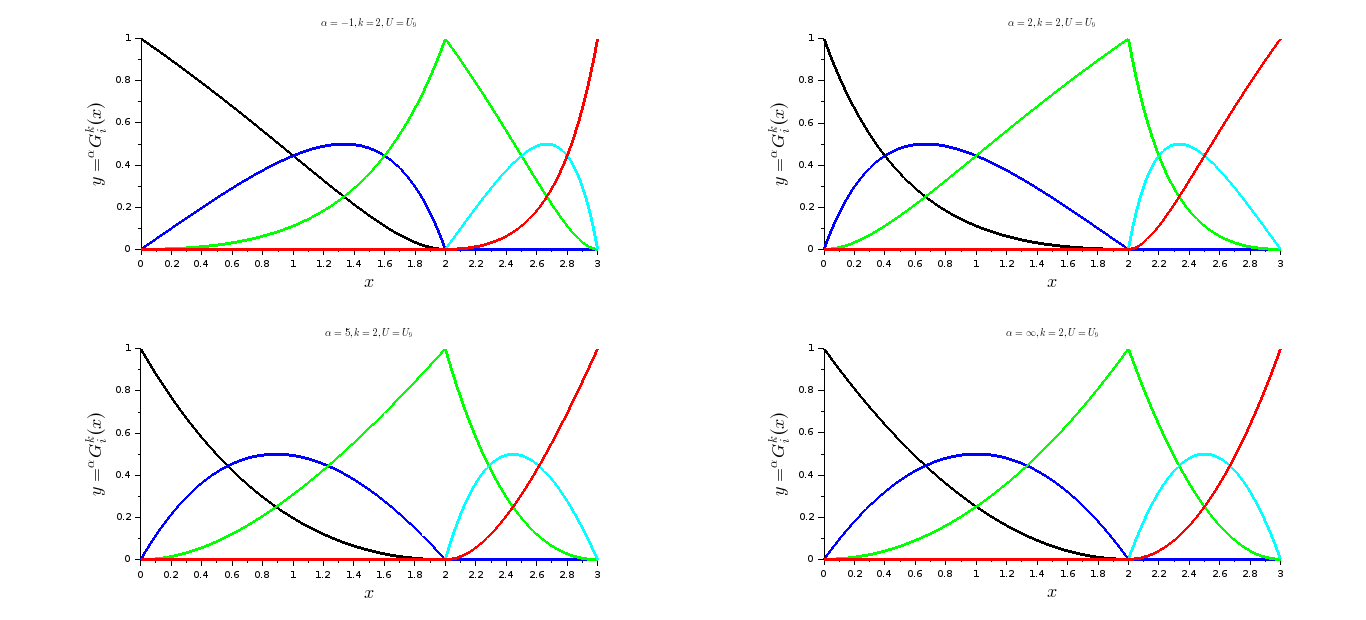}
\caption{The B-spline basis $\displaystyle{\Bspline{i}{2}{\alpha}}$
of knot vector $U_8$}
\label{figBaseBspline13}
\end{center}
\end{figure}

The figures \ref{figBaseBspline9} to  \ref{figBaseBspline13} illustrate abundantly
the properties of the proposition \ref{PropoContinu} especially those of values
at extreme nodes.

The figures \ref{figBaseBspline9} and \ref{figBaseBspline10} depict the behaviors 
of basis generated respectively by $U_4$ and $U_5$ which are symmetric knot vectors.
One can observe that for all $x \in \segment{t_{0}}{t_{7}}$, we have
$$
\begin{array}{l}
\displaystyle{
\Bspline{i}{2}{-1}  (t_{0}+t_{7}-x)  = \Bspline{4-i}{2}{2}  (x) 
}\\
\displaystyle{
\Bspline{i}{2}{2}  (t_{0}+t_{7}-x)  = \Bspline{4-i}{2}{-1}  (x) 
}\\
\displaystyle{
\Bspline{i}{2}{\infty}  (t_{0}+t_{7}-x)  = \Bspline{4-i}{2}{\infty}  (x) 
}\\
\end{array}
$$

For the non-uniform open knot vector $U_6$, $U_7$ and $U_8$ we observe 
a large diversity of behaviors of generated basis.
\end{castest}

\begin{castest}
The B-spline basis of degree $2$ we are illustrating explore the existing relation
between the regularity and the multiplicity of an interior node of an open knot vector
in the following cases:

\begin{tabular}{l}
$\displaystyle{U_9=\left(0,0,0,3/4,6/4,9/4,3,3,3 \right)}$\\
$\displaystyle{U_{10}=\left(0,0,0,3/4,3/4,9/4,3,3,3 \right)}$\\
$\displaystyle{U_{11}=\left(0,0,0,3/4,3/4,3/4,3,3,3 \right)}$\\
$\displaystyle{U_{12}=\left(0,0,0,3/4,9/4,9/4,3,3,3 \right)}$\\
$\displaystyle{U_{13}=\left(0,0,0,9/4,9/4,9/4,3,3,3 \right)}$\\
\end{tabular}

\begin{figure}[h!]
\begin{center}
\includegraphics[width=11cm]{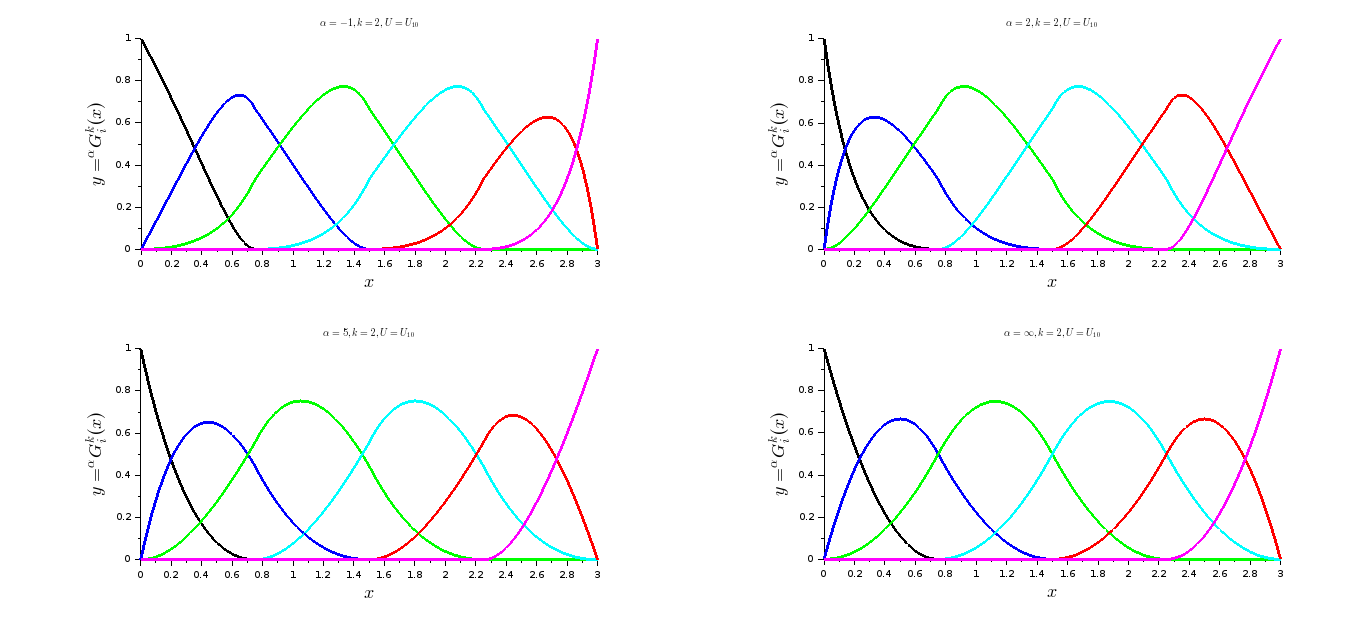}
\caption{The B-spline basis $\displaystyle{\Bspline{i}{2}{\alpha}}$
of knot vector $U_{9}$}
\label{figBaseBspline14}
\end{center}
\end{figure}

\begin{figure}[h!]
\begin{center}
\includegraphics[width=11cm]{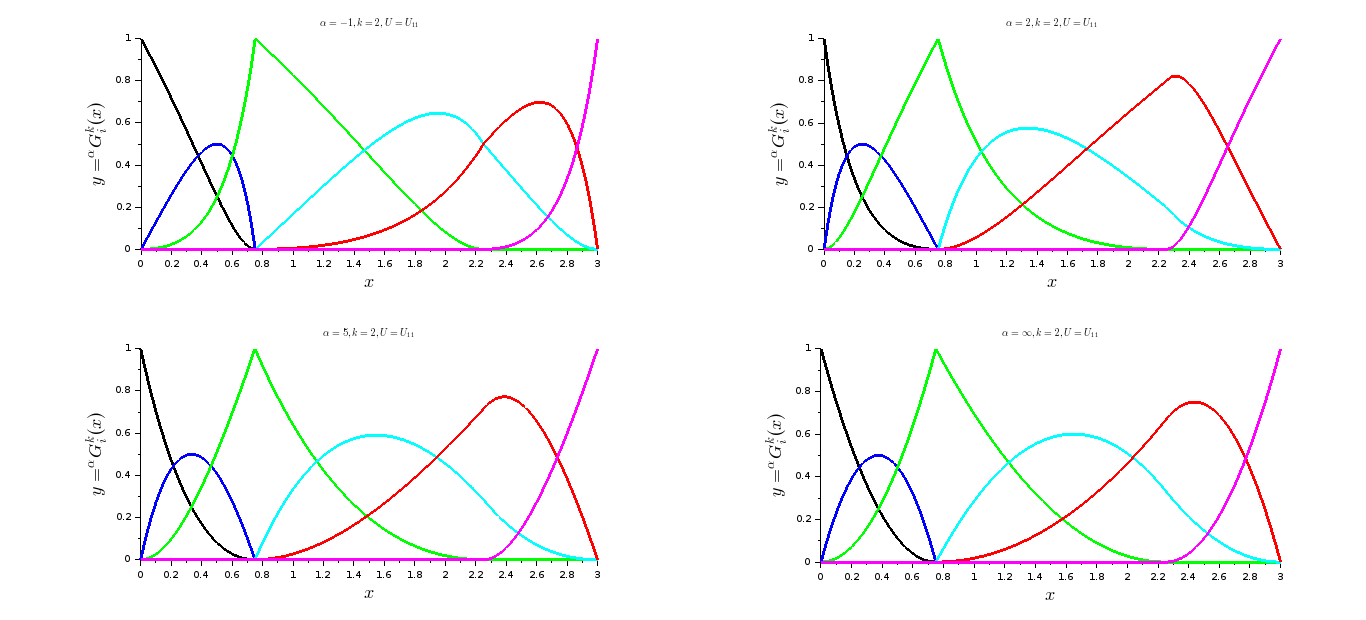}
\caption{The B-spline basis $\displaystyle{\Bspline{i}{2}{\alpha}}$
of knot vector $U_{10}$}
\label{figBaseBspline15}
\end{center}
\end{figure}

\begin{figure}[h!]
\begin{center}
\includegraphics[width=11cm]{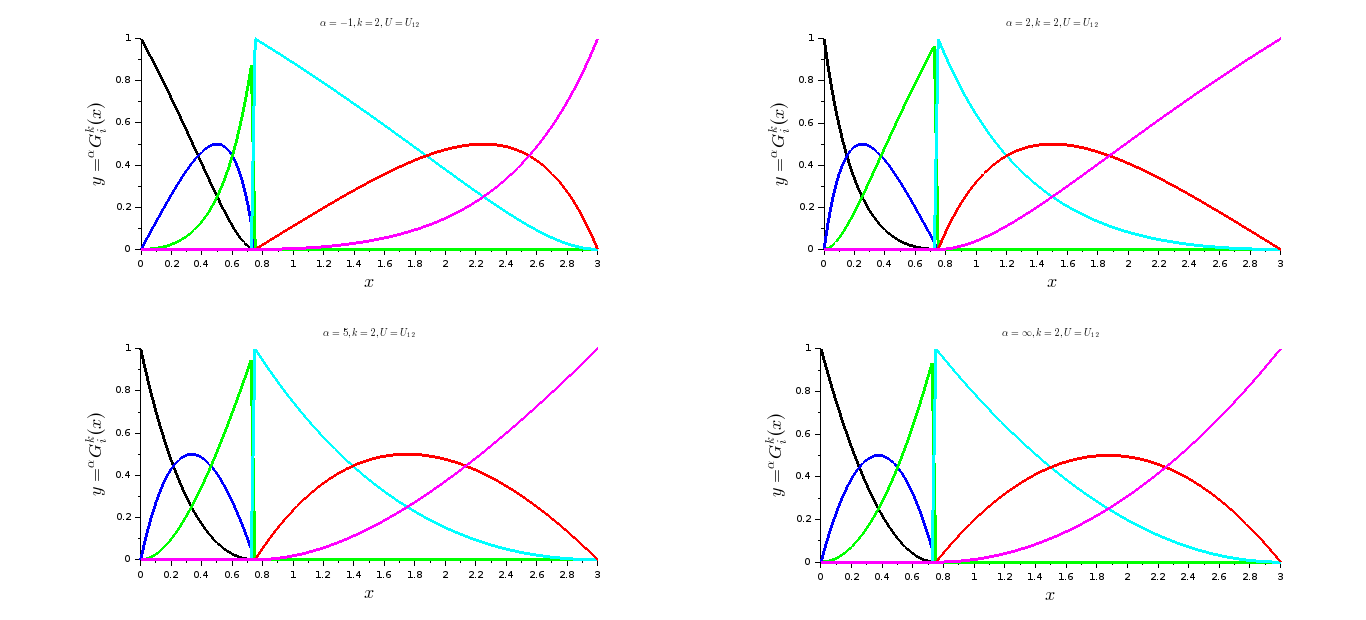}
\caption{The B-spline basis $\displaystyle{\Bspline{i}{2}{\alpha}}$
of knot vector $U_{11}$}
\label{figBaseBspline16}
\end{center}
\end{figure}

\begin{figure}[h!]
\begin{center}
\includegraphics[width=11cm]{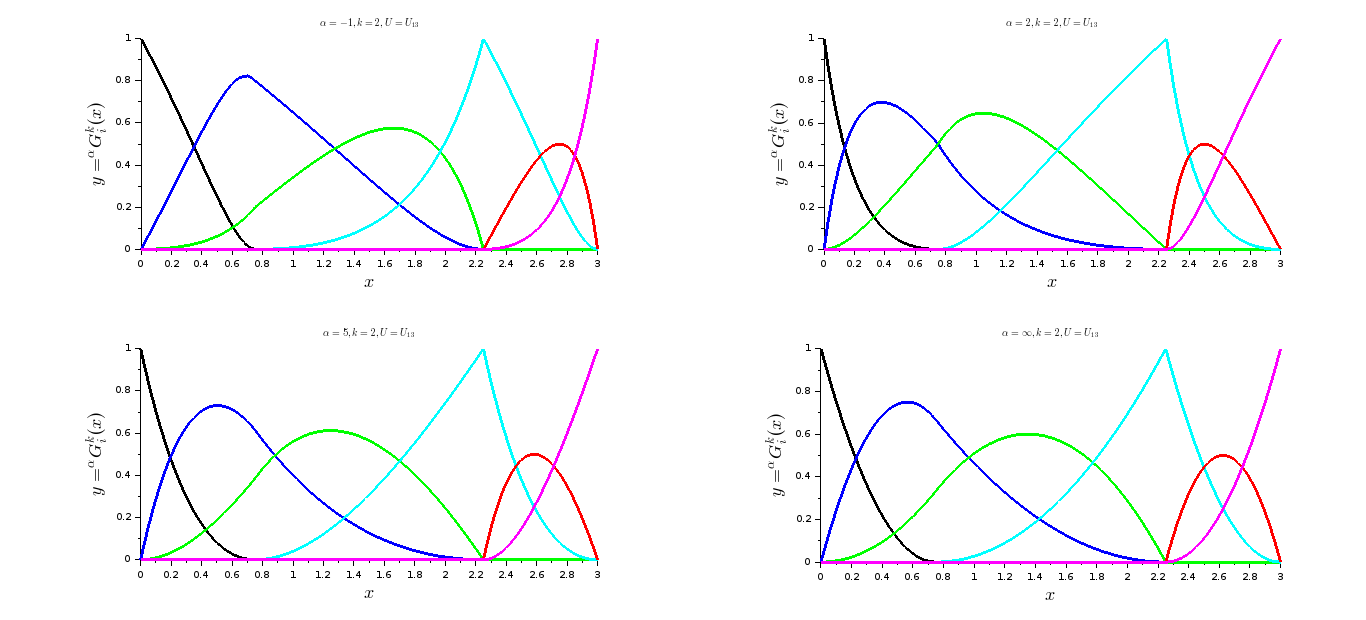}
\caption{The B-spline basis $\displaystyle{\Bspline{i}{2}{\alpha}}$
of knot vector $U_{12}$}
\label{figBaseBspline17}
\end{center}
\end{figure}

\begin{figure}[h!]
\begin{center}
\includegraphics[width=11cm]{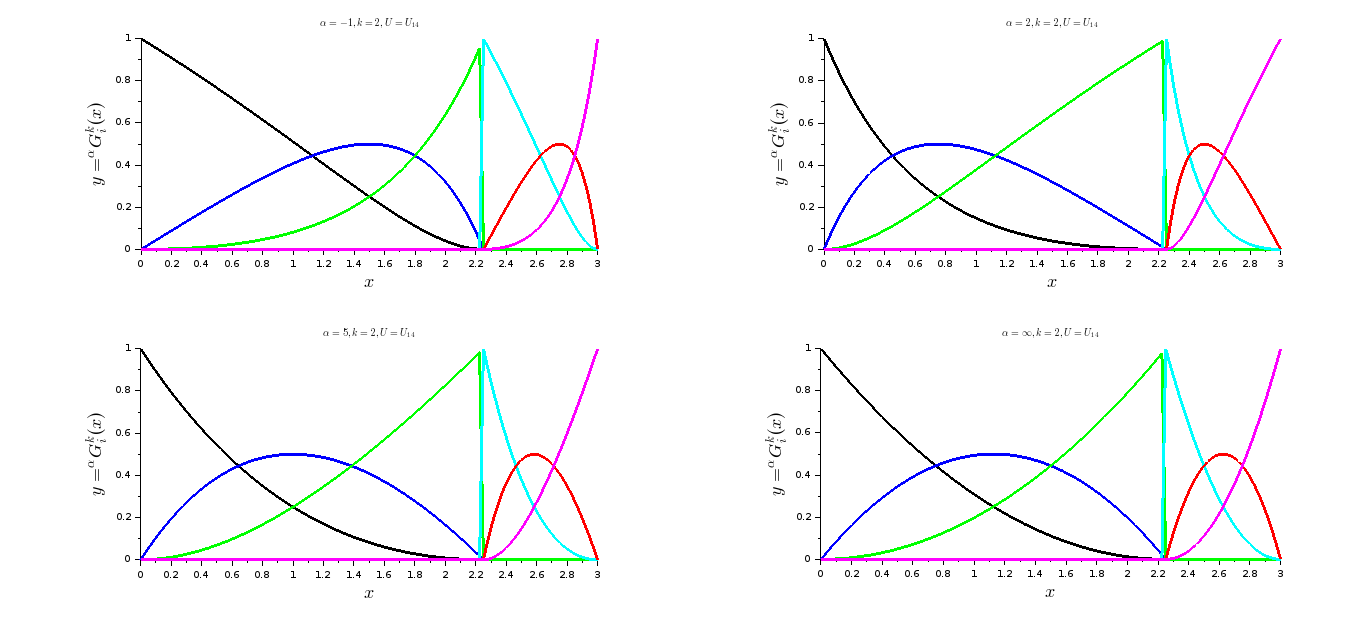}
\caption{The B-spline basis $\displaystyle{\Bspline{i}{2}{\alpha}}$
of knot vector $U_{13}$}
\label{figBaseBspline18}
\end{center}
\end{figure}

The knot vector $U_{9}$ is uniform with interior nodes of multiplicity $1$ and we 
observe in figure \ref{figBaseBspline14} that the generated 
basis confirms the behaviors we already observed with $U_4$.
We can state their regularity of ${\cal C}^0$ as well as the left and right 
differentiability at any interior node as provided in proposition \ref{PropoRegular}.

Each of the knot vectors $U_{10}$ and $U_{12}$ has one interior node with multiplicity $2$. 
The analysis of figures \ref{figBaseBspline15} and \ref{figBaseBspline17} shows that  
the associated basis $\displaystyle{\Bspline{i}{2}{\alpha}}$ are at least of ${\cal C}^0$
with the existence of a left and right derivatives at any interior node even at a double node
confirming the results in proposition \ref{PropoRegular}.

Each of the knot vectors $U_{11}$ and  $U_{13}$ has one interior triple node $t_3=t_4=t_5$. 
We must expect a first type of discontinuity for the elements
$\displaystyle{\Bspline{2}{2}{\alpha}}$ and $\displaystyle{\Bspline{3}{2}{\alpha}}$ 
of the associated basis as 
$\displaystyle{\Support{\Bspline{2}{2}{\alpha}}=\segment{t_2}{t_5}}$ and 
$\displaystyle{\Support{\Bspline{3}{2}{\alpha}}=\segment{t_3}{t_6}}$.
The other elements of the basis keep the regularity of 
${\cal C}^0$ with the existence of a left and right 
derivatives at any interior node. This is confirmed by the analysis of 
figures \ref{figBaseBspline16} and \ref{figBaseBspline18}.

\end{castest}

\begin{remarque}
Either the knot vector is periodic or open, 
uinform or not, we observe in all the cases that 
$
\displaystyle{
\Bspline{i}{2}{\infty}  \approx \Bspline{i}{2}{5} 
}
$
and the conjecture \ref{ConjecMaximum} is verified.
\end{remarque}

\subsection{The new class of rational B-spline curves}

Let us have a look on some examples showing the behavior 
of new B-spline curves under the effect of various parameter appearing in their definition. 

Amongst some parameters we can refer to index $\alpha$, the degree $k$, the knot vector
 $U$ and the control polygon $\Pi$.

\begin{castest}
\label{CastestAlpha}
Let begin with the new parameter which is the index $\alpha$.
We fix the degree to $3$ on the uniform and open knot vector $U$ and the control polygon 
$\Pi$ as follows:

\begin{tabular}{l}
$\displaystyle{U=\left( 0,0,0,0,1,2,3,4,5,5,5,5 \right)}$\\
 $
\displaystyle{
\Pi=
\left\lbrace 
\couple{0}{2}, \couple{1.5}{5}, \couple{2.5}{4}, \couple{ 3}{1},
\couple{ 5}{4}, \couple{ 7}{1}, \couple{ 8}{4}, \couple{10}{4}
\right\rbrace
}
$\\
\end{tabular}

We will go through
$\displaystyle{\alpha \in \lbrace -\infty,-4,-1/2,-1/5,-1/7 \rbrace}$, 
as well as its conjugated $1-\alpha$.

\begin{figure}[h!]
\begin{center}
\includegraphics[width=11cm]{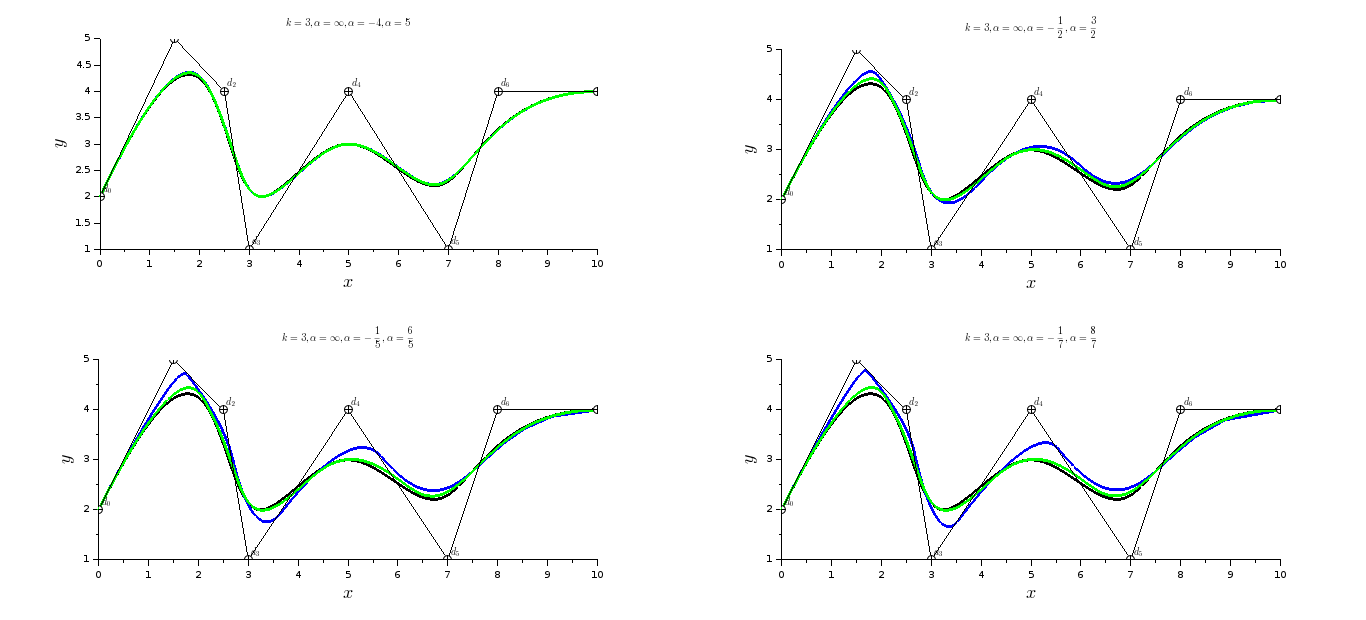}
\caption{Influence of $\alpha$  to $k=3$, $U$ uniform and open with fixed $\Pi$ }
\label{figEffetAlfa}
\end{center}
\end{figure}

A quick analysis of figure \ref{figEffetAlfa} reveals:
\begin{enumerate}
\item
For
$\displaystyle{\alpha \leq -4}$ and $\displaystyle{\alpha \geq 5}$, the
B-spline curve $\displaystyle{G_\alpha}$ of degree $k$ and index $\alpha$ is a good
 approximationof the standard polynomial B-spline curve $\displaystyle{G_\infty}$ 
generated by the same control polygon $\Pi$.
\item
When $\alpha$ tends to $0^{-}$ or to $1^{+}$,
the curve $\displaystyle{G_\alpha}$ is really separated from
the standard curve $\displaystyle{G_\infty}$. The effect seems more viewed
 at the neighborhood of
$0$ but the question is still to be tackled later on.
\item
We reach a conclusion that the B-spline curves family becomes more interesting.

\end{enumerate}

\end{castest}

\begin{castest}
\label{CastestDegre}
The second important parameter is the degree  $k$  of the basis which generates the
 B-spline curve.
We will observe its influence on two examples discribed by the following data where 
the control polygon $\Pi_i$ has been fixed with a uniform and open knot vector $U_{i,k}$
giving the degree $k$ as follows:
 
\begin{enumerate}
\item 
\textbf{Example 1} 

\begin{tabular}{l}
$\displaystyle{
\Pi_1=
\left\lbrace 
\couple{0}{0}, \couple{3}{9}, \couple{6}{3}, \couple{9}{6}
\right\rbrace
}$\\
$\displaystyle{U_{1,1}=\left( 0,0,1,2,3,3\right)}$\\
$\displaystyle{U_{1,2}=\left( 0,0,0,1.5,3,3,3\right)}$\\
$\displaystyle{U_{1,3}=\left( 0,0,0,0,3,3,3,3 \right)}$\\
\end{tabular}

\item
\textbf{Example 2} 

\begin{tabular}{l}
$
\displaystyle{
\Pi_2=
\left\lbrace 
\couple{1}{3}, \couple{0}{5}, \couple{5}{5}, 
\couple{3}{0}, \couple{8}{0}, \couple{7}{3}
\right\rbrace
}$\\
$\displaystyle{U_{2,1}=\left( 0,0,1,2,3,4,5,5\right)}$\\
$\displaystyle{U_{2,2}=\left( 0,0,0,5/4,5/2,15/4,5,5,5\right)}$\\
$\displaystyle{U_{2,3}=\left( 0,0,0,0,5/3,10/3,5,5,5,5\right)}$\\
$\displaystyle{U_{2,4}=\left( 0,0,0,0,0,5/2,5,5,5,5,5\right)}$\\
$\displaystyle{U_{2,5}=\left( 0,0,0,0,0,0,5,5,5,5,5,5\right)}$\\
\end{tabular}

\end{enumerate}

\begin{figure}[h!]
\begin{center}
\includegraphics[width=11cm]{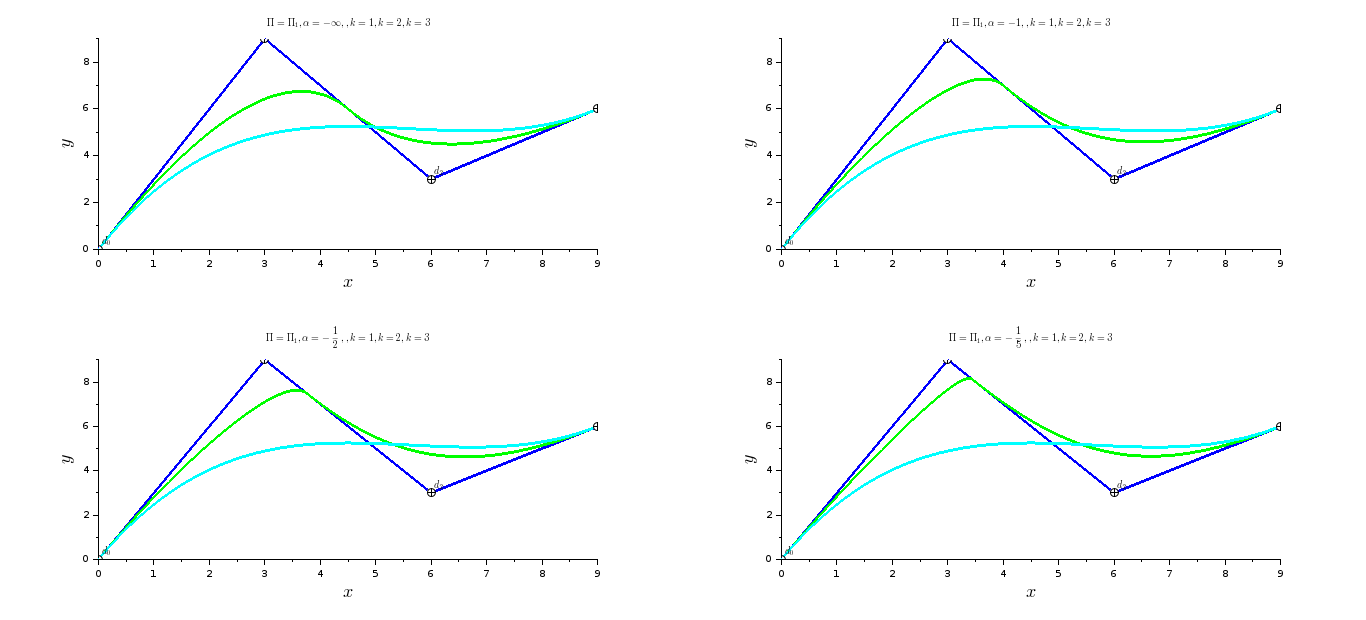}
\caption{Influence of degree $k$, $U$ uniform and open at $\alpha$ with fixed $\Pi$ }
\label{figEffetDegre1}
\end{center}
\end{figure}

\begin{figure}[h!]
\begin{center}
\includegraphics[width=11cm]{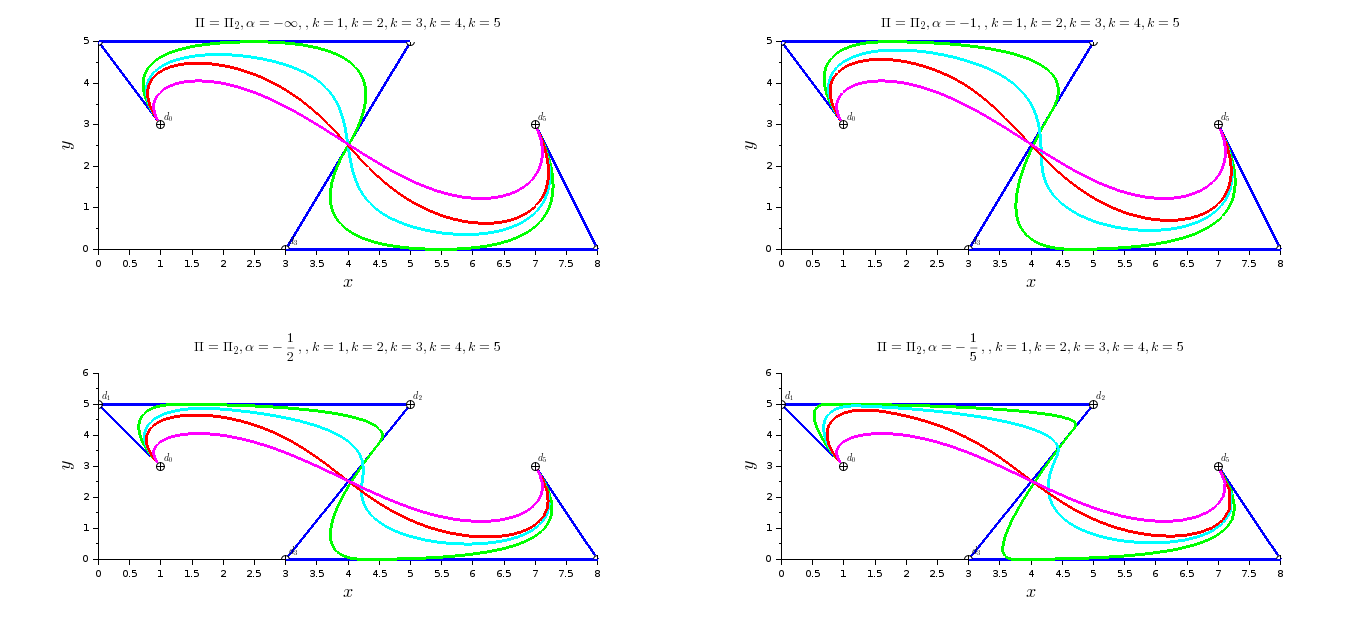}
\caption{Influence of degree $k$, $U$ uniform and open at $\alpha$ with fixed $\Pi$ }
\label{figEffetDegre2}
\end{center}
\end{figure}

The figure \ref{figEffetDegre1} summarizes example 1 and show on one hand that 
independently from $\alpha$, the degree $k=1$ yields the control polygon $\Pi$.
On the other hand, $k=3$ corresponds to a knot vector without any interior node and the 
obtained B-spline curve $\displaystyle{G_\alpha}$ is independent from $\alpha$. Only 
the degree $k=2$ between the extremes undergo the influence of index $\alpha$ with some
 highlight when $\alpha$ tends to $0$.

 The results of example 2 shown in figure \ref{figEffetDegre2} confirm 
above observations.

 The degree $k=1$ yields the control polygon $\Pi_2$ and the degree
 $k=5$ which corresponds to a knot vector with no interior node 
 does not have any influence under $\alpha$. For the intermediate degrees $k$ the index
 $\alpha$ has an incresing influence when $\alpha$ tends to $0$. 
 
\end{castest}

\begin{castest}
\label{CastestControlLocal}
Now we intend to look at the influence of control polygon $\Pi$ on the local behavior
of a B-spline curve. We fix the degree to $3$ on the uniform and open knot vector $U$
by varing only one point of the control polygon as follows:
 
\begin{tabular}{l}
$\displaystyle{U=\left( 0,0,0,0,1,2,3,4,4,4,4 \right)}$\\
$
\displaystyle{
\Pi_1=
\left\lbrace 
\couple{0}{4}, \couple{5}{4}, \couple{5}{8}, \couple{11}{7.5},
\couple{6}{2}, \couple{12}{0}, \couple{2}{0}
\right\rbrace
}
$\\
$
\displaystyle{
\Pi_2=
\left\lbrace 
\couple{0}{4}, \couple{5}{4}, \couple{5}{8}, \couple{11}{7.5},
\couple{9}{3}, \couple{12}{0}, \couple{2}{0}
\right\rbrace
}
$\\
$
\displaystyle{
\Pi_3=
\left\lbrace 
\couple{0}{4}, \couple{5}{4}, \couple{5}{8}, \couple{11}{7.5},
\couple{12}{4}, \couple{12}{0}, \couple{2}{0}
\right\rbrace
}
$\\
\end{tabular}

We take
$\displaystyle{\alpha \in \lbrace -\infty,-4,-1/2,-1/5,-1/7 \rbrace}$, 
as well as its conjugated $1-\alpha$.

\begin{figure}[h!]
\begin{center}
\includegraphics[width=11cm]{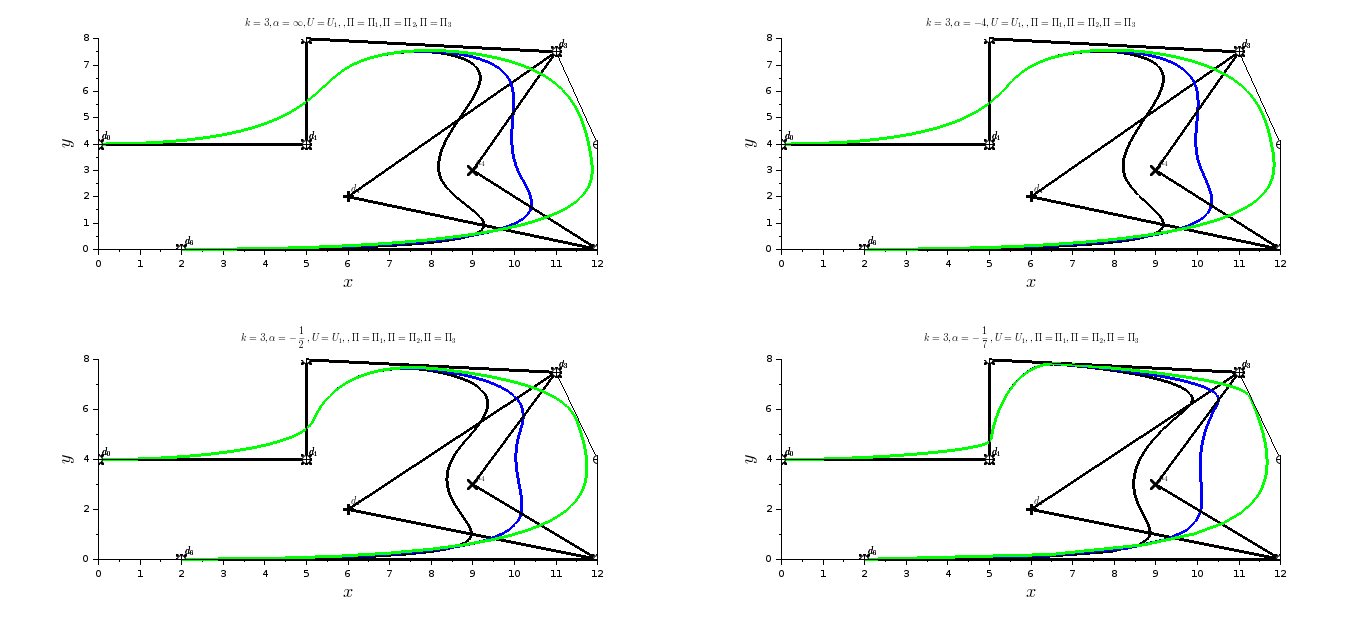}
\caption{Influence of the variation of a point of $\Pi$  at $k=3$, $U$ uniform and open and
 $\alpha \in \lbrace -\infty,-4,-1/2,-1/5,-1/7 \rbrace$  }
\label{figControleLocal1}
\end{center}
\end{figure}

\begin{figure}[h!]
\begin{center}
\includegraphics[width=11cm]{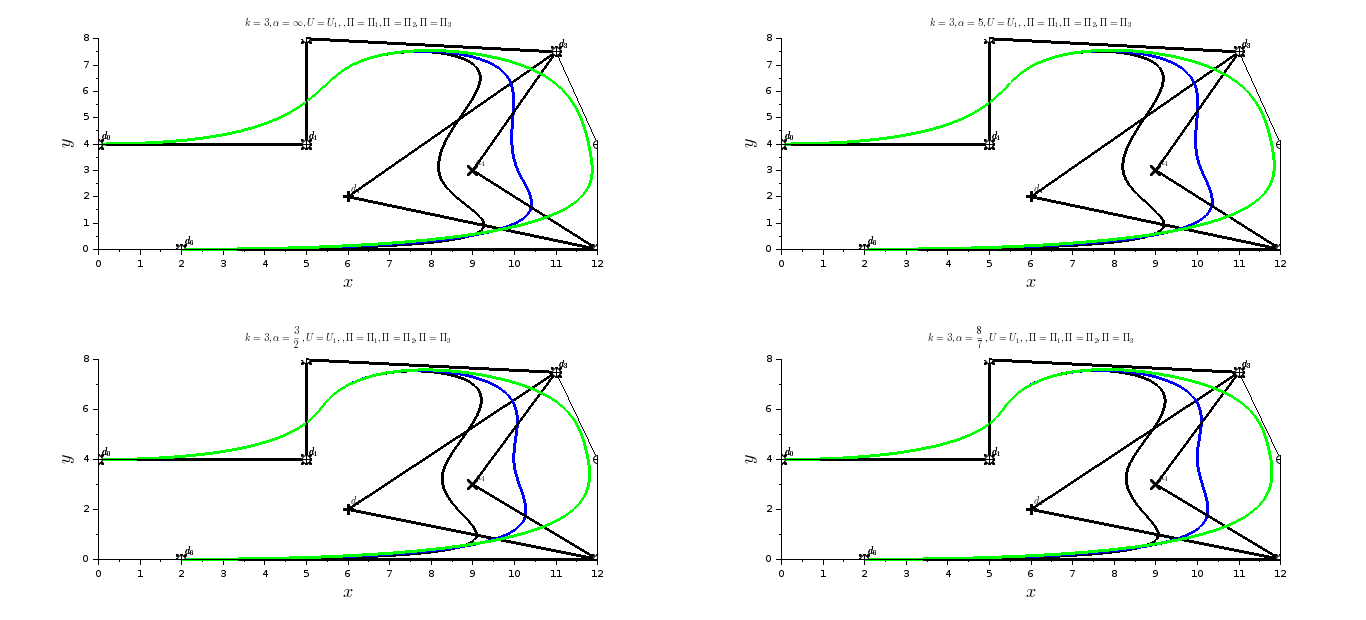}
\caption{Influence of the variation of a point of $\Pi$  at $k=3$, $U$ uniform and open and
 $\alpha \in \lbrace \infty,5,3/2,6/5,8/7 \rbrace$  }
\label{figControleLocal2}
\end{center}
\end{figure}

Figures \ref{figControleLocal1} and \ref{figControleLocal2} let us to state that
each curve $\displaystyle{G_\alpha}$ is made up of three
segments where the second one is under the motion of the fifth endpoint of 
the control polygon  $\Pi$. As we have noted so far, the influence of $\alpha$
is not so remarkable for $\alpha \leq -4$ and $\alpha \geq 5$ as one can note in 
polynomial case that is to say $\displaystyle{G_\alpha \approx G_\infty}$.

In the deformation region of the curve
$\displaystyle{G_\alpha}$ at the neighborhood of a segment
$\displaystyle{\segment{d_{i}}{d_{i+1}}}$ 
of control polygon $\Pi_j$, the deformation moves towards
the point $\displaystyle{d_{i}}$ when $\displaystyle{\alpha\in(-1\,,\,0)}$
and towards the point $\displaystyle{d_{i+1}}$ when $\displaystyle{\alpha\in(1\,,\,2)}$
as shown in figures 
 \ref{figControleLocal1} and \ref{figControleLocal2} respectively.
In all cases, the curve $\displaystyle{G_\alpha}$ belongs to the convex envelop of the 
control polygon $\Pi_j$.

\end{castest}

\begin{remarque}
Through the figure \ref{figEffetAlfa} of illustration \ref{CastestAlpha} and
figures \ref{figEffetDegre1} and \ref{figEffetDegre2} 
of illustration \ref{CastestDegre}
as well as
 figures \ref{figControleLocal1} and \ref{figControleLocal2} 
of illustration \ref{CastestControlLocal}, 
 we realize that the property of convex envelop is widely verified.
\end{remarque}

\begin{castest}
\label{CastestSymetrie}
In this test case, we will explore the property of symmetry proved in proposition \ref{PropoSymetrie} 
through seven contexts where we restrict ourselves to an axis of symmetry parallel to the 
coordinate axes which does not reduce generality. The data are as follow:

\begin{enumerate}

\item  Axial symmetry of $\Pi$ with axis parallel to Oy with no multiple point 

\begin{tabular}{l}
$
\displaystyle{
\Pi_1=
\left\lbrace 
\begin{array}{l}
\couple{4}{0}, \couple{0}{11}, \couple{6}{14}, \\
\couple{10}{14}, \couple{16}{11}, \couple{12}{0}\\
\end{array}
\right\rbrace
}
$\\
$\displaystyle{U_1=\left(0,0,0,0,1,2,3,3,3,3 \right)}$\\
\end{tabular}

\item  Axial symmetry of $\Pi$ with axis parallel to Oy with one double point

\begin{tabular}{l}
$
\displaystyle{
\Pi_2=
\left\lbrace 
\begin{array}{l}
\couple{4}{0}, \couple{0}{11}, \couple{8}{14},  \\
\couple{8}{14}, \couple{16}{11}, \couple{12}{0} \\
\end{array}
\right\rbrace
}
$\\
$\displaystyle{U_2=\left(0,0,0,0,1,2,3,3,3,3 \right)}$\\
\end{tabular}

\item  Axial symmetry of $\Pi$ with axis parallel to Oy with double point and double node

\begin{tabular}{l}
$
\displaystyle{
\Pi_3=
\left\lbrace 
\begin{array}{l}
\couple{4}{0}, \couple{0}{11}, \couple{8}{14},  \\
\couple{8}{14}, \couple{16}{11}, \couple{12}{0} \\
\end{array}
\right\rbrace
}
$\\
$\displaystyle{U_3=\left(0,0,0,0,2,2,4,4,4,4 \right)}$\\
\end{tabular}

\item  Axial symmetry of $\Pi$ with axis parallel to Ox with no multiple point

\begin{tabular}{l}
$
\displaystyle{
\Pi_4=
\left\lbrace 
\begin{array}{l}
\couple{0}{5}, \couple{0}{4}, \couple{1}{4},  \\
\couple{2}{4}, \couple{2}{6}, \couple{4}{6},\couple{5}{5}, \\
\couple{5}{1}, \couple{4}{0}, \couple{2}{0},  \\
\couple{2}{2}, \couple{1}{2},\couple{0}{2}, \couple{0}{1} \\
\end{array}
\right\rbrace
}
$\\
$\displaystyle{U_4=\left(0,0,0,0,1,2,3,4,5,6,7,8,9,10,11,11,11,11 \right)}$\\
\end{tabular}

\item  Axial symmetry of $\Pi$ with axis parallel to Ox with double point

\begin{tabular}{l}
$
\displaystyle{
\Pi_5=
\left\lbrace 
\begin{array}{l}
\couple{0}{5}, \couple{0}{4}, \couple{1}{4},  \\
\couple{2}{4}, \couple{2}{6}, \couple{4}{6},\couple{5}{3}, \\
\couple{5}{3}, \couple{4}{0}, \couple{2}{0},  \\
\couple{2}{2}, \couple{1}{2},\couple{0}{2}, \couple{0}{1} \\
\end{array}
\right\rbrace
}
$\\
$\displaystyle{U_5=\left(0,0,0,0,1,2,3,4,5,6,7,8,9,10,11,11,11,11 \right)}$\\
\end{tabular}

\item  Axial symmetry of $\Pi$ with axis parallel to Ox with double point and double node 

\begin{tabular}{l}
$
\displaystyle{
\Pi_6=
\left\lbrace 
\begin{array}{l}
\couple{0}{5}, \couple{0}{4}, \couple{1}{4},  \\
\couple{2}{4}, \couple{2}{6}, \couple{4}{6},\couple{5}{3}, \\
\couple{5}{3}, \couple{4}{0}, \couple{2}{0},  \\
\couple{2}{2}, \couple{1}{2},\couple{0}{2}, \couple{0}{1} \\
\end{array}
\right\rbrace
}
$\\
$\displaystyle{U_6=\left(0,0,0,0,1,2,3,4,5,5,6,7,8,9,10,10,10,10 \right)}$\\
\end{tabular}

\item Double axial  symmetry  of $\Pi$ with one double point

\begin{tabular}{l}
$
\displaystyle{
\Pi_7=
\left\lbrace 
\begin{array}{l}
\couple{0}{2}, \couple{0}{3}, \couple{1}{4}, \\
\couple{3}{4}, \couple{5}{4}, \couple{6}{3}, \\
\couple{6}{2}, \couple{6}{1}, \couple{5}{0},  \\
\couple{3}{0}, \couple{1}{0}, \couple{0}{1},\couple{0}{2} \\
\end{array}
\right\rbrace
}
$\\
$\displaystyle{U_7=\left(0,0,0,0,1,2,3,4,5,6,7,8,9,10,10,10,10 \right)}$\\
\end{tabular}
\end{enumerate}

\begin{figure}[h!]
\begin{center}
\includegraphics[width=11cm]{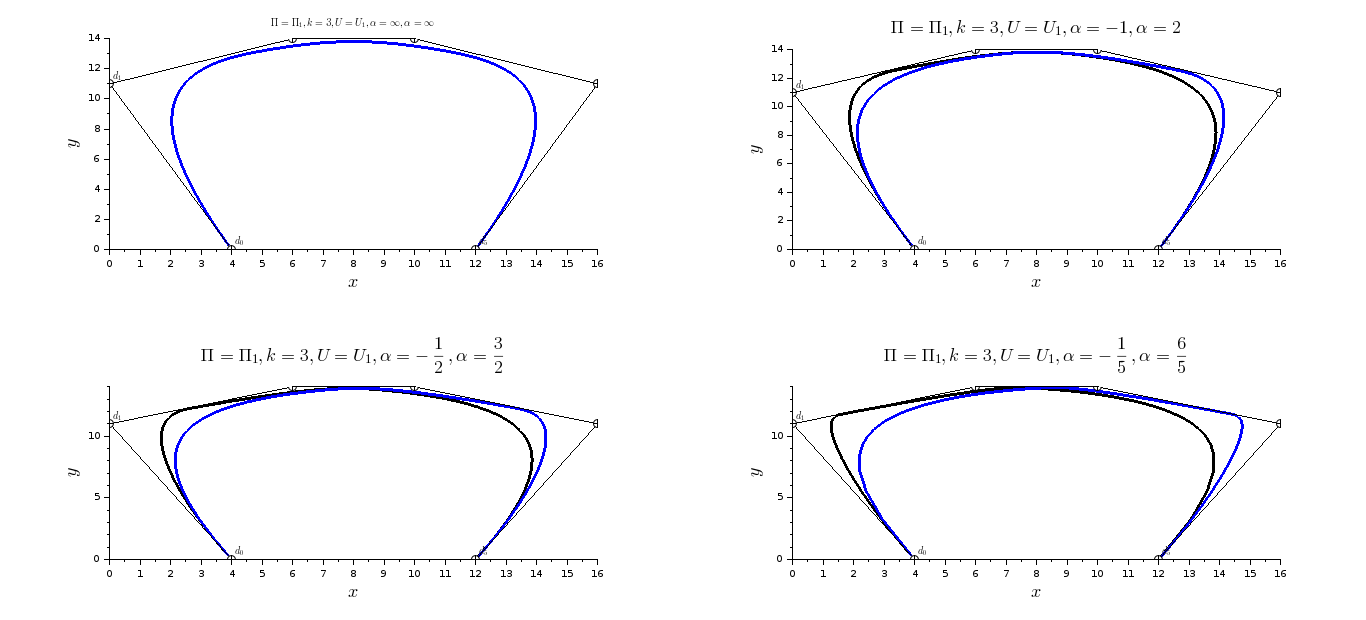}
\caption{$\displaystyle{G_\alpha}$ curves of degree $k=3$,
$U_1$ uniform and open, $\Pi_1$ symmetric with no multiple point and  
$\alpha \in \lbrace \infty,-1,-1/2,-1/5 \rbrace$    }
\label{figSymetrie1}
\end{center}
\end{figure}

\begin{figure}[h!]
\begin{center}
\includegraphics[width=11cm]{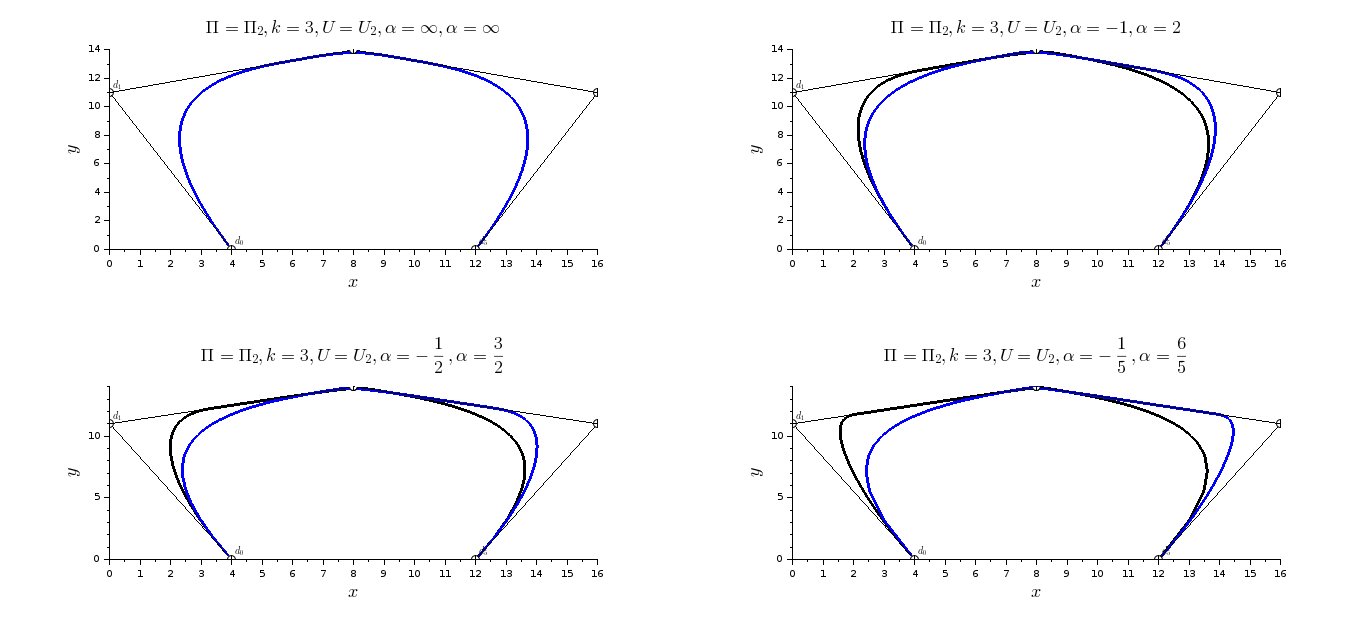}
\caption{$\displaystyle{G_\alpha}$ curves of degree $k=3$,
$U_2$ uniform and open, $\Pi_2$ symmetric with double point and 
$\alpha \in \lbrace \infty,-1,-1/2,-1/5 \rbrace$    }
\label{figSymetrie2}
\end{center}
\end{figure}

\begin{figure}[h!]
\begin{center}
\includegraphics[width=11cm]{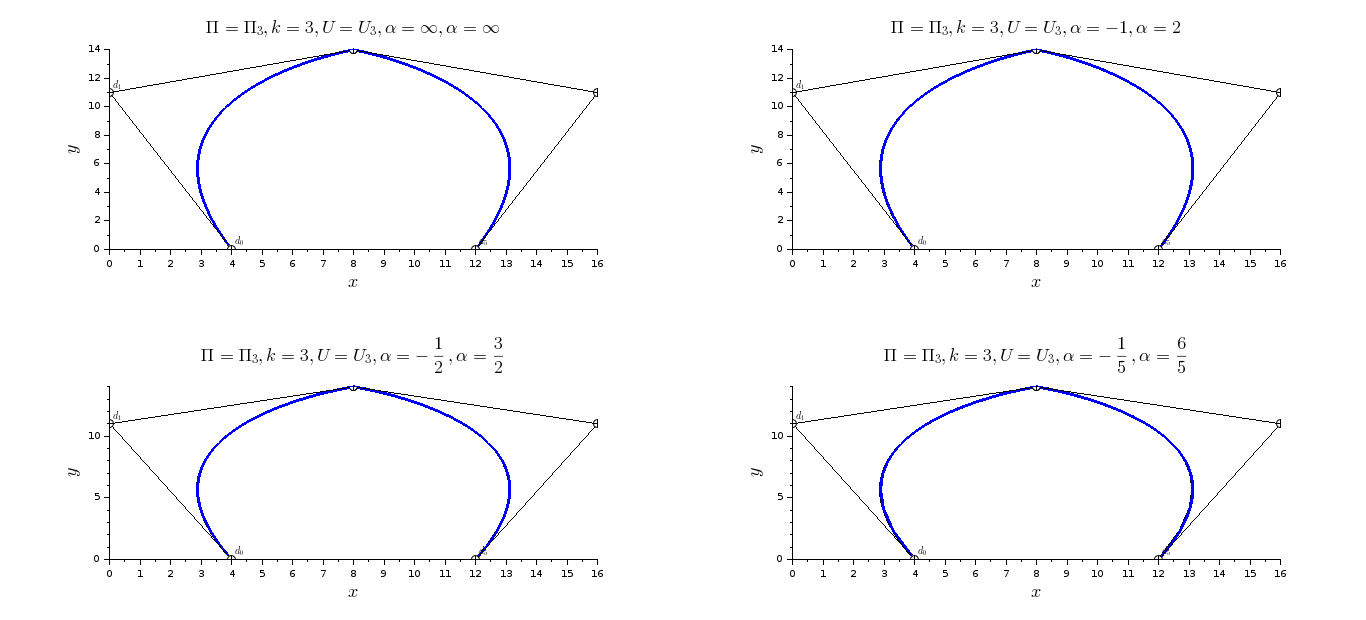}
\caption{$\displaystyle{G_\alpha}$ curves of degree $k=3$,
$U_3$ symmetric and open with double node, $\Pi_3$ symmetric with double point and   
$\alpha \in \lbrace \infty,-1,-1/2,-1/5 \rbrace$    }
\label{figSymetrie3}
\end{center}
\end{figure}

\begin{figure}[h!]
\begin{center}
\includegraphics[width=11cm]{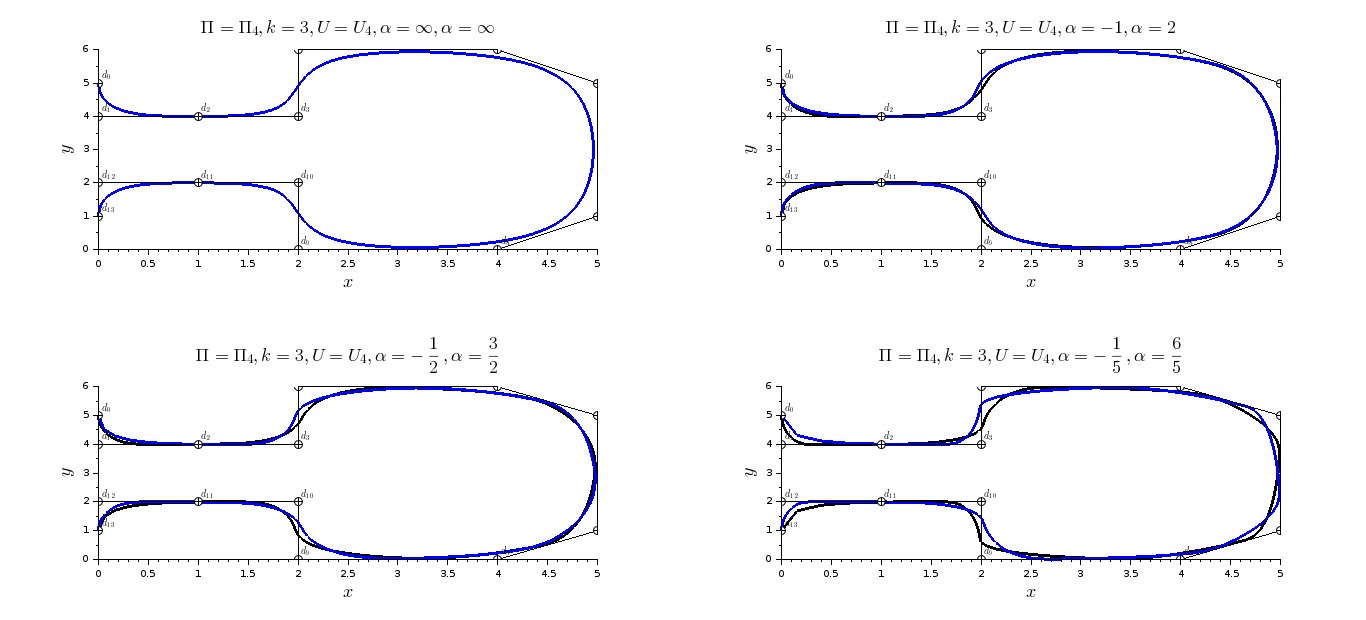}
\caption{ $\displaystyle{G_\alpha}$ curves of degree $k=3$,
$U_4$ uniform and open, $\Pi_4$ symmetric with no multiple point and   
$\alpha \in \lbrace \infty,-1,-1/2,-1/5 \rbrace$    }
\label{figSymetrie4}
\end{center}
\end{figure}

\begin{figure}[h!]
\begin{center}
\includegraphics[width=11cm]{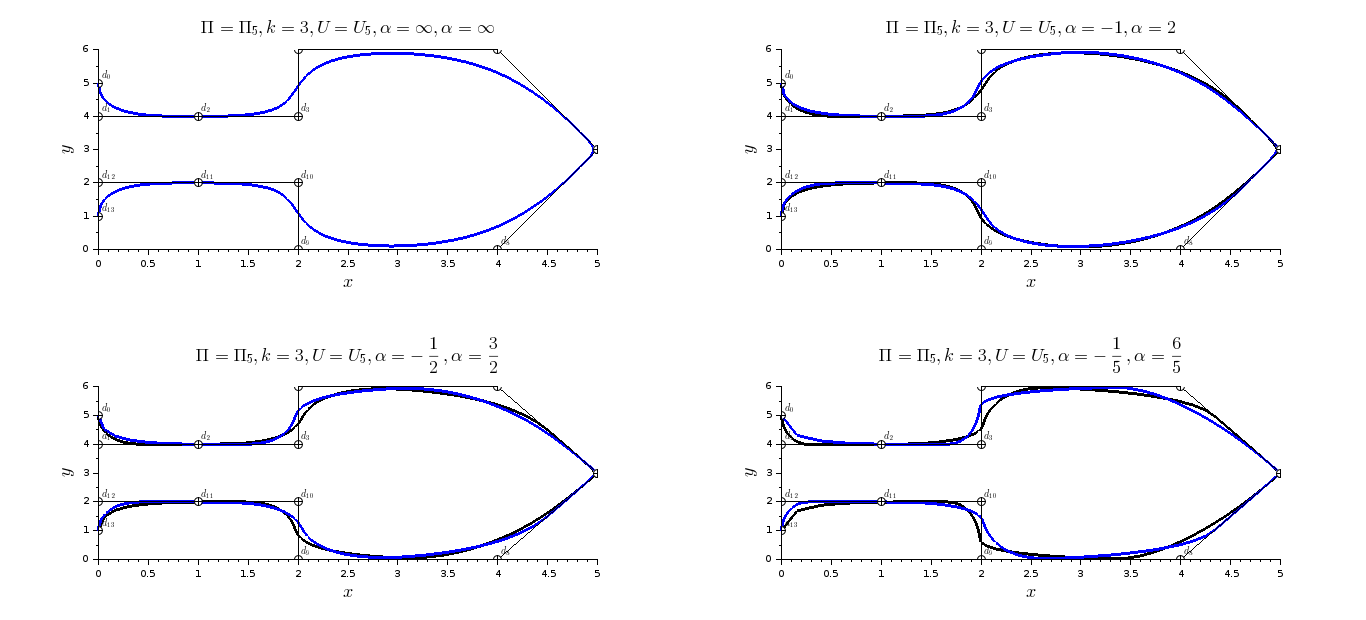}
\caption{ $\displaystyle{G_\alpha}$ curves of degree $k=3$,
$U_5$ uniform and open, $\Pi_5$ symmetric with double point and   
$\alpha \in \lbrace \infty,-1,-1/2,-1/5 \rbrace$    }
\label{figSymetrie5}
\end{center}
\end{figure}

\begin{figure}[h!]
\begin{center}
\includegraphics[width=11cm]{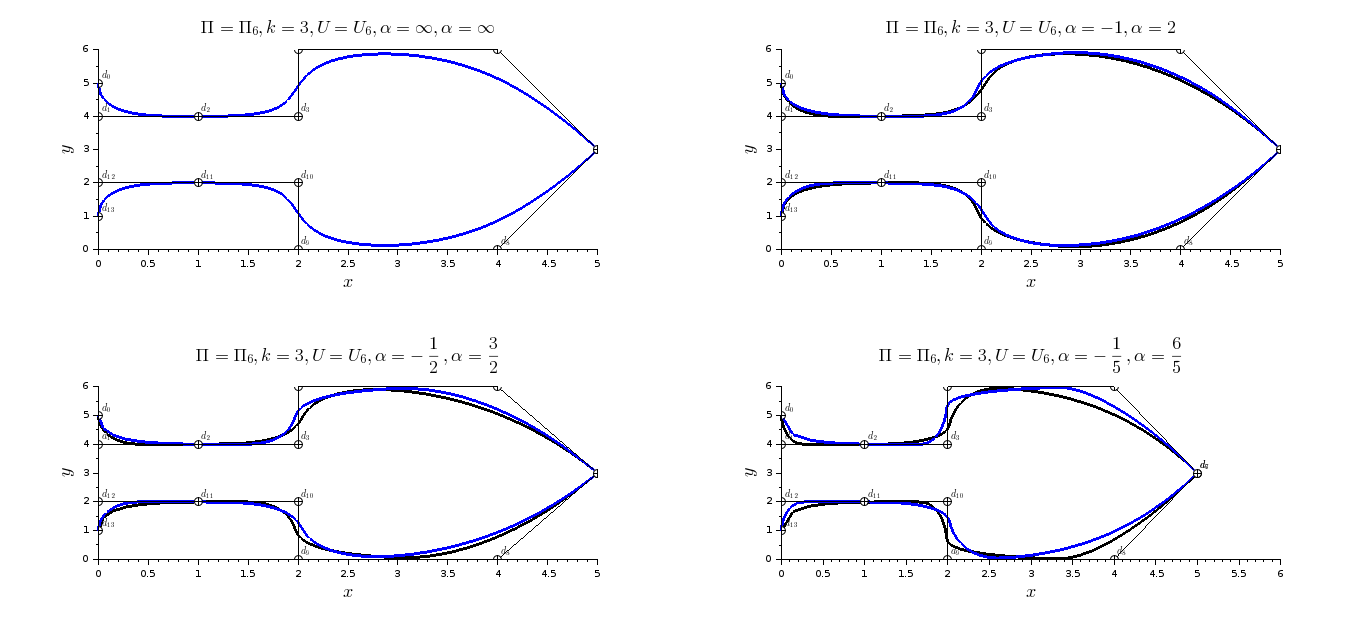}
\caption{ $\displaystyle{G_\alpha}$ curves of degree $k=3$,
$U_6$ symmetric and open with double node, $\Pi_6$ symmetric with double point and   
$\alpha \in \lbrace \infty,-1,-1/2,-1/5 \rbrace$    }
\label{figSymetrie6}
\end{center}
\end{figure}

\begin{figure}[h!]
\begin{center}
\includegraphics[width=11cm]{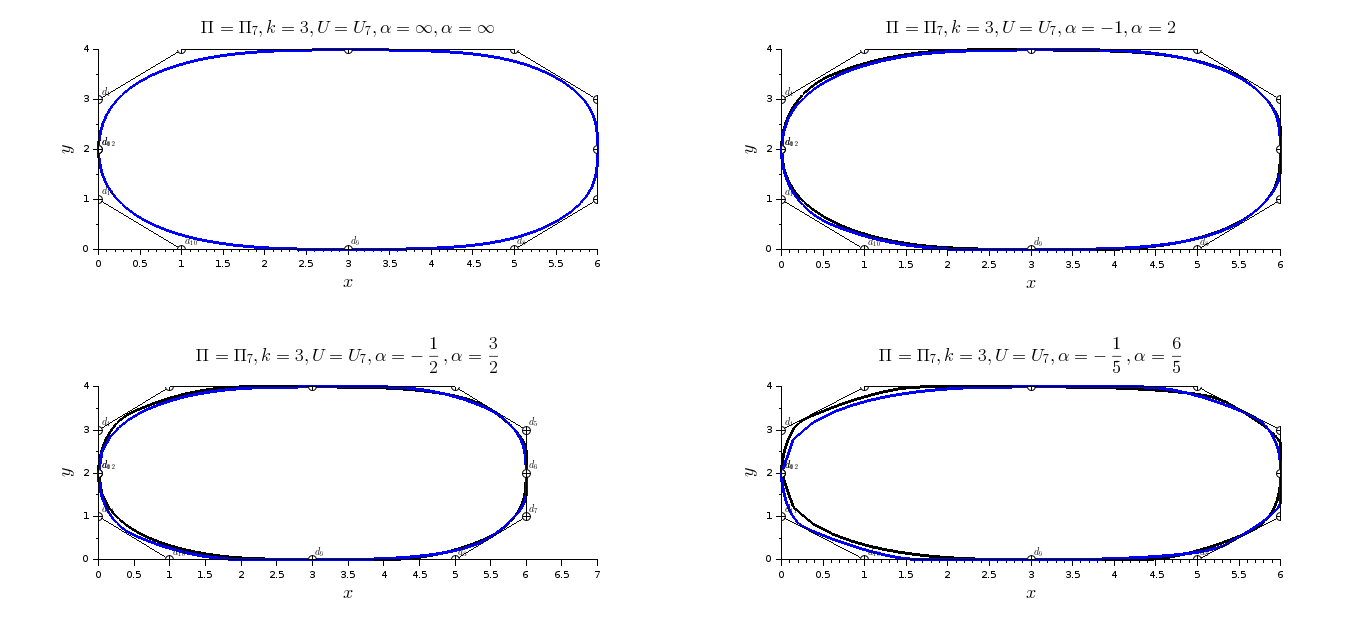}
\caption{ $\displaystyle{G_\alpha}$ curves of degree $k=3$,
$U_7$ uniform and open, $\Pi_7$ symmetric with double point and   
$\alpha \in \lbrace \infty,-1,-1/2,-1/5 \rbrace$    }
\label{figSymetrie7}
\end{center}
\end{figure}

Based on figures from \ref{figSymetrie1} to  \ref{figSymetrie7}, it can be drawn that
the curves  $\displaystyle{G_\alpha}$  and 
$\displaystyle{G_{1-\alpha}}$ are symmetric with respect to the perpendicular bisector 
of extreme points of the control polygon $\Pi$. As stated above,
the effect àf index $\alpha$ is very remarkable for
$\displaystyle{\alpha \in (-1\,,\,0) \cup  (1\,,\,2) }$.

The multiplicity of a node acts on the geometrical regularity of curves
$\displaystyle{G_\alpha}$  and $\displaystyle{G_{1-\alpha}}$. In the presence of a double control point,
the curves $\displaystyle{G_\alpha}$  and
$\displaystyle{G_{1-\alpha}}$ adhere to this point.

The figure \ref{figSymetrie3} shows however a singular case which we will light upon later on since 
$\alpha$ seems  to have no influence on it. 

\end{castest}

\section{Conclusion}
The class of parametrization we developed allows us to construct a family of 
rational B-spline basis depending on a parameter $\alpha$ which generalizes all
including polynomial B-spline basis.
This new family of B-spline basis possesses all the classical fundamental properties
such as positivity, unit partition property and linear independence. Some symmetry property 
has been established.

We have proved that the family of B-spline curves we obtained is larger than the 
polynomial B-spline curves one and globally extend their properties.
Illustrations are given to explain more the properties we proved with the desire of 
the extension to practical computation algorithms of curves (deBoor algorithm) in future work.

It is left with the exploration in more details of the effect of this new parametrization on 
Bernstein functions and the resulting B\'ezier curves.



\newpage
\section*{Appendix}
This section shows results  extracted from  outputs processed by a  Maxima code.

\newpage
\begin{center}
B-spline basis of degree 2 and its derivative
for the knot vector\\
$U=\left( x_{0},x_{0}, x_{0}, x_{1}, x_{2}, x_{2}, x_{2} \right)$
\end{center}

$$
\Bspline{i}{2}{\alpha} (t)=
\left\lbrace
\begin{array}{ll}
{{\left(\alpha-1\right)^2\,\left({\it x_1}-t\right)^
 2}\over{\left(\alpha\,{\it x_1}-{\it x_1}-\alpha\,{\it x_0}+t\right)
 ^2}}
&\textrm{if } i=0 \textrm{ and } t\in [x_{0}\, , \, x_{1}[\\ 
 -\frac{\left(\alpha-1\right)\,\alpha\,\left({\it x_0}-t
 \right)}{{\left(\alpha\,
 {\it x_1}-{\it x_1}-\alpha\,{\it x_0}+t\right)^2\,\left(\alpha\,
 {\it x_2}-{\it x_2}-\alpha\,{\it x_0}+t\right)}}\times
 &\\
 \left[2\,\alpha\,{\it x_1}\,{\it x_2}-2\,{\it x_1}\,
 {\it x_2}-\alpha\,{\it x_0}\,{\it x_2}
 \right.
 &\\
 \left.
 -\alpha\,t\,{\it x_2}+2\,t\,
 {\it x_2}-\alpha\,{\it x_0}\,{\it x_1}-\alpha\,t\,{\it x_1}
 \right.
 &\\
 \left.
 +2\,t\, {\it x_1}+2\,\alpha\,t\,{\it x_0}-2\,t^2\right]
&\textrm{if } i=1 \textrm{ and } t\in [x_{0}\, , \, x_{1}[\\ 
 {{\left(\alpha-1
 \right)^2\,\left({\it x_2}-t\right)^2}\over{\left(\alpha\,{\it x_2}-
 {\it x_2}-\alpha\,{\it x_0}+t\right)\,\left(\alpha\,{\it x_2}-
 {\it x_2}-\alpha\,{\it x_1}+t\right)}}
&\textrm{if } i=1 \textrm{ and } t\in [x_{1}\, , \, x_{2}[\\ 
 {{\alpha^2\,\left({\it x_0}
 -t\right)^2}\over{\left(\alpha\,{\it x_1}-{\it x_1}-\alpha\,
 {\it x_0}+t\right)\,\left(\alpha\,{\it x_2}-{\it x_2}-\alpha\,
 {\it x_0}+t\right)}}
&\textrm{if } i=2 \textrm{ and } t\in [x_{0}\, , \, x_{1}[\\ 
 -\frac{\left(\alpha-1\right)\,\alpha\,\left(
 {\it x_2}-t\right)}{{\left(\alpha\,{\it x_2}-{\it x_2}-\alpha\,
 {\it x_0}+t\right)\,\left(\alpha\,{\it x_2}-{\it x_2}-\alpha\,
 {\it x_1}+t\right)^2}}\times
 &\\
 \left(\alpha\,{\it x_1}\,{\it x_2}-{\it x_1}\,
 {\it x_2}+\alpha\,{\it x_0}\,{\it x_2}-{\it x_0}\,{\it x_2}
 \right.
 &\\
 \left.
 -2\,
 \alpha\,t\,{\it x_2}+2\,t\,{\it x_2}-2\,\alpha\,{\it x_0}\,{\it x_1}
 +\alpha\,t\,{\it x_1}+t\,{\it x_1}
  \right.
 &\\
 \left.
+\alpha\,t\,{\it x_0}+t\,{\it x_0}
 -2\,t^2\right)
&\textrm{if } i=2 \textrm{ and } t\in [x_{1}\, , \, x_{2}[\\ 
 {{\alpha^2\,\left({\it x_1}-t\right)^2
 }\over{\left(\alpha\,{\it x_2}-{\it x_2}-\alpha\,{\it x_1}+t\right)^
 2}}
&\textrm{if } i=3 \textrm{ and } t\in [x_{1}\, , \, x_{2}[\\ 
 0
&\textrm{otherwise } \\ 
\end{array}
\right.
 $$


 $$
\frac{d}{dt}\Bspline{i}{2}{\alpha} (t)=
\left\lbrace
\begin{array}{ll}
\displaystyle{
 -{{2\,\left(\alpha-1\right)^2\,\alpha\,\left(
 {\it x_1}-t\right)\,\left({\it x_1}-{\it x_0}\right)}\over{\left(
 \alpha\,{\it x_1}-{\it x_1}-\alpha\,{\it x_0}+t\right)^3}}
 }
&\textrm{if } i=0 \textrm{ and } t\in [x_{0}\, , \, x_{1}[\\ 
\displaystyle{
 \frac{\left(\alpha-1\right)\,\alpha}{{\left(
 \alpha\,{\it x_1}-{\it x_1}-\alpha\,{\it x_0}+t\right)^3\,\left(
 \alpha\,{\it x_2}-{\it x_2}-\alpha\,{\it x_0}+t\right)^2}}\times}%
 &\\
 \left[
 2\,\alpha^3\,{\it x_1}^2\,
 {\it x_2}^2-6\,\alpha^2\,{\it x_1}^2\,{\it x_2}^2+6\,\alpha\,
 {\it x_1}^2\,{\it x_2}^2
 \right.
 &\\
 \left.
 -2\,{\it x_1}^2\,{\it x_2}^2-2\,\alpha^3\,
 {\it x_0}\,{\it x_1}\,{\it x_2}^2+6\,\alpha^2\,{\it x_0}\,{\it x_1}
 \,{\it x_2}^2
 \right.
 &\\
 \left.
 -6\,\alpha\,{\it x_0}\,{\it x_1}\,{\it x_2}^2+2\,
 {\it x_0}\,{\it x_1}\,{\it x_2}^2-2\,\alpha^3\,t\,{\it x_1}\,
 {\it x_2}^2
 \right.
 &\\
 \left.
 +6\,\alpha^2\,t\,{\it x_1}\,{\it x_2}^2-6\,\alpha\,t\,
 {\it x_1}\,{\it x_2}^2+2\,t\,{\it x_1}\,{\it x_2}^2
 \right.
 &\\
 \left.
 +2\,\alpha^3\,t\,
 {\it x_0}\,{\it x_2}^2-6\,\alpha^2\,t\,{\it x_0}\,{\it x_2}^2+6\,
 \alpha\,t\,{\it x_0}\,{\it x_2}^2
 \right.
 &\\
 \left.
 -2\,t\,{\it x_0}\,{\it x_2}^2
 -2\, \alpha^3\,{\it x_0}\,{\it x_1}^2\,{\it x_2}+4\,\alpha^2\,{\it x_0}\,
 {\it x_1}^2\,{\it x_2}
 \right.
 &\\
 \left.
 -2\,\alpha\,{\it x_0}\,{\it x_1}^2\,{\it x_2}
 - 2\,\alpha^3\,t\,{\it x_1}^2\,{\it x_2}+8\,\alpha^2\,t\,{\it x_1}^2\,
 {\it x_2}
  \right.
 &\\
 \left.
 -10\,\alpha\,t\,{\it x_1}^2\,{\it x_2}+4\,t\,{\it x_1}^2\,
 {\it x_2}-3\,\alpha^2\,{\it x_0}^2\,{\it x_1}\,{\it x_2}
  \right.
 &\\
 \left.
+3\,\alpha\,
 {\it x_0}^2\,{\it x_1}\,{\it x_2}+8\,\alpha^3\,t\,{\it x_0}\,
 {\it x_1}\,{\it x_2}-14\,\alpha^2\,t\,{\it x_0}\,{\it x_1}\,
 {\it x_2}
  \right.
 &\\
 \left.
+10\,\alpha\,t\,{\it x_0}\,{\it x_1}\,{\it x_2}-4\,t\,
 {\it x_0}\,{\it x_1}\,{\it x_2}-7\,\alpha^2\,t^2\,{\it x_1}\,
 {\it x_2}
  \right.
 &\\
 \left.
+11\,\alpha\,t^2\,{\it x_1}\,{\it x_2}-4\,t^2\,{\it x_1}\,
 {\it x_2}+2\,\alpha^3\,{\it x_0}^3\,{\it x_2}-\alpha^2\,{\it x_0}^3
 \,{\it x_2}
  \right.
 &\\
 \left.
-6\,\alpha^3\,t\,{\it x_0}^2\,{\it x_2}+6\,\alpha^2\,t\,
 {\it x_0}^2\,{\it x_2}
-3\,\alpha\,t\,{\it x_0}^2\,{\it x_2}
   \right.
 &\\
 \left.
+7\, \alpha^2\,t^2\,{\it x_0}\,{\it x_2}-8\,\alpha\,t^2\,{\it x_0}\,
 {\it x_2}+4\,t^2\,{\it x_0}\,{\it x_2}
   \right.
 &\\
 \left.
 -\alpha\,t^3\,{\it x_2}+\alpha
 ^2\,{\it x_0}^2\,{\it x_1}^2-\alpha\,{\it x_0}^2\,{\it x_1}^2
 +2\, \alpha^3\,t\,{\it x_0}\,{\it x_1}^2
   \right.
 &\\
 \left.
-6\,\alpha^2\,t\,{\it x_0}\,
 {\it x_1}^2+4\,\alpha\,t\,{\it x_0}\,{\it x_1}^2-\alpha^2\,t^2\,
 {\it x_1}^2
 \right.
 &\\
 \left.
+3\,\alpha\,t^2\,{\it x_1}^2-2\,t^2\,{\it x_1}^2+2\,
 \alpha^3\,{\it x_0}^3\,{\it x_1}-\alpha^2\,{\it x_0}^3\,{\it x_1}
   \right.
 &\\
 \left.
 -6 \,\alpha^3\,t\,{\it x_0}^2\,{\it x_1}+4\,\alpha^2\,t\,{\it x_0}^2\,
 {\it x_1}-\alpha\,t\,{\it x_0}^2\,{\it x_1}
   \right.
 &\\
 \left.
+9\,\alpha^2\,t^2\,
 {\it x_0}\,{\it x_1}-8\,\alpha\,t^2\,{\it x_0}\,{\it x_1} +2\,t^2\,
 {\it x_0}\,{\it x_1}-3\,\alpha\,t^3\,{\it x_1}
\right.
 &\\
 \left.
+2\,t^3\,{\it x_1}-2\,
 \alpha^3\,{\it x_0}^4+4\,\alpha^3\,t\,{\it x_0}^3
 +2\,\alpha^2\,t\,
 {\it x_0}^3
\right.
 &\\
 \left.
-8\,\alpha^2\,t^2\,{\it x_0}^2+2\,\alpha\,t^2\,{\it x_0}^
 2+4\,\alpha\,t^3\,{\it x_0}-2\,t^3\,{\it x_0}\right]
&\textrm{if } i=1 \textrm{ and } t\in [x_{0}\, , \, x_{1}[\\ 
\end{array}
\right.
 $$

 $$
\frac{d}{dt}\Bspline{i}{2}{\alpha} (t)=
\left\lbrace
\begin{array}{ll}
\displaystyle{
 -\frac{\left(\alpha-1\right)^2\,\alpha\,\left({\it x_2}-t\right)}{{\left(\alpha\,{\it x_2}-
 {\it x_2}-\alpha\,{\it x_0}+t\right)^2\,\left(\alpha\,{\it x_2}-
 {\it x_2}-\alpha\,{\it x_1}+t\right)^2}}\times
 }
&\\
 \left[
 2\, \alpha\,{\it x_2}^2-2\,{\it x_2}^2-2\,\alpha\,{\it x_1}\,{\it x_2}+
 {\it x_1}\,{\it x_2}-2\,\alpha\,{\it x_0}\,{\it x_2}+{\it x_0}\,
 {\it x_2}
   \right.
 &\\
 \left.
+2\,t\,{\it x_2}+2\,\alpha\,{\it x_0}\,{\it x_1}-t\,
 {\it x_1}-t\,{\it x_0}\right]
&\textrm{if } i=1 \textrm{ and } t\in [x_{1}\, , \, x_{2}[\\ 
\displaystyle{
 -\frac{\left(\alpha-1\right)
 \,\alpha^2\,\left({\it x_0}-t\right)}{{\left(\alpha\,{\it x_1}-{\it x_1}-\alpha\,
 {\it x_0}+t\right)^2\,\left(\alpha\,{\it x_2}-{\it x_2}-\alpha\,
 {\it x_0}+t\right)^2}}\times
 }
 &\\
 \left[2\,\alpha\,{\it x_1}\,
 {\it x_2}-2\,{\it x_1}\,{\it x_2}-2\,\alpha\,{\it x_0}\,{\it x_2}+
 {\it x_0}\,{\it x_2}
 \right.
 &\\
 \left.
 +t\,{\it x_2}-2\,\alpha\,{\it x_0}\,{\it x_1}+
 {\it x_0}\,{\it x_1}+t\,{\it x_1}+2\,\alpha\,{\it x_0}^2-2\,t\,
 {\it x_0}\right]
&\textrm{if } i=2 \textrm{ and } t\in [x_{0}\, , \, x_{1}[\\ 
\displaystyle{
 \frac{\left(\alpha-1\right)\,\alpha}{{\left(\alpha\,{\it x_2}-
 {\it x_2}-\alpha\,{\it x_0}+t\right)^2\,\left(\alpha\,{\it x_2}-
 {\it x_2}-\alpha\,{\it x_1}+t\right)^3}}\times
 }
 &\\
 \left[
 2\,
 \alpha^3\,{\it x_2}^4-6\,\alpha^2\,{\it x_2}^4+6\,\alpha\,{\it x_2}^4
 -2\,{\it x_2}^4-2\,\alpha^3\,{\it x_1}\,{\it x_2}^3
 \right.
 &\\
 \left.
+5\,\alpha^2\,
 {\it x_1}\,{\it x_2}^3 -4\,\alpha\,{\it x_1}\,{\it x_2}^3
+{\it x_1}\,
 {\it x_2}^3-2\,\alpha^3\,{\it x_0}\,{\it x_2}^3
  \right.
 &\\
 \left.
+5\,\alpha^2\,
 {\it x_0}\,{\it x_2}^3-4\,\alpha\,{\it x_0}\,{\it x_2}^3+{\it x_0}\,
 {\it x_2}^3-4\,\alpha^3\,t\,{\it x_2}^3
  \right.
 &\\
 \left.
 +14\,\alpha^2\,t\,{\it x_2}^3-16\,\alpha\,t\,{\it x_2}^3+6\,t\,{\it x_2}^3
 +\alpha^2\,{\it x_1}^2  \,{\it x_2}^2
  \right.
 &\\
 \left.
-\alpha\,{\it x_1}^2\,{\it x_2}^2-3\,\alpha^2\,
 {\it x_0}\,{\it x_1}\,{\it x_2}^2+3\,\alpha\,{\it x_0}\,{\it x_1}\,
 {\it x_2}^2+6\,\alpha^3\,t\,{\it x_1}\,{\it x_2}^2
  \right.
 &\\
 \left.
-14\,\alpha^2\,t\,
 {\it x_1}\,{\it x_2}^2+11\,\alpha\,t\,{\it x_1}\,{\it x_2}^2-3\,t\,
 {\it x_1}\,{\it x_2}^2
+6\,\alpha^3\,t\,{\it x_0}\,{\it x_2}^2
\right.
 &\\
 \left.
-12\,
 \alpha^2\,t\,{\it x_0}\,{\it x_2}^2+9\,\alpha\,t\,{\it x_0}\,
 {\it x_2}^2-3\,t\,{\it x_0}\,{\it x_2}^2-8\,\alpha^2\,t^2\,{\it x_2}
 ^2
  \right.
 &\\
 \left.
+14\,\alpha\,t^2\,{\it x_2}^2-6\,t^2\,{\it x_2}^2+2\,\alpha^3\,
 {\it x_0}\,{\it x_1}^2\,{\it x_2}-2\,\alpha^3\,t\,{\it x_1}^2\,{\it x_2}
  \right.
 &\\
 \left.
+2\,\alpha\,t\,
 {\it x_1}^2\,{\it x_2}+2\,\alpha^3\,{\it x_0}^2\,{\it x_1}\,
 {\it x_2}-8\,\alpha^3\,t\,{\it x_0}\,{\it x_1}\,{\it x_2}
  \right.
 &\\
 \left.
+10\,\alpha
 ^2\,t\,{\it x_0}\,{\it x_1}\,{\it x_2}-6\,\alpha\,t\,{\it x_0}\,
 {\it x_1}\,{\it x_2}+9\,\alpha^2\,t^2\,{\it x_1}\,{\it x_2}
  \right.
 &\\
 \left.
-10\, \alpha\,t^2\,{\it x_1}\,{\it x_2}+3\,t^2\,{\it x_1}\,{\it x_2}
-2\, \alpha^3\,t\,{\it x_0}^2\,{\it x_2}+7\,\alpha^2\,t^2\,{\it x_0}\,
 {\it x_2}
  \right.
 &\\
 \left.
-6\,\alpha\,t^2\,{\it x_0}\,{\it x_2}+3\,t^2\,{\it x_0}\,
 {\it x_2}-4\,\alpha\,t^3\,{\it x_2}+2\,t^3\,{\it x_2}
  \right.
 &\\
 \left.
-2\,\alpha^3\,
 {\it x_0}^2\,{\it x_1}^2+2\,\alpha^3\,t\,{\it x_0}\,{\it x_1}^2+2\,
 \alpha^2\,t\,{\it x_0}\,{\it x_1}^2 -\alpha^2\,t^2\,{\it x_1}^2
  \right.
 &\\
 \left.
- \alpha\,t^2\,{\it x_1}^2+2\,\alpha^3\,t\,{\it x_0}^2\,{\it x_1}
-7\, \alpha^2\,t^2\,{\it x_0}\,{\it x_1}
  \right.
 &\\
 \left.
+3\,\alpha\,t^2\,{\it x_0}\,
 {\it x_1}
+3\,\alpha\,t^3\,{\it x_1}-t^3\,{\it x_1}+\alpha\,t^3\,
 {\it x_0}-t^3\,{\it x_0}\right]
&\textrm{if } i=2 \textrm{ and } t\in [x_{1}\, , \, x_{2}[\\ 
\displaystyle{
 -{{2\,\left(\alpha-1
 \right)\,\alpha^2\,\left({\it x_1}-t\right)\,\left({\it x_2}-
 {\it x_1}\right)}\over{\left(\alpha\,{\it x_2}-{\it x_2}-\alpha\,
 {\it x_1}+t\right)^3}}
 }
&\textrm{if } i=3 \textrm{ and } t\in [x_{1}\, , \, x_{2}[\\ 
 0
&\textrm{otherwise } \\ 
\end{array}
\right.
 $$

$$
\lim_{t \to x_{1}}\Bspline{i}{2}{\alpha} (t)=
\left\lbrace
\begin{array}{ll}
{{\left(\alpha-1\right)\,\left({\it x_2}-
 {\it x_1}\right)}\over{\alpha\,{\it x_2}-{\it x_2}+{\it x_1}-\alpha
 \,{\it x_0}}}
&\textrm{if } i=1 \textrm{ and } t\in [x_{0}\, , \, x_{1}[\\ 
 {{\left(\alpha-1\right)\,\left({\it x_2}-{\it x_1}
 \right)}\over{\alpha\,{\it x_2}-{\it x_2}+{\it x_1}-\alpha\,
 {\it x_0}}}
 &\textrm{if } i=1 \textrm{ and } t\in [x_{1}\, , \, x_{2}[\\ 
{{\alpha\,\left({\it x_1}-{\it x_0}\right)}\over{
 \alpha\,{\it x_2}-{\it x_2}+{\it x_1}-\alpha\,{\it x_0}}}
&\textrm{if } i=2 \textrm{ and } t\in [x_{0}\, , \, x_{1}[\\ 
 {{\alpha
 \,\left({\it x_1}-{\it x_0}\right)}\over{\alpha\,{\it x_2}-{\it x_2}
 +{\it x_1}-\alpha\,{\it x_0}}}
&\textrm{if } i=2 \textrm{ and } t\in [x_{1}\, , \, x_{2}[\\ 
0
&\textrm{otherwise } \\ 
\end{array}
\right.
 $$
$$
\lim_{t \to x_{1}}\frac{d}{dt}\Bspline{i}{2}{\alpha} (t)=
\left\lbrace
\begin{array}{ll}
-{{\left(\alpha-1\right)\,\left(2\,\alpha\,
 {\it x_2}-{\it x_2}+{\it x_1}-2\,\alpha\,{\it x_0}\right)}\over{
 \left(\alpha\,{\it x_2}-{\it x_2}+{\it x_1}-\alpha\,{\it x_0}\right)
 ^2}}
&\textrm{if } i=1 \textrm{ and } t\in [x_{0}\, , \, x_{1}[\\ 
 -{{\alpha\,\left(2\,\alpha\,{\it x_2}-2\,{\it x_2}+{\it x_1}-
 2\,\alpha\,{\it x_0}+{\it x_0}\right)}\over{\left(\alpha\,{\it x_2}-
 {\it x_2}+{\it x_1}-\alpha\,{\it x_0}\right)^2}}
&\textrm{if } i=1 \textrm{ and } t\in [x_{1}\, , \, x_{2}[\\ 
 {{\left(\alpha-1
 \right)\,\left(2\,\alpha\,{\it x_2}-{\it x_2}+{\it x_1}-2\,\alpha\,
 {\it x_0}\right)}\over{\left(\alpha\,{\it x_2}-{\it x_2}+{\it x_1}-
 \alpha\,{\it x_0}\right)^2}}
&\textrm{if } i=2 \textrm{ and } t\in [x_{0}\, , \, x_{1}[\\ 
 {{\alpha\,\left(2\,\alpha\,{\it x_2}-
 2\,{\it x_2}+{\it x_1}-2\,\alpha\,{\it x_0}+{\it x_0}\right)}\over{
 \left(\alpha\,{\it x_2}-{\it x_2}+{\it x_1}-\alpha\,{\it x_0}\right)
 ^2}}
&\textrm{if } i=2 \textrm{ and } t\in [x_{1}\, , \, x_{2}[\\ 
0
&\textrm{otherwise }\\ 
\end{array}
\right.
 $$


\newpage
\begin{center}
B-spline basis of degree 2 and its derivative
for the knot vector\\
$U=\left( x_{0},x_{0}, x_{0}, x_{1}, x_{2}, x_{3}, x_{3}, x_{3} \right)$
\end{center}

$$
\Bspline{i}{2}{\alpha}(t) =
\left\lbrace
\begin{array}{ll}
\displaystyle{
{{\left(\alpha-1\right)^2\,\left({\it x_1}-t\right)^
 2}\over{\left(\alpha\,{\it x_1}-{\it x_1}-\alpha\,{\it x_0}+t\right)
 ^2}} 
}
&\textrm{if } i=0 \textrm{ and } t\in [x_{0}\, , \, x_{1}[\\ 
 &\\
\displaystyle{
 -\frac{\left(\alpha-1\right)\,\alpha\,\left({\it x_0}-t
 \right)}{\left(\alpha\,
 {\it x_1}-{\it x_1}-\alpha\,{\it x_0}+t\right)^2\,\left(\alpha\,
 {\it x_2}-{\it x_2}-\alpha\,{\it x_0}+t\right)}\times %
}
&\\
 \,\left[2\,\alpha\,{\it x_1}\,{\it x_2}-2\,{\it x_1}\,
 {\it x_2}-\alpha\,{\it x_0}\,{\it x_2}-\alpha\,t\,{\it x_2}+2\,t\,
 {\it x_2}
 \right.
 &\\
 \left.
 -\alpha\,{\it x_0}\,{\it x_1}-\alpha\,t\,{\it x_1}+2\,t\,
 {\it x_1}+2\,\alpha\,t\,{\it x_0}-2\,t^2\right]
 &\textrm{if } i=1 \textrm{ and } t\in [x_{0}\, , \, x_{1}[\\
 &\\
 \displaystyle{
{{\left(\alpha-1
 \right)^2\,\left({\it x_2}-t\right)^2}\over{\left(\alpha\,{\it x_2}-
 {\it x_2}-\alpha\,{\it x_0}+t\right)\,\left(\alpha\,{\it x_2}-
 {\it x_2}-\alpha\,{\it x_1}+t\right)}} 
}
&\textrm{if } i=1 \textrm{ and } t\in [x_{1}\, , \, x_{2}[\\
\end{array}
\right.
$$


$$
\Bspline{i}{2}{\alpha}(t) =
\left\lbrace
\begin{array}{ll}
\displaystyle{
 {{\alpha^2\,\left({\it x_0}
 -t\right)^2}\over{\left(\alpha\,{\it x_1}-{\it x_1}-\alpha\,
 {\it x_0}+t\right)\,\left(\alpha\,{\it x_2}-{\it x_2}-\alpha\,
 {\it x_0}+t\right)}} 
}
&\textrm{if } i=2 \textrm{ and } t\in [x_{0}\, , \, x_{1}[\\
 &\\
\displaystyle{
 -\frac{\left(\alpha-1\right)\,\alpha}{\left(\alpha\,{\it x_2}-{\it x_2}-\alpha\,{\it x_0}+t
 \right)\,\left(\alpha\,{\it x_2}-{\it x_2}-\alpha\,{\it x_1}+t
 \right)}\times %
} &\\
\displaystyle{
 \frac{1}{\left(\alpha\,{\it x_3}-{\it x_3}-\alpha\,{\it x_1}+t
 \right)}\times %
} &\\
 \left[
 \alpha\,{\it x_1}\,{\it x_2}\,{\it x_3}-{\it x_1}\,{\it x_2}\,
 {\it x_3}+\alpha\,{\it x_0}\,{\it x_2}\,{\it x_3}-{\it x_0}\,
 {\it x_2}\,{\it x_3}
 \right.
 &\\
 \left.
 -2\,\alpha\,t\,{\it x_2}\,{\it x_3}+2\,t\,
 {\it x_2}\,{\it x_3}-\alpha\,{\it x_0}\,{\it x_1}\,{\it x_3}+t\,
 {\it x_1}\,{\it x_3}+t\,{\it x_0}\,{\it x_3}
 \right.
 &\\
 \left.
 +\alpha\,t^2\,{\it x_3}-
 2\,t^2\,{\it x_3}-\alpha\,{\it x_0}\,{\it x_1}\,{\it x_2}
 +t\,{\it x_1}\,{\it x_2}+t\,{\it x_0}\,{\it x_2}
 \right.
 &\\
 \left.
 +\alpha\,t^2\,{\it x_2}-
 2\,t^2\,{\it x_2}+2\,\alpha\,t\,{\it x_0}\,{\it x_1}
 \right.
 &\\
 \left.
 -\alpha\,t^2\,
 {\it x_1}-t^2\,{\it x_1}-\alpha\,t^2\,{\it x_0}-t^2\,{\it x_0}+2\,t^
 3\right] %
 &\textrm{if } i=2 \textrm{ and } t\in [x_{1}\, , \, x_{2}[\\
 &\\
\displaystyle{
 {{\left(\alpha-1\right)^2\,\left({\it x_3}-t\right)^2
 }\over{\left(\alpha\,{\it x_3}-{\it x_3}-\alpha\,{\it x_1}+t\right)
 \,\left(\alpha\,{\it x_3}-{\it x_3}-\alpha\,{\it x_2}+t\right)}} 
}
&\textrm{if } i=2 \textrm{ and } t\in [x_{2}\, , \, x_{3}[\\
\end{array}
\right.
$$


$$
\Bspline{i}{2}{\alpha}(t) =
\left\lbrace
\begin{array}{ll}
\displaystyle{
 {{\alpha^2\,\left({\it x_1}-t\right)^2}\over{\left(\alpha\,{\it x_2}
 -{\it x_2}-\alpha\,{\it x_1}+t\right)\,\left(\alpha\,{\it x_3}-
 {\it x_3}-\alpha\,{\it x_1}+t\right)}} 
}
&\textrm{if } i=3 \textrm{ and } t\in [x_{1}\, , \, x_{2}[\\
 &\\
\displaystyle{
 -\frac{\left(\alpha-1\right)\,
 \alpha\,\left({\it x_3}-t\right)}{\left(\alpha\,{\it x_3}-
 {\it x_3}-\alpha\,{\it x_1}+t\right)\,\left(\alpha\,{\it x_3}-
 {\it x_3}-\alpha\,{\it x_2}+t\right)^2}\times %
}
&\\
 \,\left[\alpha\,{\it x_2}\,{\it x_3}
 -{\it x_2}\,{\it x_3}+\alpha\,{\it x_1}\,{\it x_3}-{\it x_1}\,
 {\it x_3}
 \right.
 &\\
 \left.
 -2\,\alpha\,t\,{\it x_3}+2\,t\,{\it x_3}-2\,\alpha\,
 {\it x_1}\,{\it x_2}
 \right.
 &\\
 \left.
 +\alpha\,t\,{\it x_2}+t\,{\it x_2}+\alpha\,t\,
 {\it x_1}+t\,{\it x_1}-2\,t^2\right]%
 &\textrm{if } i=3 \textrm{ and } t\in [x_{2}\, , \, x_{3}[\\
 &\\
\displaystyle{
 {{\alpha^2\,\left(
 {\it x_2}-t\right)^2}\over{\left(\alpha\,{\it x_3}-{\it x_3}-\alpha
 \,{\it x_2}+t\right)^2}} 
}
&\textrm{if } i=4 \textrm{ and } t\in [x_{2}\, , \, x_{3}[\\
 &\\
 0 &\textrm{otherwise}\\
\end{array}
\right.
$$


 $$
\frac{d}{dt}\Bspline{i}{2}{\alpha}(t) =
\left\lbrace
\begin{array}{ll}
\displaystyle{
-{{2\,\left(\alpha-1\right)^2\,\alpha\,\left(
 {\it x_1}-t\right)\,\left({\it x_1}-{\it x_0}\right)}\over{\left(
 \alpha\,{\it x_1}-{\it x_1}-\alpha\,{\it x_0}+t\right)^3}}
} 
 &\textrm{if } i=0 \textrm{ and } t\in [x_{0}\, , \, x_{1}[\\ 
 &\\
\displaystyle{
 \frac{\left(\alpha-1\right)\,\alpha}{\left(
 \alpha\,{\it x_1}-{\it x_1}-\alpha\,{\it x_0}+t\right)^3\,\left(
 \alpha\,{\it x_2}-{\it x_2}-\alpha\,{\it x_0}+t\right)^2}\times%
}
&\\
  \,\left[2\,\alpha^3\,{\it x_1}^2\,
 {\it x_2}^2-6\,\alpha^2\,{\it x_1}^2\,{\it x_2}^2+6\,\alpha\,
 {\it x_1}^2\,{\it x_2}^2
 \right.
 &\\
 \left.
 -2\,{\it x_1}^2\,{\it x_2}^2-2\,\alpha^3\,
 {\it x_0}\,{\it x_1}\,{\it x_2}^2+6\,\alpha^2\,{\it x_0}\,{\it x_1}
 \,{\it x_2}^2
 \right.
 &\\
 \left.
 -6\,\alpha\,{\it x_0}\,{\it x_1}\,{\it x_2}^2+2\,
 {\it x_0}\,{\it x_1}\,{\it x_2}^2-2\,\alpha^3\,t\,{\it x_1}\,
 {\it x_2}^2
 \right.
 &\\
 \left.
 +6\,\alpha^2\,t\,{\it x_1}\,{\it x_2}^2-6\,\alpha\,t\,
 {\it x_1}\,{\it x_2}^2+2\,t\,{\it x_1}\,{\it x_2}^2+2\,\alpha^3\,t\,
 {\it x_0}\,{\it x_2}^2
 \right.
 &\\
 \left.
 -6\,\alpha^2\,t\,{\it x_0}\,{\it x_2}^2+6\,
 \alpha\,t\,{\it x_0}\,{\it x_2}^2-2\,t\,{\it x_0}\,{\it x_2}^2
 \right.
 &\\
 \left.
 -2\, \alpha^3\,{\it x_0}\,{\it x_1}^2\,{\it x_2}+4\,\alpha^2\,{\it x_0}\,
 {\it x_1}^2\,{\it x_2}-2\,\alpha\,{\it x_0}\,{\it x_1}^2\,{\it x_2}
 \right.
 &\\
 \left.
 - 2\,\alpha^3\,t\,{\it x_1}^2\,{\it x_2}
 \right.
 &\\
 \left.
 +8\,\alpha^2\,t\,{\it x_1}^2\,
 {\it x_2}-10\,\alpha\,t\,{\it x_1}^2\,{\it x_2}+4\,t\,{\it x_1}^2\,
 {\it x_2}-3\,\alpha^2\,{\it x_0}^2\,{\it x_1}\,{\it x_2}
 \right.
 &\\
 \left.
 +3\,\alpha\,
 {\it x_0}^2\,{\it x_1}\,{\it x_2}+8\,\alpha^3\,t\,{\it x_0}\,
 {\it x_1}\,{\it x_2}-14\,\alpha^2\,t\,{\it x_0}\,{\it x_1}\,
 {\it x_2}
  \right.
 &\\
 \left.
+10\,\alpha\,t\,{\it x_0}\,{\it x_1}\,{\it x_2}
 \right.
 &\\
 \left.
 -4\,t\,
 {\it x_0}\,{\it x_1}\,{\it x_2}-7\,\alpha^2\,t^2\,{\it x_1}\,
 {\it x_2}+11\,\alpha\,t^2\,{\it x_1}\,{\it x_2}-4\,t^2\,{\it x_1}\,
 {\it x_2}
 \right.
 &\\
 \left.
 +2\,\alpha^3\,{\it x_0}^3\,{\it x_2}-\alpha^2\,{\it x_0}^3
 \,{\it x_2}-6\,\alpha^3\,t\,{\it x_0}^2\,{\it x_2}+6\,\alpha^2\,t\,
 {\it x_0}^2\,{\it x_2}
 \right.
 &\\
 \left.
 -3\,\alpha\,t\,{\it x_0}^2\,{\it x_2}+7\,
 \alpha^2\,t^2\,{\it x_0}\,{\it x_2}-8\,\alpha\,t^2\,{\it x_0}\,
 {\it x_2}+4\,t^2\,{\it x_0}\,{\it x_2}
 \right.
 &\\
 \left.
 -\alpha\,t^3\,{\it x_2}
 \right.
 &\\
 \left.
 +\alpha^2\,{\it x_0}^2\,{\it x_1}^2-\alpha\,{\it x_0}^2\,{\it x_1}^2+2\,
 \alpha^3\,t\,{\it x_0}\,{\it x_1}^2-6\,\alpha^2\,t\,{\it x_0}\,
 {\it x_1}^2
 \right.
 &\\
 \left.
 +4\,\alpha\,t\,{\it x_0}\,{\it x_1}^2-\alpha^2\,t^2\,
 {\it x_1}^2+3\,\alpha\,t^2\,{\it x_1}^2-2\,t^2\,{\it x_1}^2+2\,
 \alpha^3\,{\it x_0}^3\,{\it x_1}
 \right.
 &\\
 \left.
 -\alpha^2\,{\it x_0}^3\,{\it x_1}-6
 \,\alpha^3\,t\,{\it x_0}^2\,{\it x_1}+4\,\alpha^2\,t\,{\it x_0}^2\,
 {\it x_1}-\alpha\,t\,{\it x_0}^2\,{\it x_1}
 \right.
 &\\
 \left.
 +9\,\alpha^2\,t^2\,
 {\it x_0}\,{\it x_1}-8\,\alpha\,t^2\,{\it x_0}\,{\it x_1}+2\,t^2\,
 {\it x_0}\,{\it x_1}-3\,\alpha\,t^3\,{\it x_1}+2\,t^3\,{\it x_1}
 \right.
 &\\
 \left.
 -2\, \alpha^3\,{\it x_0}^4+4\,\alpha^3\,t\,{\it x_0}^3+2\,\alpha^2\,t\,
 {\it x_0}^3-8\,\alpha^2\,t^2\,{\it x_0}^2
 \right.
 &\\
 \left.
 +2\,\alpha\,t^2\,{\it x_0}^
 2+4\,\alpha\,t^3\,{\it x_0}-2\,t^3\,{\it x_0}\right]%
 &\textrm{if } i=1 \textrm{ and } t\in [x_{0}\, , \, x_{1}[\\ 
 &\\
\displaystyle{
 -\frac{\left(\alpha-1\right)^2\,\alpha\,\left({\it x_2}-t\right)}{\left(\alpha\,{\it x_2}-
 {\it x_2}-\alpha\,{\it x_0}+t\right)^2\,\left(\alpha\,{\it x_2}-
 {\it x_2}-\alpha\,{\it x_1}+t\right)^2}\times%
}
&\\
  \,\left[2\,
 \alpha\,{\it x_2}^2-2\,{\it x_2}^2-2\,\alpha\,{\it x_1}\,{\it x_2}+
 {\it x_1}\,{\it x_2}-2\,\alpha\,{\it x_0}\,{\it x_2}+{\it x_0}\,
 {\it x_2}
 \right.
 &\\
 \left.
 +2\,t\,{\it x_2}+2\,\alpha\,{\it x_0}\,{\it x_1}-t\,
 {\it x_1}-t\,{\it x_0}\right] %
 &\textrm{if } i=1 \textrm{ and } t\in [x_{1}\, , \, x_{2}[\\ 
 &\\
\displaystyle{
 -\frac{\left(\alpha-1\right)
 \,\alpha^2\,\left({\it x_0}-t\right)}{\left(\alpha\,{\it x_1}-{\it x_1}-\alpha\,
 {\it x_0}+t\right)^2\,\left(\alpha\,{\it x_2}-{\it x_2}-\alpha\,
 {\it x_0}+t\right)^2}\times%
}
&\\
 \,\left[2\,\alpha\,{\it x_1}\,
 {\it x_2}-2\,{\it x_1}\,{\it x_2}-2\,\alpha\,{\it x_0}\,{\it x_2}+
 {\it x_0}\,{\it x_2}+t\,{\it x_2}-2\,\alpha\,{\it x_0}\,{\it x_1}
 \right.
 &\\
 \left.
 + {\it x_0}\,{\it x_1}+t\,{\it x_1}+2\,\alpha\,{\it x_0}^2-2\,t\,
 {\it x_0}\right] %
 &\textrm{if } i=2 \textrm{ and } t\in [x_{0}\, , \, x_{1}[\\ 
\end{array}
\right.
$$

\begin{tiny}

 $$
\frac{d}{dt}\Bspline{i}{2}{\alpha}(t) =
\left\lbrace
\begin{array}{ll}
\displaystyle{
 \frac{\left(\alpha-1\right)\,\alpha}{\left(
 \alpha\,{\it x_2}-{\it x_2}-\alpha\,{\it x_0}+t\right)^2\,\left(
 \alpha\,{\it x_2}-{\it x_2}-\alpha\,{\it x_1}+t\right)^2\,\left(
 \alpha\,{\it x_3}-{\it x_3}-\alpha\,{\it x_1}+t\right)^2}\times%
}
&\\
 \,\left[2\,
 \alpha^4\,{\it x_2}^3\,{\it x_3}^2-8\,\alpha^3\,{\it x_2}^3\,
 {\it x_3}^2+12\,\alpha^2\,{\it x_2}^3\,{\it x_3}^2-8\,\alpha\,
 {\it x_2}^3\,{\it x_3}^2
 \right.
 &\\
 \left.
 +2\,{\it x_2}^3\,{\it x_3}^2-2\,\alpha^4\,
 {\it x_1}\,{\it x_2}^2\,{\it x_3}^2+7\,\alpha^3\,{\it x_1}\,
 {\it x_2}^2\,{\it x_3}^2
 \right.
 &\\
 \left.
 -9\,\alpha^2\,{\it x_1}\,{\it x_2}^2\,
 {\it x_3}^2+5\,\alpha\,{\it x_1}\,{\it x_2}^2\,{\it x_3}^2-{\it x_1}
 \,{\it x_2}^2\,{\it x_3}^2
 \right.
 &\\
 \left.
 -2\,\alpha^4\,{\it x_0}\,{\it x_2}^2\,
 {\it x_3}^2+7\,\alpha^3\,{\it x_0}\,{\it x_2}^2\,{\it x_3}^2-9\,
 \alpha^2\,{\it x_0}\,{\it x_2}^2\,{\it x_3}^2
 \right.
 &\\
 \left.
 +5\,\alpha\,{\it x_0}\,
 {\it x_2}^2\,{\it x_3}^2-{\it x_0}\,{\it x_2}^2\,{\it x_3}^2-2\,
 \alpha^4\,t\,{\it x_2}^2\,{\it x_3}^2+10\,\alpha^3\,t\,{\it x_2}^2\,
 {\it x_3}^2
 \right.
 &\\
 \left.
 -18\,\alpha^2\,t\,{\it x_2}^2\,{\it x_3}^2+14\,\alpha\,t
 \,{\it x_2}^2\,{\it x_3}^2-4\,t\,{\it x_2}^2\,{\it x_3}^2
 \right.
 &\\
 \left.
 +2\,\alpha^4\,{\it x_0}\,{\it x_1}\,{\it x_2}\,{\it x_3}^2-6\,\alpha^3\,
 {\it x_0}\,{\it x_1}\,{\it x_2}\,{\it x_3}^2+6\,\alpha^2\,{\it x_0}
 \,{\it x_1}\,{\it x_2}\,{\it x_3}^2
 \right.
 &\\
 \left.
 -2\,\alpha\,{\it x_0}\,{\it x_1}
 \,{\it x_2}\,{\it x_3}^2+2\,\alpha^4\,t\,{\it x_1}\,{\it x_2}\,
 {\it x_3}^2-8\,\alpha^3\,t\,{\it x_1}\,{\it x_2}\,{\it x_3}^2
 \right.
 &\\
 \left.
 +12\, \alpha^2\,t\,{\it x_1}\,{\it x_2}\,{\it x_3}^2-8\,\alpha\,t\,
 {\it x_1}\,{\it x_2}\,{\it x_3}^2+2\,t\,{\it x_1}\,{\it x_2}\,
 {\it x_3}^2
 \right.
 &\\
 \left.
 +2\,\alpha^4\,t\,{\it x_0}\,{\it x_2}\,{\it x_3}^2-8\,
 \alpha^3\,t\,{\it x_0}\,{\it x_2}\,{\it x_3}^2+12\,\alpha^2\,t\,
 {\it x_0}\,{\it x_2}\,{\it x_3}^2
 \right.
 &\\
 \left.
 -8\,\alpha\,t\,{\it x_0}\,{\it x_2}
 \,{\it x_3}^2+2\,t\,{\it x_0}\,{\it x_2}\,{\it x_3}^2-2\,\alpha^3\,t
 ^2\,{\it x_2}\,{\it x_3}^2
 \right.
 &\\
 \left.
 +6\,\alpha^2\,t^2\,{\it x_2}\,{\it x_3}^2-
 6\,\alpha\,t^2\,{\it x_2}\,{\it x_3}^2+2\,t^2\,{\it x_2}\,{\it x_3}^
 2-2\,\alpha^4\,t\,{\it x_0}\,{\it x_1}\,{\it x_3}^2
 \right.
 &\\
 \left.
 +6\,\alpha^3\,t\,
 {\it x_0}\,{\it x_1}\,{\it x_3}^2-6\,\alpha^2\,t\,{\it x_0}\,
 {\it x_1}\,{\it x_3}^2+2\,\alpha\,t\,{\it x_0}\,{\it x_1}\,{\it x_3}
 ^2
 \right.
 &\\
 \left.
 +\alpha^3\,t^2\,{\it x_1}\,{\it x_3}^2-3\,\alpha^2\,t^2\,{\it x_1}
 \,{\it x_3}^2+3\,\alpha\,t^2\,{\it x_1}\,{\it x_3}^2-t^2\,{\it x_1}
 \,{\it x_3}^2
 \right.
 &\\
 \left.
 +\alpha^3\,t^2\,{\it x_0}\,{\it x_3}^2-3\,\alpha^2\,t^2
 \,{\it x_0}\,{\it x_3}^2+3\,\alpha\,t^2\,{\it x_0}\,{\it x_3}^2-t^2
 \,{\it x_0}\,{\it x_3}^2
 \right.
 &\\
 \left.
 -2\,\alpha^4\,{\it x_1}\,{\it x_2}^3\,
 {\it x_3}+6\,\alpha^3\,{\it x_1}\,{\it x_2}^3\,{\it x_3}-6\,\alpha^2
 \,{\it x_1}\,{\it x_2}^3\,{\it x_3}+2\,\alpha\,{\it x_1}\,{\it x_2}^
 3\,{\it x_3}
 \right.
 &\\
 \left.
 -2\,\alpha^4\,t\,{\it x_2}^3\,{\it x_3}+10\,\alpha^3\,t
 \,{\it x_2}^3\,{\it x_3}-18\,\alpha^2\,t\,{\it x_2}^3\,{\it x_3}+14
 \,\alpha\,t\,{\it x_2}^3\,{\it x_3}
 \right.
 &\\
 \left.
 -4\,t\,{\it x_2}^3\,{\it x_3}+2\,
 \alpha^4\,{\it x_1}^2\,{\it x_2}^2\,{\it x_3}-5\,\alpha^3\,{\it x_1}
 ^2\,{\it x_2}^2\,{\it x_3}+4\,\alpha^2\,{\it x_1}^2\,{\it x_2}^2\,
 {\it x_3}
 \right.
 &\\
 \left.
 -\alpha\,{\it x_1}^2\,{\it x_2}^2\,{\it x_3}-2\,\alpha^3\,
 {\it x_0}\,{\it x_1}\,{\it x_2}^2\,{\it x_3}+4\,\alpha^2\,{\it x_0}
 \,{\it x_1}\,{\it x_2}^2\,{\it x_3}
 \right.
 &\\
 \left.
 -2\,\alpha\,{\it x_0}\,{\it x_1}
 \,{\it x_2}^2\,{\it x_3}+6\,\alpha^4\,t\,{\it x_1}\,{\it x_2}^2\,
 {\it x_3}-20\,\alpha^3\,t\,{\it x_1}\,{\it x_2}^2\,{\it x_3}
 \right.
 &\\
 \left.
 +24\, \alpha^2\,t\,{\it x_1}\,{\it x_2}^2\,{\it x_3}-12\,\alpha\,t\,
 {\it x_1}\,{\it x_2}^2\,{\it x_3}+2\,t\,{\it x_1}\,{\it x_2}^2\,
 {\it x_3}
 \right.
 &\\
 \left.
 +4\,\alpha^4\,t\,{\it x_0}\,{\it x_2}^2\,{\it x_3}-12\,
 \alpha^3\,t\,{\it x_0}\,{\it x_2}^2\,{\it x_3}+14\,\alpha^2\,t\,
 {\it x_0}\,{\it x_2}^2\,{\it x_3}
 \right.
 &\\
 \left.
 -8\,\alpha\,t\,{\it x_0}\,{\it x_2}
 ^2\,{\it x_3}+2\,t\,{\it x_0}\,{\it x_2}^2\,{\it x_3}-9\,\alpha^3\,t
 ^2\,{\it x_2}^2\,{\it x_3}+26\,\alpha^2\,t^2\,{\it x_2}^2\,{\it x_3}
 \right.
 &\\
 \left.
 -25\,\alpha\,t^2\,{\it x_2}^2\,{\it x_3}+8\,t^2\,{\it x_2}^2\,
 {\it x_3}+2\,\alpha^3\,{\it x_0}\,{\it x_1}^2\,{\it x_2}\,{\it x_3}-
 2\,\alpha^2\,{\it x_0}\,{\it x_1}^2\,{\it x_2}\,{\it x_3}
 \right.
 &\\
 \left.
 -4\,\alpha^4\,t\,{\it x_1}^2\,{\it x_2}\,{\it x_3}+8\,\alpha^3\,t\,{\it x_1}^2
 \,{\it x_2}\,{\it x_3}-6\,\alpha^2\,t\,{\it x_1}^2\,{\it x_2}\,
 {\it x_3}+2\,\alpha\,t\,{\it x_1}^2\,{\it x_2}\,{\it x_3}
 \right.
 &\\
 \left.
 +2\,\alpha^4\,{\it x_0}^2\,{\it x_1}\,{\it x_2}\,{\it x_3}-2\,\alpha^3\,
 {\it x_0}^2\,{\it x_1}\,{\it x_2}\,{\it x_3}-8\,\alpha^4\,t\,
 {\it x_0}\,{\it x_1}\,{\it x_2}\,{\it x_3}
 \right.
 &\\
 \left.
 +16\,\alpha^3\,t\,
 {\it x_0}\,{\it x_1}\,{\it x_2}\,{\it x_3}-16\,\alpha^2\,t\,
 {\it x_0}\,{\it x_1}\,{\it x_2}\,{\it x_3}+8\,\alpha\,t\,{\it x_0}\,
 {\it x_1}\,{\it x_2}\,{\it x_3}
 \right.
 &\\
 \left.
 +12\,\alpha^3\,t^2\,{\it x_1}\,
 {\it x_2}\,{\it x_3}-22\,\alpha^2\,t^2\,{\it x_1}\,{\it x_2}\,
 {\it x_3}+14\,\alpha\,t^2\,{\it x_1}\,{\it x_2}\,{\it x_3}-4\,t^2\,
 {\it x_1}\,{\it x_2}\,{\it x_3}
 \right.
 &\\
 \left.
 -2\,\alpha^4\,t\,{\it x_0}^2\,
 {\it x_2}\,{\it x_3}+2\,\alpha^3\,t\,{\it x_0}^2\,{\it x_2}\,
 {\it x_3}+10\,\alpha^3\,t^2\,{\it x_0}\,{\it x_2}\,{\it x_3}
 \right.
 &\\
 \left.
 -18\, \alpha^2\,t^2\,{\it x_0}\,{\it x_2}\,{\it x_3}+12\,\alpha\,t^2\,
 {\it x_0}\,{\it x_2}\,{\it x_3}-4\,t^2\,{\it x_0}\,{\it x_2}\,
 {\it x_3}-8\,\alpha^2\,t^3\,{\it x_2}\,{\it x_3}
 \right.
 &\\
 \left.
 +12\,\alpha\,t^3\,
 {\it x_2}\,{\it x_3}-4\,t^3\,{\it x_2}\,{\it x_3}-2\,\alpha^4\,
 {\it x_0}^2\,{\it x_1}^2\,{\it x_3}+\alpha^3\,{\it x_0}^2\,{\it x_1}
 ^2\,{\it x_3}
 \right.
 &\\
 \left.
 +4\,\alpha^4\,t\,{\it x_0}\,{\it x_1}^2\,{\it x_3}-4\,
 \alpha^3\,t\,{\it x_0}\,{\it x_1}^2\,{\it x_3}+2\,\alpha^2\,t\,
 {\it x_0}\,{\it x_1}^2\,{\it x_3}-2\,\alpha^3\,t^2\,{\it x_1}^2\,
 {\it x_3}
 \right.
 &\\
 \left.
 +2\,\alpha^2\,t^2\,{\it x_1}^2\,{\it x_3}-\alpha\,t^2\,
 {\it x_1}^2\,{\it x_3}+2\,\alpha^4\,t\,{\it x_0}^2\,{\it x_1}\,
 {\it x_3}-10\,\alpha^3\,t^2\,{\it x_0}\,{\it x_1}\,{\it x_3}
 \right.
 &\\
 \left.
 +12\, \alpha^2\,t^2\,{\it x_0}\,{\it x_1}\,{\it x_3}-6\,\alpha\,t^2\,
 {\it x_0}\,{\it x_1}\,{\it x_3}+4\,\alpha^2\,t^3\,{\it x_1}\,
 {\it x_3}-4\,\alpha\,t^3\,{\it x_1}\,{\it x_3}
 \right.
 &\\
 \left.
 +2\,t^3\,{\it x_1}\,
 {\it x_3}-\alpha^3\,t^2\,{\it x_0}^2\,{\it x_3}+4\,\alpha^2\,t^3\,
 {\it x_0}\,{\it x_3}-4\,\alpha\,t^3\,{\it x_0}\,{\it x_3}+2\,t^3\,
 {\it x_0}\,{\it x_3}
 \right.
 &\\
 \left.
 -\alpha\,t^4\,{\it x_3}+\alpha^3\,{\it x_1}^2\,
 {\it x_2}^3-2\,\alpha^2\,{\it x_1}^2\,{\it x_2}^3+\alpha\,{\it x_1}^
 2\,{\it x_2}^3+2\,\alpha^4\,t\,{\it x_1}\,{\it x_2}^3
 \right.
 &\\
 \left.
 -8\,\alpha^3\,t
 \,{\it x_1}\,{\it x_2}^3+10\,\alpha^2\,t\,{\it x_1}\,{\it x_2}^3-4\,
 \alpha\,t\,{\it x_1}\,{\it x_2}^3-\alpha^3\,t^2\,{\it x_2}^3+4\,
 \alpha^2\,t^2\,{\it x_2}^3
 \right.
 &\\
 \left.
 -5\,\alpha\,t^2\,{\it x_2}^3+2\,t^2\,
 {\it x_2}^3-\alpha^3\,{\it x_1}^3\,{\it x_2}^2+\alpha^2\,{\it x_1}^3
 \,{\it x_2}^2+2\,\alpha^4\,{\it x_0}\,{\it x_1}^2\,{\it x_2}^2
 \right.
 &\\
 \left.
 -3\, \alpha^3\,{\it x_0}\,{\it x_1}^2\,{\it x_2}^2+\alpha^2\,{\it x_0}\,
 {\it x_1}^2\,{\it x_2}^2-4\,\alpha^4\,t\,{\it x_1}^2\,{\it x_2}^2+8
 \,\alpha^3\,t\,{\it x_1}^2\,{\it x_2}^2
 \right.
 &\\
 \left.
 -2\,\alpha^2\,t\,{\it x_1}^2
 \,{\it x_2}^2-2\,\alpha\,t\,{\it x_1}^2\,{\it x_2}^2-4\,\alpha^4\,t
 \,{\it x_0}\,{\it x_1}\,{\it x_2}^2+8\,\alpha^3\,t\,{\it x_0}\,
 {\it x_1}\,{\it x_2}^2
 \right.
 &\\
 \left.
 -6\,\alpha^2\,t\,{\it x_0}\,{\it x_1}\,
 {\it x_2}^2+2\,\alpha\,t\,{\it x_0}\,{\it x_1}\,{\it x_2}^2+10\,
 \alpha^3\,t^2\,{\it x_1}\,{\it x_2}^2-22\,\alpha^2\,t^2\,{\it x_1}\,
 {\it x_2}^2
 \right.
 &\\
 \left.
 +13\,\alpha\,t^2\,{\it x_1}\,{\it x_2}^2-t^2\,{\it x_1}\,
 {\it x_2}^2+2\,\alpha^3\,t^2\,{\it x_0}\,{\it x_2}^2-4\,\alpha^2\,t^
 2\,{\it x_0}\,{\it x_2}^2
 \right.
 &\\
 \left.
 +3\,\alpha\,t^2\,{\it x_0}\,{\it x_2}^2-t^2
 \,{\it x_0}\,{\it x_2}^2-4\,\alpha^2\,t^3\,{\it x_2}^2+8\,\alpha\,t^
 3\,{\it x_2}^2-4\,t^3\,{\it x_2}^2
 \right.
 &\\
 \left.
 -2\,\alpha^4\,{\it x_0}\,{\it x_1}
 ^3\,{\it x_2}+2\,\alpha^3\,{\it x_0}\,{\it x_1}^3\,{\it x_2}+2\,
 \alpha^4\,t\,{\it x_1}^3\,{\it x_2}-2\,\alpha^2\,t\,{\it x_1}^3\,
 {\it x_2}
 \right.
 &\\
 \left.
 -2\,\alpha^4\,{\it x_0}^2\,{\it x_1}^2\,{\it x_2}+\alpha^3
 \,{\it x_0}^2\,{\it x_1}^2\,{\it x_2}+6\,\alpha^4\,t\,{\it x_0}\,
 {\it x_1}^2\,{\it x_2}-4\,\alpha^3\,t\,{\it x_0}\,{\it x_1}^2\,
 {\it x_2}
 \right.
 &\\
 \left.
 -10\,\alpha^3\,t^2\,{\it x_1}^2\,{\it x_2}+8\,\alpha^2\,t^2
 \,{\it x_1}^2\,{\it x_2}+\alpha\,t^2\,{\it x_1}^2\,{\it x_2}+2\,
 \alpha^4\,t\,{\it x_0}^2\,{\it x_1}\,{\it x_2}
 \right.
 &\\
 \left.
 -12\,\alpha^3\,t^2\,
 {\it x_0}\,{\it x_1}\,{\it x_2}+14\,\alpha^2\,t^2\,{\it x_0}\,
 {\it x_1}\,{\it x_2}-6\,\alpha\,t^2\,{\it x_0}\,{\it x_1}\,{\it x_2}
 +12\,\alpha^2\,t^3\,{\it x_1}\,{\it x_2}
 \right.
 &\\
 \left.
 -12\,\alpha\,t^3\,{\it x_1}
 \,{\it x_2}+2\,t^3\,{\it x_1}\,{\it x_2}-\alpha^3\,t^2\,{\it x_0}^2
 \,{\it x_2}+4\,\alpha^2\,t^3\,{\it x_0}\,{\it x_2}-4\,\alpha\,t^3\,
 {\it x_0}\,{\it x_2}
 \right.
 &\\
 \left.
 +2\,t^3\,{\it x_0}\,{\it x_2}-3\,\alpha\,t^4\,
 {\it x_2}+2\,t^4\,{\it x_2}+2\,\alpha^4\,{\it x_0}^2\,{\it x_1}^3-2
 \,\alpha^4\,t\,{\it x_0}\,{\it x_1}^3
 \right.
 &\\
 \left.
 -2\,\alpha^3\,t\,{\it x_0}\,
 {\it x_1}^3+\alpha^3\,t^2\,{\it x_1}^3+\alpha^2\,t^2\,{\it x_1}^3-2
 \,\alpha^4\,t\,{\it x_0}^2\,{\it x_1}^2-2\,\alpha^3\,t\,{\it x_0}^2
 \,{\it x_1}^2
 \right.
 &\\
 \left.
 +9\,\alpha^3\,t^2\,{\it x_0}\,{\it x_1}^2-\alpha^2\,t^2
 \,{\it x_0}\,{\it x_1}^2-4\,\alpha^2\,t^3\,{\it x_1}^2+2\,\alpha^3\,
 t^2\,{\it x_0}^2\,{\it x_1}
 \right.
 &\\
 \left.
 -8\,\alpha^2\,t^3\,{\it x_0}\,{\it x_1}+4
 \,\alpha\,t^3\,{\it x_0}\,{\it x_1}+3\,\alpha\,t^4\,{\it x_1}-t^4\,
 {\it x_1}+\alpha\,t^4\,{\it x_0}-t^4\,{\it x_0}\right]%
 &\textrm{if } i=2 \textrm{ and } t\in [x_{1}\, , \, x_{2}[\\ 
\end{array}
\right.
$$
\end{tiny}


 $$
\frac{d}{dt}\Bspline{i}{2}{\alpha}(t) =
\left\lbrace
\begin{array}{ll}
\displaystyle{
 -\frac{\left(\alpha-1\right)^2\,\alpha\,\left({\it x_3}-t\right)}{\left(\alpha\,{\it x_3}-
 {\it x_3}-\alpha\,{\it x_1}+t\right)^2\,\left(\alpha\,{\it x_3}-
 {\it x_3}-\alpha\,{\it x_2}+t\right)^2}\times%
}
&\\
 \,\left[2\,
 \alpha\,{\it x_3}^2-2\,{\it x_3}^2-2\,\alpha\,{\it x_2}\,{\it x_3}+
 {\it x_2}\,{\it x_3}-2\,\alpha\,{\it x_1}\,{\it x_3}
 \right.
 &\\
 \left.
 +{\it x_1}\, {\it x_3}+2\,t\,{\it x_3}+2\,\alpha\,{\it x_1}\,{\it x_2}-t\,
 {\it x_2}-t\,{\it x_1}\right]%
 &\textrm{if } i=2 \textrm{ and } t\in [x_{2}\, , \, x_{3}[\\ 
 &\\
\displaystyle{
 -\frac{\left(\alpha-1\right)
 \,\alpha^2\,\left({\it x_1}-t\right)}{\left(\alpha\,{\it x_2}-{\it x_2}-\alpha\,
 {\it x_1}+t\right)^2\,\left(\alpha\,{\it x_3}-{\it x_3}-\alpha\,
 {\it x_1}+t\right)^2}\times%
}
&\\
 \,\left[2\,\alpha\,{\it x_2}\,
 {\it x_3}-2\,{\it x_2}\,{\it x_3}-2\,\alpha\,{\it x_1}\,{\it x_3}+
 {\it x_1}\,{\it x_3}+t\,{\it x_3}
 \right.
 &\\
 \left.
 -2\,\alpha\,{\it x_1}\,{\it x_2}+
 {\it x_1}\,{\it x_2}+t\,{\it x_2}+2\,\alpha\,{\it x_1}^2-2\,t\,
 {\it x_1}\right]%
 &\textrm{if } i=3 \textrm{ and } t\in [x_{1}\, , \, x_{2}[\\ 
 &\\
\displaystyle{
 \frac{\left(\alpha-1\right)\,\alpha}{\left(\alpha\,{\it x_3}-
 {\it x_3}-\alpha\,{\it x_1}+t\right)^2\,\left(\alpha\,{\it x_3}-
 {\it x_3}-\alpha\,{\it x_2}+t\right)^3}\times%
}
&\\
 \,\left[2\,
 \alpha^3\,{\it x_3}^4-6\,\alpha^2\,{\it x_3}^4+6\,\alpha\,{\it x_3}^
 4-2\,{\it x_3}^4-2\,\alpha^3\,{\it x_2}\,{\it x_3}^3
  \right.
 &\\
 \left.
+5\,\alpha^2\,
 {\it x_2}\,{\it x_3}^3
 \right.
 &\\
 \left.
 -4\,\alpha\,{\it x_2}\,{\it x_3}^3+{\it x_2}\,
 {\it x_3}^3-2\,\alpha^3\,{\it x_1}\,{\it x_3}^3+5\,\alpha^2\,
 {\it x_1}\,{\it x_3}^3-4\,\alpha\,{\it x_1}\,{\it x_3}^3
 \right.
 &\\
 \left.
 +{\it x_1}\, {\it x_3}^3-4\,\alpha^3\,t\,{\it x_3}^3+14\,\alpha^2\,t\,{\it x_3}^3
 -16\,\alpha\,t\,{\it x_3}^3+6\,t\,{\it x_3}^3
 \right.
 &\\
 \left.
 +\alpha^2\,{\it x_2}^2
 \,{\it x_3}^2
 \right.
 &\\
 \left.
 -\alpha\,{\it x_2}^2\,{\it x_3}^2-3\,\alpha^2\,
 {\it x_1}\,{\it x_2}\,{\it x_3}^2+3\,\alpha\,{\it x_1}\,{\it x_2}\,
 {\it x_3}^2+6\,\alpha^3\,t\,{\it x_2}\,{\it x_3}^2
 \right.
 &\\
 \left.
 -14\,\alpha^2\,t\,
 {\it x_2}\,{\it x_3}^2+11\,\alpha\,t\,{\it x_2}\,{\it x_3}^2-3\,t\,
 {\it x_2}\,{\it x_3}^2+6\,\alpha^3\,t\,{\it x_1}\,{\it x_3}^2
 \right.
 &\\
 \left.
 -12\, \alpha^2\,t\,{\it x_1}\,{\it x_3}^2+9\,\alpha\,t\,{\it x_1}\,
 {\it x_3}^2-3\,t\,{\it x_1}\,{\it x_3}^2-8\,\alpha^2\,t^2\,{\it x_3}^2
 \right.
 &\\
 \left.
 +14\,\alpha\,t^2\,{\it x_3}^2
 \right.
 &\\
 \left.
 -6\,t^2\,{\it x_3}^2+2\,\alpha^3\,
 {\it x_1}\,{\it x_2}^2\,{\it x_3}-2\,\alpha^2\,{\it x_1}\,{\it x_2}^
 2\,{\it x_3}-2\,\alpha^3\,t\,{\it x_2}^2\,{\it x_3}
 \right.
 &\\
 \left.
 +2\,\alpha\,t\, {\it x_2}^2\,{\it x_3}
 \right.
 &\\
 \left.
 +2\,\alpha^3\,{\it x_1}^2\,{\it x_2}\,
 {\it x_3}-8\,\alpha^3\,t\,{\it x_1}\,{\it x_2}\,{\it x_3}+10\,\alpha
 ^2\,t\,{\it x_1}\,{\it x_2}\,{\it x_3}
 \right.
 &\\
 \left.
 -6\,\alpha\,t\,{\it x_1}\, {\it x_2}\,{\it x_3}
 \right.
 &\\
 \left.
 +9\,\alpha^2\,t^2\,{\it x_2}\,{\it x_3}-10\,
 \alpha\,t^2\,{\it x_2}\,{\it x_3}+3\,t^2\,{\it x_2}\,{\it x_3}-2\,
 \alpha^3\,t\,{\it x_1}^2\,{\it x_3}
 \right.
 &\\
 \left.
 +7\,\alpha^2\,t^2\,{\it x_1}\,{\it x_3}
 \right.
 &\\
 \left.
 -6\,\alpha\,t^2\,{\it x_1}\,{\it x_3}+3\,t^2\,{\it x_1}\,
 {\it x_3}-4\,\alpha\,t^3\,{\it x_3}+2\,t^3\,{\it x_3}
 \right.
 &\\
 \left.
 -2\,\alpha^3\, {\it x_1}^2\,{\it x_2}^2
 \right.
 &\\
 \left.
 +2\,\alpha^3\,t\,{\it x_1}\,{\it x_2}^2+2\,
 \alpha^2\,t\,{\it x_1}\,{\it x_2}^2-\alpha^2\,t^2\,{\it x_2}^2-
 \alpha\,t^2\,{\it x_2}^2
 \right.
 &\\
 \left.
 +2\,\alpha^3\,t\,{\it x_1}^2\,{\it x_2}
 \right.
 &\\
 \left.
 -7\, \alpha^2\,t^2\,{\it x_1}\,{\it x_2}+3\,\alpha\,t^2\,{\it x_1}\,
 {\it x_2}+3\,\alpha\,t^3\,{\it x_2}-t^3\,{\it x_2}
 \right.
 &\\
 \left.
 +\alpha\,t^3\, {\it x_1}-t^3\,{\it x_1}\right] %
 &\textrm{if } i=3 \textrm{ and } t\in [x_{2}\, , \, x_{3}[\\ 
 &\\
\displaystyle{
 -{{2\,\left(\alpha-1
 \right)\,\alpha^2\,\left({\it x_2}-t\right)\,\left({\it x_3}-
 {\it x_2}\right)}\over{\left(\alpha\,{\it x_3}-{\it x_3}-\alpha\,
 {\it x_2}+t\right)^3}} 
}
&\textrm{if } i=4 \textrm{ and } t\in [x_{2}\, , \, x_{3}[\\ 
 &\\
 0 &\textrm{otherwise}\\ 
\end{array}
\right.
$$



$$
\lim_{t \to x_{1}}\Bspline{i}{2}{\alpha} (t)=
\left\lbrace
\begin{array}{ll}
{{\left(\alpha-1\right)\,\left({\it x_2}-
 {\it x_1}\right)}\over{\alpha\,{\it x_2}-{\it x_2}+{\it x_1}-\alpha
 \,{\it x_0}}}
&\textrm{if } i=1 \textrm{ and } t\in [x_{0}\, , \, x_{1}[\\ 
 {{\left(\alpha-1\right)\,\left({\it x_2}-{\it x_1}
 \right)}\over{\alpha\,{\it x_2}-{\it x_2}+{\it x_1}-\alpha\,
 {\it x_0}}}
&\textrm{if } i=1 \textrm{ and } t\in [x_{1}\, , \, x_{2}[\\ 
 {{\alpha\,\left({\it x_1}-{\it x_0}\right)}\over{
 \alpha\,{\it x_2}-{\it x_2}+{\it x_1}-\alpha\,{\it x_0}}}
&\textrm{if } i=2 \textrm{ and } t\in [x_{0}\, , \, x_{1}[\\ 
 {{\alpha
 \,\left({\it x_1}-{\it x_0}\right)}\over{\alpha\,{\it x_2}-{\it x_2}
 +{\it x_1}-\alpha\,{\it x_0}}}
&\textrm{if } i=2 \textrm{ and } t\in [x_{1}\, , \, x_{2}[\\ 
 0
&\textrm{otherwise }\\ 
\end{array} 
\right.
 $$

 $$
\lim_{t \to x_{1}}\frac{d}{dt}\Bspline{i}{2}{\alpha} (t)=
\left\lbrace
\begin{array}{ll}
-{{\left(\alpha-1\right)\,\left(2\,\alpha\,
 {\it x_2}-{\it x_2}+{\it x_1}-2\,\alpha\,{\it x_0}\right)}\over{
 \left(\alpha\,{\it x_2}-{\it x_2}+{\it x_1}-\alpha\,{\it x_0}\right)
 ^2}}
&\textrm{if } i=1 \textrm{ and } t\in [x_{0}\, , \, x_{1}[\\ 
 -{{\alpha\,\left(2\,\alpha\,{\it x_2}-2\,{\it x_2}+{\it x_1}-
 2\,\alpha\,{\it x_0}+{\it x_0}\right)}\over{\left(\alpha\,{\it x_2}-
 {\it x_2}+{\it x_1}-\alpha\,{\it x_0}\right)^2}}
&\textrm{if } i=1 \textrm{ and } t\in [x_{1}\, , \, x_{2}[\\ 
 {{\left(\alpha-1
 \right)\,\left(2\,\alpha\,{\it x_2}-{\it x_2}+{\it x_1}-2\,\alpha\,
 {\it x_0}\right)}\over{\left(\alpha\,{\it x_2}-{\it x_2}+{\it x_1}-
 \alpha\,{\it x_0}\right)^2}}
&\textrm{if } i=2 \textrm{ and } t\in [x_{0}\, , \, x_{1}[\\ 
 {{\alpha\,\left(2\,\alpha\,{\it x_2}-
 2\,{\it x_2}+{\it x_1}-2\,\alpha\,{\it x_0}+{\it x_0}\right)}\over{
 \left(\alpha\,{\it x_2}-{\it x_2}+{\it x_1}-\alpha\,{\it x_0}\right)
 ^2}}
&\textrm{if } i=2 \textrm{ and } t\in [x_{1}\, , \, x_{2}[\\ 
 0
&\textrm{otherwise }\\ 
\end{array} 
\right.
 $$


$$
\lim_{t \to x_{2}}\Bspline{i}{2}{\alpha} (t)=
\left\lbrace
\begin{array}{ll}
 {{\left(
 \alpha-1\right)\,\left({\it x_3}-{\it x_2}\right)}\over{\alpha\,
 {\it x_3}-{\it x_3}+{\it x_2}-\alpha\,{\it x_1}}}
&\textrm{if } i=2 \textrm{ and } t\in [x_{1}\, , \, x_{2}[\\ 
 {{\left(\alpha-1
 \right)\,\left({\it x_3}-{\it x_2}\right)}\over{\alpha\,{\it x_3}-
 {\it x_3}+{\it x_2}-\alpha\,{\it x_1}}}
&\textrm{if } i=2 \textrm{ and } t\in [x_{2}\, , \, x_{3}[\\ 
 {{\alpha\,\left({\it x_2}-
 {\it x_1}\right)}\over{\alpha\,{\it x_3}-{\it x_3}+{\it x_2}-\alpha
 \,{\it x_1}}}
&\textrm{if } i=3 \textrm{ and } t\in [x_{1}\, , \, x_{2}[\\ 
 {{\alpha\,\left({\it x_2}-{\it x_1}\right)}\over{
 \alpha\,{\it x_3}-{\it x_3}+{\it x_2}-\alpha\,{\it x_1}}}
&\textrm{if } i=3 \textrm{ and } t\in [x_{2}\, , \, x_{3}[\\ 
 0
&\textrm{otherwise }\\ 
\end{array} 
\right.
 $$

 $$
\lim_{t \to x_{2}}\frac{d}{dt}\Bspline{i}{2}{\alpha} (t)=
\left\lbrace
\begin{array}{ll}
 -{{\left(\alpha-1\right)\,\left(2\,\alpha\,{\it x_3}-
 {\it x_3}+{\it x_2}-2\,\alpha\,{\it x_1}\right)}\over{\left(\alpha\,
 {\it x_3}-{\it x_3}+{\it x_2}-\alpha\,{\it x_1}\right)^2}}
&\textrm{if } i=2 \textrm{ and } t\in [x_{1}\, , \, x_{2}[\\ 
 -{{
 \alpha\,\left(2\,\alpha\,{\it x_3}-2\,{\it x_3}+{\it x_2}-2\,\alpha
 \,{\it x_1}+{\it x_1}\right)}\over{\left(\alpha\,{\it x_3}-{\it x_3}
 +{\it x_2}-\alpha\,{\it x_1}\right)^2}}
&\textrm{if } i=2 \textrm{ and } t\in [x_{2}\, , \, x_{3}[\\ 
 {{\left(\alpha-1\right)\,
 \left(2\,\alpha\,{\it x_3}-{\it x_3}+{\it x_2}-2\,\alpha\,{\it x_1}
 \right)}\over{\left(\alpha\,{\it x_3}-{\it x_3}+{\it x_2}-\alpha\,
 {\it x_1}\right)^2}}
&\textrm{if } i=3 \textrm{ and } t\in [x_{1}\, , \, x_{2}[\\ 
 {{\alpha\,\left(2\,\alpha\,{\it x_3}-2\,
 {\it x_3}+{\it x_2}-2\,\alpha\,{\it x_1}+{\it x_1}\right)}\over{
 \left(\alpha\,{\it x_3}-{\it x_3}+{\it x_2}-\alpha\,{\it x_1}\right)
 ^2}}
&\textrm{if } i=3 \textrm{ and } t\in [x_{2}\, , \, x_{3}[\\ 
 0
&\textrm{otherwise }\\ 
\end{array} 
\right.
 $$

\newpage
\begin{center}
B-spline basis of degree 2 and its derivative
for the knot vector\\
$U=\left( x_{0},x_{0}, x_{0}, x_{1}, x_{1}, x_{2}, x_{2}, x_{2} \right)$
\end{center}

$$
\Bspline{i}{2}{\alpha} (t)=
\left\lbrace
\begin{array}{ll}
{{\left(\alpha-1\right)^2\,\left({\it x_1}-t\right)^
 2}\over{\left(\alpha\,{\it x_1}-{\it x_1}-\alpha\,{\it x_0}+t\right)
 ^2}}
&\textrm{if } i=0 \textrm{ and } t\in [x_{0}\, , \, x_{1}[\\ 
-{{2\,\left(\alpha-1\right)\,\alpha\,\left({\it x_0}-t
 \right)\,\left({\it x_1}-t\right)}\over{\left(\alpha\,{\it x_1}-
 {\it x_1}-\alpha\,{\it x_0}+t\right)^2}}
&\textrm{if } i=1 \textrm{ and } t\in [x_{0}\, , \, x_{1}[\\ 
{{\alpha^2\,\left(
 {\it x_0}-t\right)^2}\over{\left(\alpha\,{\it x_1}-{\it x_1}-\alpha
 \,{\it x_0}+t\right)^2}}
&\textrm{if } i=2 \textrm{ and } t\in [x_{0}\, , \, x_{1}[\\ 
{{\left(\alpha-1\right)^2\,\left(
 {\it x_2}-t\right)^2}\over{\left(\alpha\,{\it x_2}-{\it x_2}-\alpha
 \,{\it x_1}+t\right)^2}}
&\textrm{if } i=2 \textrm{ and } t\in [x_{1}\, , \, x_{2}[\\ 
-{{2\,\left(\alpha-1\right)\,\alpha\,
 \left({\it x_1}-t\right)\,\left({\it x_2}-t\right)}\over{\left(
 \alpha\,{\it x_2}-{\it x_2}-\alpha\,{\it x_1}+t\right)^2}}
&\textrm{if } i=3 \textrm{ and } t\in [x_{1}\, , \, x_{2}[\\ 
{{
 \alpha^2\,\left({\it x_1}-t\right)^2}\over{\left(\alpha\,{\it x_2}-
 {\it x_2}-\alpha\,{\it x_1}+t\right)^2}}
&\textrm{if } i=4 \textrm{ and } t\in [x_{1}\, , \, x_{2}[\\ 
0 
&\textrm{otherwise } \\ 
\end{array}
\right.
$$

$$
\frac{d}{dt}\Bspline{i}{2}{\alpha} (t)=
\left\lbrace
\begin{array}{ll}
-{{2\,\left(\alpha-1\right)^2\,\alpha\,\left(
 {\it x_1}-t\right)\,\left({\it x_1}-{\it x_0}\right)}\over{\left(
 \alpha\,{\it x_1}-{\it x_1}-\alpha\,{\it x_0}+t\right)^3}}
&\textrm{if } i=0 \textrm{ and } t\in [x_{0}\, , \, x_{1}[\\ 
{{2
 \,\left(\alpha-1\right)\,\alpha\,\left({\it x_1}-{\it x_0}\right)\,
 \left(\alpha\,{\it x_1}-{\it x_1}+\alpha\,{\it x_0}-2\,\alpha\,t+t
 \right)}\over{\left(\alpha\,{\it x_1}-{\it x_1}-\alpha\,{\it x_0}+t
 \right)^3}}
&\textrm{if } i=1 \textrm{ and } t\in [x_{0}\, , \, x_{1}[\\ 
-{{2\,\left(\alpha-1\right)\,\alpha^2\,\left(
 {\it x_0}-t\right)\,\left({\it x_1}-{\it x_0}\right)}\over{\left(
 \alpha\,{\it x_1}-{\it x_1}-\alpha\,{\it x_0}+t\right)^3}}
&\textrm{if } i=2 \textrm{ and } t\in [x_{0}\, , \, x_{1}[\\ 
-{{
 2\,\left(\alpha-1\right)^2\,\alpha\,\left({\it x_2}-t\right)\,\left(
 {\it x_2}-{\it x_1}\right)}\over{\left(\alpha\,{\it x_2}-{\it x_2}-
 \alpha\,{\it x_1}+t\right)^3}}
 &\textrm{if } i=2 \textrm{ and } t\in [x_{1}\, , \, x_{2}[\\ 
 {{2\,\left(\alpha-1\right)\,
 \alpha\,\left({\it x_2}-{\it x_1}\right)\,\left(\alpha\,{\it x_2}-
 {\it x_2}+\alpha\,{\it x_1}-2\,\alpha\,t+t\right)}\over{\left(\alpha
 \,{\it x_2}-{\it x_2}-\alpha\,{\it x_1}+t\right)^3}}
&\textrm{if } i=3 \textrm{ and } t\in [x_{1}\, , \, x_{2}[\\ 
-{{2\,
 \left(\alpha-1\right)\,\alpha^2\,\left({\it x_1}-t\right)\,\left(
 {\it x_2}-{\it x_1}\right)}\over{\left(\alpha\,{\it x_2}-{\it x_2}-
 \alpha\,{\it x_1}+t\right)^3}}
&\textrm{if } i=4 \textrm{ and } t\in [x_{1}\, , \, x_{2}[\\ 
0 
&\textrm{otherwise } \\ 
\end{array}
\right.
$$

$$
\lim_{t\to x_{1}}\Bspline{i}{2}{\alpha} (t)=
\left\lbrace
\begin{array}{ll}
1
&\textrm{if } i=2 \textrm{ and } t\in [x_{0}\, , \, x_{1}[\\ 
1
&\textrm{if } i=2 \textrm{ and } t\in [x_{1}\, , \, x_{2}[\\ 
0
&\textrm{otherwise } \\ 
\end{array}
\right.
$$

$$
\lim_{t\to x_{1}}\frac{d}{dt}\Bspline{i}{2}{\alpha} (t)=
\left\lbrace
\begin{array}{ll}
-{{2\,\left(\alpha-1\right)}\over{\alpha\,
 \left({\it x_1}-{\it x_0}\right)}}
&\textrm{if } i=1 \textrm{ and } t\in [x_{0}\, , \, x_{1}[\\ 
{{2\,\left(\alpha-1\right)
 }\over{\alpha\,\left({\it x_1}-{\it x_0}\right)}}
&\textrm{if } i=2 \textrm{ and } t\in [x_{0}\, , \, x_{1}[\\ 
-{{2\,\alpha
 }\over{\left(\alpha-1\right)\,\left({\it x_2}-{\it x_1}\right)}}
&\textrm{if } i=2 \textrm{ and } t\in [x_{1}\, , \, x_{2}[\\ 
 {{2\,\alpha}\over{\left(\alpha-1\right)\,\left({\it x_2}-
 {\it x_1}\right)}}
&\textrm{if } i=3 \textrm{ and } t\in [x_{1}\, , \, x_{2}[\\ 
0 
&\textrm{otherwise } \\ 
\end{array}
\right.
$$

\end{document}